\documentclass[12pt]{article}

\usepackage[margin=1in]{geometry}
\usepackage{setspace}
\usepackage{amsmath,amssymb,amsthm}
\newtheorem{theorem}{Theorem}[section]
\newtheorem{lemma}{Lemma}
\newtheorem{proposition}{Proposition}
\newtheorem{definition}{Definition}[section]
\newtheorem{assumption}{Assumption}

\usepackage{booktabs}
\usepackage{enumitem}
\usepackage{graphicx}
\usepackage[para,online,flushleft]{threeparttable}
\usepackage{float}
\usepackage{algorithm}
\usepackage{algpseudocode}
\usepackage[section]{placeins}

\usepackage{xcolor}
\definecolor{sangre}{rgb}{0.6,0.18,0.19}
\definecolor{darkblue}{rgb}{0,0,0.6}
\definecolor{violet}{cmyk}{0.79,0.88,0,0}
\usepackage{natbib}
\usepackage[
    colorlinks=true,
    urlcolor=darkblue,
    citecolor=sangre,
    linkcolor=blue
]{hyperref}

\makeatletter
\newcommand\EightPtClose{\@setfontsize\EightPtClose\@viiipt{9}}
\newcommand\notesize{\EightPtClose}
\newenvironment{figurenotes}[1][Note]
  {\begin{minipage}[t]{\linewidth}\notesize{\itshape#1: }}
  {\end{minipage}}
\makeatother

\newcommand{\E}{\mathbb{E}}
\newcommand{\R}{\mathbb{R}}
\newcommand{\convp}{\xrightarrow{p}}

\newcommand{\bk}{\boldsymbol{\kappa}}
\newcommand{\bkz}{\boldsymbol{\kappa}^{0}}
\newcommand{\bks}{\boldsymbol{\kappa}^{*}}
\newcommand{\bksw}{\boldsymbol{\kappa}^{**}}
\newcommand{\cT}{\mathcal{T}}
\newcommand{\cC}{\mathcal{C}}
\newcommand{\fTpre}{\mathbb{T}_{\mathrm{pre}}}
\newcommand{\fTpost}{\mathbb{T}_{\mathrm{post}}}
\newcommand{\Wmat}{\mathcal{W}}
\newcommand{\What}{\widehat{\mathcal{W}}}
\newcommand{\G}{\mathbf{G}}
\newcommand{\Gw}{\mathbf{G}_{w}}
\newcommand{\Gamt}{\boldsymbol{\Gamma}_{mt}}
\newcommand{\mumt}{\boldsymbol{\varpi}_{m}}
\newcommand{\LamT}{\boldsymbol{\Lambda}^{T}}
\newcommand{\delt}{\boldsymbol{\delta}_{mt}}
\newcommand{\dbar}{\bar{\boldsymbol{\delta}}}
\newcommand{\dtil}{\widetilde{\boldsymbol{\delta}}}
\newcommand{\gbar}{\bar{g}}
\newcommand{\gtil}{\widetilde{g}}
\newcommand{\mup}{\mu_{\mathrm{GMM}}}
\newcommand{\muw}{\mu_{S-GMM}}
\usepackage{pdflscape}
\begin{document}
\title{Bridging Quasi-Experimental and Structural Approaches for Robust Evaluation of US Airline Mergers\thanks{\scriptsize We thank Dennis Carlton, Mark Israel, Ian MacSwain, and Eugene Orlov for sharing their codes. We also thank Charles Murry, Victor Aguirregabiria, Alex Gross, Chun-Yu Ho, John Kwoka, Robin Lee, Elie Tamer, Christine Zulehner, Alberto Abadie, Matthew Backus, Robert Clark, Brantly Callaway, Devesh Raval, Gloria Sheu, \'Aureo de Paula, Valentin Verdier, Xiaoxia Shi, and the audience at the IIOC 2024 and 2026, 2025 Workshop on Econometrics and Models of Strategic Interactions (Oxford),  28th Air Transport Research Society World Conference, EEA 2025, EARIE 2025, APIOC 2025, 2026 Triangle Micro Conference, 2026 UChicago Interactions Conference, Mississippi, Bristol, 2026 KU Leuven Summer IO, Warwick, 2026 Hal White Antitrust Conference, EEA-ESEM 2026, 7th DC-MD-VA Econometrics Workshop. Anirban is an Associate at Analysis Group. The views presented are those of the authors and do not represent any opinions or positions of Analysis Group.}}
\author{Gaurab Aryal\thanks{ Department of Economics, Boston University, \href{mailto:aryalg@bu.edu}{aryalg@bu.edu}}
\and Anirban Chattopadhyaya\thanks{Analysis Group, \href{mailto: anirban.chattopadhyaya@analysisgroup.com}{anirban.chattopadhyaya@analysisgroup.com}}
\and {Federico Ciliberto\thanks{ Department of Economics, University of Virginia, DIW and CEPR, \href{mailto: ciliberto@virginia.edu}{ ciliberto@virginia.edu}}
}}

\date{\today}
\maketitle
\begin{abstract}
\noindent We bridge quasi-experimental and structural approaches to merger retrospectives. First, we derive the difference-in-differences (DiD) equation as a reduced form of a demand-and-supply model, decomposing price effects into changes in efficiency and strategic conduct. We show how pre-merger conduct affects parallel trends and when DiD effects generalize beyond treated markets. Second, we introduce $\textit{Synthetic GMM}$, using synthetic-DiD weights to reweight controls, reducing bias relative to standard GMM. Monte Carlo experiments support this result. Across four airline mergers, we find no efficiency gains, no systematic conduct changes, and no price reductions. In one merger, more competitive conduct partly offsets rising costs.
\end{abstract}

\thispagestyle{empty}
\doublespacing
\setcounter{page}{0}

\section{Introduction}
Merger evaluation is an important topic of economics research. Rising concentration and firm sizes across industries \citep{AutorDornKatzPattersonReenen2020} have renewed concerns about market power and increased the importance of understanding how mergers affect competition \citep{Shapiro2018}.
 A merger can generate efficiencies that lower marginal cost \citep{Williamson1968, AshenfelterHoskenWeinberg2015, GriecoPinskeSlade2018, DemirerKaraduman2026}, but it can also increase market power by reducing competition and facilitating coordination \citep{MillerWeinberg2017, Porter2020, MillerSheuWeinberg2021}. A merger retrospective, the analysis of a merger after it has taken place, therefore has to answer three questions: whether it generated efficiencies, whether it increased market power, and what was its net effect on prices \citep{FarrellShapiro1990}.

Merger retrospectives rely on two main approaches to answer these questions. The quasi-experimental approach, in particular the difference-in-differences (DiD) design, compares price changes in merger-affected markets with prices in unaffected markets \citep[e.g.,][]{KimSingal1993, AshenfelterHosken2010, CarltonIsraelMacSwainOrlov2019}. DiD estimates the merger's net price effect under a parallel-trends assumption, but it cannot distinguish efficiency gains from changes in market power. The structural approach instead specifies demand, costs, and firm competitive behavior (i.e., conduct), and uses instruments and appropriate comparison markets to identify price effects and underlying economic mechanisms.

These approaches are complementary but have largely developed separately. DiD designs estimate effects in treated markets, while structural methods provide economic interpretation and let researchers answer questions beyond the affected markets. Bridging these strengths has long been an objective of empirical economics \citep{Chetty2009, Heckman2010}, including empirical industrial organization \citep{EinavLevin2010}. Bridging is particularly valuable in merger retrospectives, which also aim to inform future merger enforcement. Knowing the price effect alone may not be enough to determine a merger's effect on competition, particularly if it changes costs and market power, potentially offsetting each other.

We further this research agenda by building a methodological two-way bridge between these approaches to give economic content to DiD estimates and to improve structural estimation by selecting comparison markets. Our First Bridge derives the DiD equation as a reduced form of either a linear or log-linear demand-and-supply model. Because prices are equilibrium outcomes, we show that the DiD estimate of the price effect equals the merger-induced change in marginal cost scaled by the cost pass-through rate, plus any additional effect from changes in strategic conduct.  

This reduced-form derivation also makes explicit the economic restrictions implicit in the DiD design, including the role of the functional form. While \cite{RothSantanna2023} made precise how parallel trends in prices and parallel trends in log prices differ, we show how these restrictions map into the underlying economic primitives. For instance, in the linear model, differences in pre-merger conduct between treated and untreated markets can violate parallel trends even when the underlying demand and cost shocks satisfy parallel trends. DiD therefore embeds restrictions on pre-merger conduct that are otherwise implicit in the parallel-trends assumption. This matters when a merger has a coordination effect, as it was thought to be the case in the Department of Justice's challenge to the AA--US merger \citep[][Section C]{DOJ2013}.

Finally, while DiD identifies the merger effect on prices in treated markets,
it may not inform about price effects in other markets.
In the linear model, the price effect varies across markets, and when the merger affects conduct, the DiD estimates do not extend outside the treated markets. In the log-linear model the estimate extends because the constant elasticity implies a homogeneous merger effect across markets. More generally, using the marginal-treatment-effect framework \citep{HeckmanVytlacil2005}, we characterize
the additional restrictions required to extrapolate the DiD effect beyond the treated markets
when pass-through and conduct effects are heterogeneous. 

The structural approach addresses these limitations by identifying the underlying demand and supply parameters using excluded demand and cost shifters. It separately estimates merger-induced changes in marginal cost and strategic conduct, allowing us to recover their contributions on the price. It also permits the evaluation of merger effects beyond treated markets, provided the estimated demand and supply parameters apply to those markets.

Our Second Bridge draws on recent advances in quasi-experimental approaches to improve the structural approach. The structural approach offers a richer evaluation of merger effects, but its credibility depends on the choice of comparison markets. Merger applications typically have many more untreated than treated markets. However, being unaffected by the merger does not guarantee that a market has valid information for structural estimation. Researchers often justify instrument validity in treated markets without establishing it throughout the much larger pool of untreated markets.\footnote{We use \emph{untreated markets} for all markets not affected by the merger and \emph{control markets} for the untreated markets used as comparisons. The Second Bridge aims to select a set of good control markets.} Selecting appropriate comparable markets is difficult because markets similar to the treated markets may lack independent identifying variation, and those that do may be too dissimilar, i.e., the similarity--independence tension   \citep{AshenfelterHoskenWeinberg2009}. 

We propose using synthetic DiD (SDiD) weights \citep{ArkhangelskyAtheyHirshbergImbensWager2021} based on price dynamics to choose comparison markets. We use synthetic DiD weights because they select comparison markets based on price dynamics while allowing for persistent differences in price levels. This is useful in our setting because identification relies on changes in prices. Consequently, we replace the uniform weights standard in GMM with data-driven SDiD weights in the moment conditions, yielding \emph{Synthetic GMM}.

We establish the properties of Synthetic GMM for the linear model with linear factor representation \citep[e.g.,][]{AbadieDiamondHainmueller2010} of the cost shocks. The factor structure provides a tractable way to parameterize failures of the moment conditions in \emph{un}treated markets and relate those failures to price dynamics.\footnote{\cite{Chen2023} shows that the synthetic control method can be interpreted as repeatedly choosing weights to minimize past prediction errors. This interpretation further motivates our weighting approach.} In this way, we formalize the above-mentioned similarity--independence tradeoff. If the instruments are such that the moment conditions hold in all untreated markets, both standard GMM and synthetic GMM are consistent. When they fail for some untreated markets, standard GMM is biased, whereas synthetic GMM reduces this bias, which vanishes as the pre-merger sample grows. 

We validate these results using Monte Carlo exercises by varying the share
of untreated markets where the supply-side moment conditions fail and assessing the performance of Synthetic GMM and standard GMM.
As this share increases, standard GMM becomes increasingly biased,
while synthetic GMM has substantially smaller bias.\footnote{We focus on only supply-side and not demand-side moment conditions because the merger effects of interest, i.e., changes in marginal cost,
conduct, and cost pass-through, operate through the supply side. Furthermore, because we have only a few treated markets, for inference we use placebo inference 
\citep{ConleyTaber2011}. In the simulations, the placebo test has
rejection rates close to the nominal 5\% level.} 
We also consider a design in which price dynamics are uninformative about the validity of the
moment conditions. In this case, Synthetic GMM offers no improvement over GMM. Thus, the gains from Synthetic GMM depend on how informative weights are about instrument invalidity.

We apply our framework to four U.S. airline mergers: Delta--Northwest (DL--NW; 2008), United--Continental (UA--CO; 2010), Southwest--AirTran (WN--FL; 2011), and American--US Airways (AA--US; 2013). We study each merger separately, using nonstop and connecting markets, and estimate its effects on fares, marginal costs, and conduct.

The DiD estimates provide no evidence that any of the four mergers reduced fares. The conventional fixed-effects estimator produces mixed results, including fare reductions for some mergers and samples. Fares, however, are highly persistent within markets, so parallel trends in levels are not credible, and the first-difference estimates suggest that none of the four mergers led to a fare reduction. Instead, we find that the fares increased after the WN--FL merger by approximately \$6--\$7 per ticket (the average pre-merger fare in treated markets is \$152), although the effect is imprecisely estimated.

The structural estimates give a similar picture of the net price effects: neither provides evidence of price reductions. Furthermore, the structural estimate finds no evidence of merger-induced efficiency gains and little evidence of systematic changes in conduct. We allow the effect on conduct to depend on the number of pre-merger competitors because the scope for coordination may depend on market structure \citep{Harrington2017}. For AA--US, the estimates suggest a larger reduction in competitive conduct in markets with fewer competitors, a pattern consistent with the DOJ's concern about coordination effects of merger. However, the estimates are too imprecise to establish such an effect. 

However, the WN--FL merger led to an \emph{increase} in marginal cost by approximately \$38 using GMM and \$27 using Synthetic GMM, and even though conduct became more competitive, fares increased by \$10.\footnote{Part of the WN--FL fare increase may reflect differences in pricing practices. FL charged separately for baggage and reservation changes, which can total up to \$12.70 per trip, while WN bundled them in the fare.\label{footnote:WN_bag}} Synthetic GMM estimates are similar to standard GMM for DL--NW and AA--US. For UA--CO, however, the weights concentrate on a small set of untreated markets, but insufficient identifying variation produces an unstable cost pass-through estimate. This illustrates the tradeoff between selecting more comparable markets and preserving sufficient variation for structural identification.

Our paper contributes to merger analysis broadly \citep{AskerNocke2021} and to airline mergers in particular \citep{Borenstein1990, KimSingal1993, Peters2006, OlleyTown2018, CarltonIsraelMacSwainOrlov2019, Das2019, LiMazurParkRobertsSweetingZhang2022, EizenbergZvuluni2024, OlsenOrchinikRemer2024, BrueggeGowrisankaranGross2024, Leyden2026, KimMazurParkYoon2026}. Empirically, we provide a common evaluation of four major airline mergers that considers both their price effects and the underlying mechanism. Methodologically, we contribute to the debate over identification and economic mechanism \citep{Bresnahan1989, NevoWhinston2010} by showing how bridging allows for a robust merger retrospective: the structural approach gives the DiD estimand an economic interpretation and permits extrapolation beyond treated markets, while synthetic DiD weights provide a way to select proper comparison markets for structural estimation.

The bridge thus also helps put merger retrospectives on stronger empirical footing. It makes explicit what each approach identifies, and uses one to improve the other. These issues are increasingly important for merger analysis as merger enforcement evolves \citep{Kwoka2014, ScottMorton2019, RoseShapiro2022, Kaplow2024}, but the core idea applies beyond mergers.

\section{Data\label{sec:data}}

We combine data from three sources. First, the U.S. Department of Transportation's Origin and Destination Survey (DB1B) is a quarterly 10\% sample of airline tickets and provides fares and passenger counts by carrier and route. Fares are measured per one-way trip, expressed in 2012Q2 dollars, and aggregated to the market-quarter level using passenger weights. Second, the Bureau of Transportation Statistics' T-100 (Form 41) data provide airline capacity. Third, the Official Airline Guide (OAG) provides nonstop flight schedules.

We study the last four major mergers in the US airline industry. Delta and Northwest Airlines (DL--NW ) announced their
merger plan on April 14, 2008; the shareholders approved it on September 26, 2008, and the DOJ approved it on October 29, 2008. United and Continental (UA--CO) announced their merger plan on May 2, 2010, which its shareholders approved on September 17, 2010. The merger was completed on October 1, 2010,
forming United Continental Holdings. However, the full operational integration
was completed only in mid-2012.
Southwest Airlines announced its
acquisition of AirTran Airways (WN--FL ) on September 27, 2010, and completed
the purchase on May 2, 2011. However, they continued to operate
separately until the FAA granted a single operating certificate on March 1,
2012, and the AirTran brand retired at the end of 2014.
Finally, the merger between American Airlines and US
Airways (AA--US  ) was announced on February 14, 2013. The DOJ filed a lawsuit against the merger on August 13, 2013, but a settlement was reached on November 12, and the merger was
completed on December 9, 2013.

We follow the literature to define pre- and post-merger periods and treated and untreated markets.
For each merger, we define the \textit{pre-} and \textit{post-merger} periods as the eight quarters preceding and following the merger event used for treatment timing, excluding the event quarter itself. Specifically,
for DL--NW, these are Q4 2006-Q3 2008 (pre) and Q1 2009-Q4 2010 (post); for
UA--CO, these are Q3 2008-Q2 2010 (pre) and Q4 2010-Q3 2012 (post); for WN--FL,
these are Q1 2010-Q4 2011 (pre) and Q2 2012-Q1 2014 (post); and for AA--US,
these are Q4 2011-Q3 2013 (pre) and Q1 2014-Q4 2015 (post).

Next, we define a market as a city pair and, unlike most papers in the literature that focus only on nonstop flights, we consider both nonstop and connecting flights.
A market is a \emph{treated market} if both merging airlines served it pre-merger, so the merger removes one of its competitors.
A treated market is a nonstop treated market if both parties
served it nonstop before the merger and if they do not have substantial connecting presence. We say that merging airlines have a \emph{substantial connecting presence} in a market if,
pre-merger, each carried at least 10\% of the market's passengers and together they carried at least 40\%. Likewise, a treated market is a connecting treated market if the
merging airlines had a substantial connecting presence and did not provide nonstop services. So, a treated market is either nonstop or connecting, but not both.

A market is classified as an \emph{untreated market} (or, when appropriate, a control market) if the merger does not affect its market structure. Specifically, an untreated market contains at most one of the merging airlines and has no substantial connecting presence by the merging airlines. It must also not be a treated market in any of the other three mergers.\footnote{A market with connecting service is classified as untreated only if (a) it carried at least 20 passengers per day in each direction during the pre-merger period, (b) at least one carrier accounting for 10\% or more of its passengers did not provide nonstop service, and (c) at most one of the merging airlines accounted for 10\% or more of its passengers. We apply the same passenger threshold to define connecting treated markets.}

Under these definitions, the DL–NW, UA–CO, WN–FL, and AA–US mergers have 7, 3, 11, and 6 nonstop treated markets, respectively, and 42, 8, 40, and 40 connecting treated markets. The corresponding numbers of untreated markets are 412, 118, 395, and 384 for nonstop service and 812, 407, 851, and 900 for connecting service.

We compare treated and untreated markets that have the same market structure pre-merger.
For instance, treated markets that went from ``three-to-two'' firms or ``two-to-one'' firms are compared with untreated markets with three or two firms pre-merger, respectively. The \emph{combined} sample pools the nonstop and
connecting treated markets.

\begin{table}[t!!]
\centering
\caption{Summary statistics by market type}
\label{tab:summary_stats}
\begin{threeparttable}
\begin{tabular}{lcccc}
\toprule
& \multicolumn{2}{c}{Nonstop} & \multicolumn{2}{c}{Combined} \\
\cmidrule(lr){2-3}\cmidrule(lr){4-5}
& Pre & Post & Pre & Post \\
\midrule
\textbf{Average Price} \\
\quad Treated    & 200.99 & 204.04 & 208.85 & 220.58 \\
\quad Untreated  & 195.72 & 198.98 & 195.72 & 198.98 \\
\quad All        & 195.83 & 199.08 & 197.13 & 201.30 \\
\textbf{Passengers} \\
\quad Treated    & 86,709 & 84,297 & 29,212 & 27,394 \\
\quad Untreated  & 70,445 & 70,228 & 70,445 & 70,228 \\
\quad All        & 70,774 & 70,513 & 66,015 & 65,631 \\
\midrule
\textit{Treated market-quarters} & 216 & 216 & 1,256 & 1,255 \\
\bottomrule
\end{tabular}
\begin{figurenotes}
Means of unweighted market-quarter
observations, pooled across the four mergers. Fares are in 2012Q2 dollars. A market is nonstop treated if
both merging airlines served it nonstop; the combined sample adds the treated markets with connection, in which both
have substantial connecting presence (as defined in the text). Untreated markets are the same in both samples, so those rows repeat.
Fares and passengers count all itineraries serving the market.
\end{figurenotes}
\end{threeparttable}
\end{table}

Table \ref{tab:summary_stats} summarizes average ticket prices and passenger
volume by market type. The table also shows the average before and after the merger, by treatment status. These averages are unweighted market-quarter observations, pooled across all mergers.

In nonstop treated markets, the mean fare rises by \$3.04, compared with \$3.26 in untreated markets, a raw difference-in-differences of about -\$0.22. Pooling nonstop and connecting treated markets, the mean fare rises by \$11.73 versus \$3.26, a raw difference-in-differences of \$8.47. Fares thus
appear to rise by more once the treated markets with connections are included, that is, where the merging airlines had competed through
connections than where they had competed nonstop.

\paragraph{Additional Data.}
To estimate a demand-and-supply model (Section \ref{sec:structural_retrospective}), we need more than prices and quantities (passenger volume); we also need demand and cost shifters.
 To this end, we supplement the DB1B data with endpoint population and income from Census/BEA county aggregates matched to airport cities.
We also use the DB1B carrier-route database and construct network connectivity
(the number of distinct markets an airline serves from a city) and the set of potential entrants (whether a carrier serves both endpoints of a market but not the market
itself); see Appendix \ref{sec:appendix_iv_stats}. We use these variables, along with market size, to construct the price and cost instruments used for model estimation (Section \ref{sec:structural_estimates}). 

\section{Quasi-Experimental Approach \label{sec:retrospective}}

We begin by studying the net effect of a merger on prices, and use the difference-in-differences (DiD) design to assess the impact of four airline mergers
(DL--NW , UA--CO, WN--FL  and AA--US ) on ticket fares.
The DiD approach compares the before-and-after change in average (market-level) prices in markets affected by the merger
against the same change in markets that were not, and interprets the difference as the average effect of the
merger on prices in the treated markets. Our preferred sample includes both nonstop and connecting services, but because mergers may affect them differently, we also estimate the effects separately for those subsamples.

Our baseline estimation equation is
\begin{eqnarray}
Y_{mt} =  \beta_0 + \beta^{\texttt{DiD}}\times I_{mt} + \digamma_m + \digamma_t + \varepsilon_{mt},\label{eq:did}
\end{eqnarray}
where $Y_{mt}$ is the outcome variable, which will be either the average price, $P_{mt}$, or the log of the average price, $\log P_{mt}$, in market $m$ and year-quarter $t$, and $I_{mt}:= \text{Treated}_m \times \text{Post-Merger}_{mt}$ is the treatment indicator with $\text{Treated}_{m}$ equals one if $m$ is a treated market and zero otherwise, and $\text{Post-Merger}_{mt}$ equals one if $t$ is after a merger and zero otherwise.
The specification includes market fixed effects, $\digamma_m$, and time fixed effects, $\digamma_t$.

Identifying the effect of a merger on prices in the treated market relies on three assumptions.
\textit{No anticipation} assumption implies that prices are not affected by anticipation of future mergers.
\textit{Stable Unit Treatment Value
Assumption} (SUTVA) requires that potential outcomes for market $m$ depend only on $m$'s treatment status, ruling
out spillover across markets. 
\textit{Parallel trends} assumption requires that the prices in the treated and the
untreated markets would have followed the same trend in the absence of the
merger. Using the standard notation in the potential-outcomes framework, the parallel trends assumption implies $\mathbb E[P_{m,\mathrm{post}}(0)-P_{m,\mathrm{pre}}(0)\mid\mathrm{Treated}_m=1]=\mathbb E[P_{m,\mathrm{post}}(0)-P_{m,\mathrm{pre}}(0)\mid\mathrm{Treated}_m=0]$, which in terms of the regression error implies  
\begin{eqnarray}
\mathbb{E}[\varepsilon_{m,\text{post}} - \varepsilon_{m,\text{pre}} \mid \text{Treated}_m = 1]
= \mathbb{E}[\varepsilon_{m,\text{post}} - \varepsilon_{m,\text{pre}} \mid \text{Treated}_m = 0].\label{eq:parallel_trends}
\end{eqnarray}
Under these assumptions, the coefficient $\beta^{\texttt{DiD}}$ in (\ref{eq:did}) recovers the average
merger effect on prices in treated markets, the average treatment effect on the treated (ATT), i.e., 
\begin{eqnarray}
\beta^{\texttt{DiD}} = \mathbb{E}(P_{t,\text{post}} -  P_{t,\text{pre}} \mid I_m=1) - \mathbb{E}( P_{t,\text{post}} -  P_{t,\text{pre}} \mid I_m=0). \label{eq:did_definition}
\end{eqnarray}
Similarly, for the effect of merger on log of fares, we can replace prices in the parallel trends assumption \citep{RothSantanna2023} and in \eqref{eq:did_definition} with log prices.

\paragraph{Fixed-Effects Estimates.}
Table \ref{tab:did_main} reports the estimated DiD coefficients with their
standard errors using both fare levels and log fares. We begin with
fare levels. Focusing on the combined
sample, our preferred sample because it includes treated markets with either
nonstop or connecting service, the baseline FE estimates imply that average
fares fell by about \$10 after DL--NW, but rose by about \$6 after UA--CO and
\$13 after WN--FL. The AA--US estimate is negative but imprecisely estimated.
Adding the demand-and-supply controls attenuates the positive estimate for WN--FL and roughly halves the estimated fare decline following DL--NW, while the estimates for UA--CO and AA--US change little.

Table \ref{tab:did_main} also presents estimates for the same specifications using log
fares as the outcome. The results give a qualitatively similar picture. In the
combined sample, the FE estimates imply that fares fell by about 5\% following
DL--NW and increased by about 9\% following WN--FL, while the effects of
UA--CO and AA--US mergers are imprecisely estimated. For comparability with the
existing airline-merger literature, including
\citet{CarltonIsraelMacSwainOrlov2019}, and for brevity, we focus the
discussion that follows on the log-fare specification.
For completeness, Table \ref{tab:did_main} reports estimates for both specifications.

\paragraph{Nonstationarity of Prices.} DiD implicitly
assumes the two groups carry no differential time trend.
However, airline fares are equilibrium objects that firms set to maximize
profits, and in the run-up to a merger the underlying cost and demand
conditions plausibly move in systematic ways such that the prices are nonstationary.
Indeed, we find that even after removing market means, log fares have first-order autocorrelation of roughly $0.6$ to $0.7$ across different samples and mergers, while their first
differences are close to white noise.\footnote{Although not reported, formal pre-trend tests reject parallel trends in levels in most every merger sample.}

\begin{table}[ht!]
\centering
\caption{DiD Estimates of Merger Effects on Fares\label{tab:did_main}}
\scalebox{0.91}{\begin{threeparttable}
\begin{tabular}{llcccc}
\toprule
& &
\multicolumn{2}{c}{Fares} &
\multicolumn{2}{c}{Log fares} \\
\cmidrule(lr){3-4}\cmidrule(lr){5-6}
Merger & & Combined & Nonstop & Combined & Nonstop \\
[0.1em]\\
\multicolumn{6}{l}{\emph{(a) Baseline}} \\
&&&&& \\
DL--NW
& FE & $-10.07$ $(4.58)$ & $-8.07$ $(5.27)$ & $-0.053$ $(0.019)$ & $-0.044$ $(0.021)$ \\
& FD & $-3.39$ $(5.79)$ & $0.14$ $(7.07)$ & $-0.014$ $(0.025)$ & $0.008$ $(0.027)$ \\
& & $[-14.27, 5.87]$ & $[-20.48, 20.09]$ & $[-0.065, 0.029]$ & $[-0.086, 0.101]$ \\[0.3em]
UA--CO
& FE & $5.76$ $(2.45)$ & $9.01$ $(2.26)$ & $-0.001$ $(0.011)$ & $0.009$ $(0.010)$ \\
& FD & $0.62$ $(2.58)$ & $0.79$ $(3.06)$ & $-0.007$ $(0.011)$ & $-0.007$ $(0.013)$ \\
& & $[-15.94, 15.14]$ & $[-20.57, 19.13]$ & $[-0.089, 0.066]$ & $[-0.112, 0.083]$ \\[0.3em]
WN--FL
& FE & $12.57$ $(1.99)$ & $9.87$ $(2.38)$ & $0.088$ $(0.011)$ & $0.077$ $(0.014)$ \\
& FD & $6.89$ $(2.18)$ & $2.94$ $(3.03)$ & $0.043$ $(0.013)$ & $0.025$ $(0.018)$ \\
& & $[-1.83, 14.48]$ & $[-10.29, 17.47]$ & $[-0.001, 0.082]$ & $[-0.045, 0.103]$ \\[0.3em]
AA--US
& FE & $-4.89$ $(5.49)$ & $-13.26$ $(8.16)$ & $-0.027$ $(0.025)$ & $-0.064$ $(0.038)$ \\
& FD & $-0.09$ $(7.11)$ & $-2.66$ $(12.71)$ & $0.001$ $(0.034)$ & $-0.009$ $(0.061)$ \\
& & $[-10.97, 9.32]$ & $[-20.87, 15.07]$ & $[-0.053, 0.047]$ & $[-0.111, 0.089]$ \\
[0.1em]\\
\multicolumn{6}{l}{\emph{(b) With additional controls}} \\
&&&&&\\
DL--NW
& FE & $-5.79$ $(5.17)$ & $-1.51$ $(6.62)$ & $-0.031$ $(0.023)$ & $-0.010$ $(0.030)$ \\
& FD & $-1.37$ $(5.95)$ & $2.79$ $(7.29)$ & $-0.004$ $(0.026)$ & $0.021$ $(0.028)$ \\
& & $[-12.28, 8.09]$ & $[-14.99, 19.84]$ & $[-0.055, 0.040]$ & $[-0.066, 0.107]$ \\[0.3em]
UA--CO
& FE & $6.46$ $(2.54)$ & $9.63$ $(2.76)$ & $0.002$ $(0.012)$ & $0.013$ $(0.014)$ \\
& FD & $0.42$ $(2.45)$ & $0.05$ $(2.82)$ & $-0.009$ $(0.010)$ & $-0.013$ $(0.012)$ \\
& & $[-15.64, 14.84]$ & $[-22.11, 19.41]$ & $[-0.087, 0.062]$ & $[-0.119, 0.079]$ \\[0.3em]
WN--FL
& FE & $9.02$ $(1.92)$ & $8.31$ $(2.53)$ & $0.071$ $(0.011)$ & $0.069$ $(0.015)$ \\
& FD & $6.38$ $(2.28)$ & $3.35$ $(3.53)$ & $0.041$ $(0.013)$ & $0.027$ $(0.021)$ \\
& & $[-2.66, 14.25]$ & $[-8.96, 16.77]$ & $[-0.003, 0.079]$ & $[-0.041, 0.102]$ \\[0.3em]
AA--US
& FE & $-5.09$ $(4.76)$ & $-10.92$ $(7.04)$ & $-0.028$ $(0.021)$ & $-0.051$ $(0.032)$ \\
& FD & $-0.15$ $(6.80)$ & $-1.13$ $(12.30)$ & $0.000$ $(0.033)$ & $-0.001$ $(0.059)$ \\
& & $[-11.40, 9.57]$ & $[-18.81, 16.46]$ & $[-0.054, 0.047]$ & $[-0.096, 0.092]$ \\
\bottomrule
\end{tabular}
\begin{figurenotes}DiD estimates of merger effects on fares and log fares defined in Equation (\ref{eq:did}), for the combined and nonstop samples. Panel (a) is passenger-weighted, includes time effects, and clusters standard errors by city-route; Panel (b) additionally includes demand-and-supply  controls. FE denotes the fixed-effects specification in levels; FD denotes the first-difference specification, with the average effect over the first eight post-merger quarters defined in Equation (\ref{eq:DiD_fd}). Standard errors are in parentheses; brackets report 95 percent placebo intervals for the FD estimates based on $B=100$ assignments of the merger to untreated markets, excluding the actual treated markets (Appendix \ref{sec:few_treated_inference}). For UA--CO, the combined sample excludes 37 connecting and 2 nonstop control markets that are two-carrier controls in the other mode but have no corresponding two-to-one merger overlap.
\end{figurenotes}
\end{threeparttable}}
\end{table}

\paragraph{First-Difference Estimates.} To implement the first-difference (FD)
approach, we start from the DiD specification
(\ref{eq:did}) and take its first difference in time before estimating.
Differencing removes the market fixed effect $\digamma_m$ and reduces the persistence of the outcome, which can improve the statistical properties of the estimator in our setting \citep{GrilichesMairesse1999}. Identification nevertheless requires restrictions on the evolution of untreated potential outcomes, now expressed through the differenced specification.

Furthermore,  in its standard form, first differencing captures only the immediate effect. The
first difference of the treatment indicator $\Delta I_{mt}$ equals one only in
the quarter immediately after the merger, so it captures only the immediate effect.
In mergers, however, integrating
operations and reoptimizing can take time and cost efficiencies unfold
gradually---a pattern documented across several industries
\citep{FocarelliPanetta2003, GuglerSiebert2007, AshenfelterHosken2010,
AshenfelterHoskenWeinberg2013} and likely to hold for airlines as well. To
capture the gradual realization of merger-induced efficiency, we estimate
\begin{eqnarray*}
Y_{mt} = \beta_0 + \sum_{k>0}^{T}\beta^{\texttt{DiD}}_k {\mathcal I}_{m,t,k} + \digamma_m + \digamma_t + \varepsilon_{mt},
\end{eqnarray*}
where, with a slight abuse of notation, $ {\mathcal I}_{m,t,k} = \text{Post-Merger}_{t,k} \times \text{Treated}_m$ and
$\text{Post-Merger}_{t,k}$ equals one if by period $t$ it has been at least $k$
periods since the merger period denoted as $t_{\text{merger}}$. Taking first differences gives us the estimating equation
\begin{eqnarray}
\Delta Y_{mt} &=& \sum_{k=1}^{T}\tilde{\beta}^{\texttt{DiD}}_k  \Delta  {\mathcal I}_{m,t,k} + \Delta \digamma_t + \Delta \varepsilon_{mt}, \label{eq:DiD_fd}
\end{eqnarray}
where $\Delta  {\mathcal I}_{m,t,k}$ is equal to one if $t = t_{\text{merger}} + k \text{ and } I_{mt} = 1$ and zero, otherwise.
We take $T = 8$ quarters (two years) post-merger.
We assume that two years is long enough for efficiency
gains to materialize and short enough to limit contamination from unrelated
market developments, while being comparable to the FE design.
When estimating, we exclude the quarter when the merger takes place, and summarize the merger effect by its \emph{eight-quarter average} defined as ${\beta}^{\texttt{DiD}}:=\sum_{k=1}^{8}\frac{9-k}{8}\tilde{\beta}^{\texttt{DiD}}_k$.

The estimated merger effects on prices are also shown in Table \ref{tab:did_main}, labeled ``FD.'' The FD estimates are imprecise; on the cluster-robust standard errors, only WN--FL's are distinguishable from zero ($\$6.89$, s.e.\ $\$2.18$, and $0.043$, s.e.\ $0.013$, in the combined sample). On the placebo intervals, each estimate contains zero, and the smallest placebo $p$-value among the FD estimates of Table \ref{tab:did_main} is $0.06$ (WN--FL, log fares, combined sample, baseline). Thus, the FD estimates provide no evidence that the mergers lowered fares. The results are similar for the connecting-only sample, although the effect of the WN--FL merger on log-fare is positive and statistically different from zero. 

\paragraph{Additional Controls.} Treated and untreated markets may differ in observable characteristics that can affect price dynamics. To account for these differences, we augment the DiD specifications with the demand and cost shifters used in the structural model, including market size and rivals' network reach on the demand side, and entry potential and rivals' presence on the cost side; see Appendix \ref{sec:appendix_iv_stats}. When controls are included, we impose \emph{conditional parallel trends}, which means only those markets with the same characteristics would follow the same trend in pricing in the absence of merger. For brevity, we do not state these conditions formally found in \cite{Abadie2005, CallawaySantAnna2021}.

We report the results in the second panel of Table \ref{tab:did_main}. Adding the controls does not materially change the results. The fixed-effects estimates are similar to the baseline estimates, while the first-difference estimates remain imprecise, except for the WN--FL merger. Overall, we still find no evidence that mergers lowered fares. 

We conclude this section with two observations. First, the quasi-experimental evidence depends on the estimator and how we define the treated sample. In our application, the FE estimates sometimes suggest fare reductions, while the FD estimates provide no evidence of lower fares. As a robustness check, the group-time estimator of
\citet{CallawaySantAnna2021} also provides no evidence that the mergers made air travel cheaper. 

The sensitivity of DiD estimates raises questions about economic restrictions implicit in the DiD
design given that prices are equilibrium outcomes. We address this
question next by deriving the DiD specification from an equilibrium 
model of demand and supply.

\section{First Bridge: Structural Interpretation of DiD\label{sec:structural_retrospective}}

The previous section showed that the conclusions from a merger retrospective based on DiD design can be sensitive to the estimator and the sample. More
fundamentally, the identifying assumptions underlying DiD are difficult to
interpret and evaluate in a merger setting because prices are equilibrium outcomes, and a merger
can change firms' costs and strategic conduct. So, the resulting price effects depend on how demand and
cost shocks are transmitted into prices. In this section, we develop a structural framework for merger retrospectives and, more importantly, show that it nests the DiD design as its reduced form.

The bridge serves two related purposes. First, it gives an economic
interpretation to the DiD estimate of the price effect of a merger. It tells us what the coefficient on the merger indicator in Equation \eqref{eq:did} is, 
in terms of efficiency and conduct. Second, it
allows us to ask whether the DiD identifies the target price
effect. If it does not, the structural model lets us determine why. 

We begin with a general model of demand, cost, and strategic conduct. To map
this model into a reduced-form price equation comparable to the DiD
specification in Equation (\ref{eq:did}), and to keep the structural model
tractable for the identification and estimation arguments developed in the
second bridge, we then consider two widely used functional forms. The first has linear
demand and marginal cost. The second has constant-elasticity demand and
log-linear marginal cost, which we refer to as the log-linear model. For each
specification, we ask the same two questions: What is the economic content of
the reduced-form merger coefficient, and under what conditions does DiD estimate the price effect of a merger?

\subsection{Demand and Supply}

To maintain consistency with our quasi-experimental approach and with the
framework used in court proceedings \citep[e.g.,][]{CarltonIsraelMacSwainOrlov2019}, we
model demand for air travel at the market level, treating it as a homogeneous
product sold by $n$ firms. Our objective in using aggregate demand at the market level is to connect market-level DiD estimates to a tractable equilibrium model of demand and supply. Let $Q=D(P, Z;\boldsymbol{\alpha})+\varphi$ denote demand, where $Q$ is quantity demanded, $P$ is price, $Z$ is an exogenous demand shifter that is excluded from the marginal cost function and uncorrelated with both the demand shock $\varphi$ and the cost shock $\vartheta$, and $\boldsymbol{\alpha}$ is the vector of demand parameters.

On the supply side, let $C=c(Q, W, I;\boldsymbol{\theta})+\vartheta$ be the marginal cost function, where $Q$ is quantity supplied,  $I\in\{0,1\}$ is a merger indicator that is equal to one post-merger if the market is treated and zero otherwise, $W$ represents an exogenous cost shifter that is excluded from the demand function and uncorrelated with both shocks, and $\boldsymbol{\theta}$ is the vector of cost parameters. We do not restrict the dependence between the demand and cost shocks beyond these orthogonality conditions.
We say that a merger generates efficiency
if $c(Q,W,1;\boldsymbol{\theta})\leq c(Q,W,0;\boldsymbol{\theta})$.

Suppressing the demand and cost parameters and with slight abuse of notation, firm $i$'s profit function is $\Pi_i=P(Q)q_i-c(q_i)q_i$, and firm $i$ chooses $q_i$ to maximize profit, where $P(Q)$ is the inverse demand. To capture competitive interactions, we follow \cite{Bresnahan1982} and introduce a strategic conduct parameter $\lambda\in[0,1]$. Suppose there are $n$ symmetric firms. If $\{q_i\}_{i=1,\ldots,n}$ is equilibrium output and $Q=\sum_i q_i$, firm $i$ chooses
$
q_i=\arg\max_{\widetilde q_i}\left\{P\left(n\lambda(\widetilde q_i-q_i)+Q\right)\widetilde q_i-c(\widetilde q_i)\widetilde q_i\right\}.
$
Thus, when firm $i$ deviates from its equilibrium output by $(\widetilde q_i-q_i)$, it believes that aggregate output changes by $n\lambda(\widetilde q_i-q_i)$.
Under symmetry, $q_i=Q/n$, the market-level marginal revenue function is
\begin{eqnarray}
MR=P+\lambda\left(\frac{\partial P}{\partial Q}Q\right)\equiv P+\lambda\times h(Q).
\label{eq:mr}
\end{eqnarray}
When $\lambda=0$, firms behave perfectly competitively; when
$\lambda=1$, firms fully internalize the effect of their output choices on
market price and behave as a monopolist, and intermediate values capture
intermediate degrees of competitive intensity.

Furthermore, we allow a merger to change
strategic conduct, i.e., $\lambda=\lambda(I)$. For example, sustaining
supracompetitive pricing may become easier after a
merger because there is one fewer firm and monitoring becomes easier. In that case, the merger increases
$\lambda$ towards one.
Letting $\lambda_{\text{pre}}=\lambda(I=0)$ and $\lambda_{\text{post}}=\lambda(I=1)$, we say a merger increases conduct and firms are less competitive after the merger if $\lambda_{\text{post}}>\lambda_{\text{pre}}$. The conduct parameter summarizes competitive behavior
in equilibrium, and a merger-induced change in conduct may
reflect unilateral or coordinated effects.

Equating marginal revenue (\ref{eq:mr}) with marginal cost gives
the equilibrium pricing condition
\begin{eqnarray}
P=c(Q,W,I;\boldsymbol{\theta})-\lambda(I)\times h(Q,Z;\boldsymbol{\alpha})+\vartheta.
\label{eq:mr_mc}
\end{eqnarray}

Equation (\ref{eq:mr_mc}) makes clear why an observed price change cannot, by
itself, be interpreted as an efficiency effect. A merger can affect equilibrium
prices through both marginal cost and strategic conduct, and the equilibrium
mapping from these changes into prices depends on demand and supply, i.e., the cost-pass-through rate.

We implicitly assume that mergers affect only supply-side relationships, not demand. This assumption reflects the principle that preferences do not change
with the merger, even though consumers adjust to the resulting price changes.
Firms may strategically modify product attributes after a merger
\citep{ChenGayle2019}, including route networks \citep{Borenstein1990}, service
quality \citep{PrinceSimon2017}, or operational frequencies. We assume that
the unobservable determinants of these adjustments are statistically
independent of the demand and cost shocks in the pricing equations.\footnote{In
estimation, we use characteristics of only the non-merging firms, and not
those of the merging firms.}

\subsection{DiD as a Reduced Form of Demand and Supply}
We next connect this equilibrium model to the DiD specification in Equation (\ref{eq:did}), which requires imposing additional functional-form restrictions. To this end, we consider both linear and log-linear models. Studying both specifications also helps clarify the role of functional-form assumptions in merger retrospectives. For each specification, we bridge the structural model and DiD in two steps. First, we derive the reduced-form price equation implied by the structural model and interpret the merger coefficient in terms of efficiency, conduct, and pass-through. Second, we derive the corresponding DiD estimand and determine the conditions under which it identifies the merger's average price effect on treated markets. We then study identification of the underlying parameters and conclude by showing how, unlike DiD, we can use the structural approach to evaluate merger effects beyond the treated markets.

\subsubsection{Linear Demand and Cost\label{sec:linear_model}}
We begin with the following linear demand and marginal cost functions 
\begin{eqnarray*}
Q_{mt}&=&D(P_{mt},Z_{mt};\boldsymbol{\alpha})+\varphi_{mt}=\alpha_0+\alpha_1P_{mt}+\alpha_2Z_{mt}+\varphi_{mt},\\
C_{mt}&=&c(Q_{mt},W_{mt},I_{mt};\boldsymbol{\theta})=\theta_0+\theta_1Q_{mt}+\theta_2W_{mt}+\theta_3I_{mt}+\vartheta_{mt},
\end{eqnarray*}
where $\alpha_1<0$.
Combining the cost equation with the equilibrium condition (\ref{eq:mr_mc}), and using $h(Q_{mt},Z_{mt};\boldsymbol{\alpha}) = \frac{Q_{mt}}{\alpha_1}$, and defining  $\tilde{\gamma}_{mt}:=\theta_1-\frac{\lambda(I_{mt})}{\alpha_1}$, we can write the equilibrium price in market $m$ and period $t$ as, 
\begin{equation}
\begin{aligned}
P_{mt}&=\theta_0+\tilde{\gamma}_{mt}Q_{mt}+\theta_2W_{mt}+\theta_3I_{mt}+\vartheta_{mt}. \label{eq:systems}
\end{aligned}
\end{equation}

We first consider the benchmark in which treated and control markets have the same strategic conduct absent the merger. Using the definition of $I_{mt}$ from Equation (\ref{eq:did}), let $\tilde{\gamma}_{mt}=\tilde{\gamma}_1$ when $I_{mt}=1$ and $\tilde{\gamma}_{mt}=\tilde{\gamma}_0$ otherwise. With a slight abuse of notation, let $\tilde{\boldsymbol{\theta}}=(\theta_0,\tilde{\gamma}_1,\tilde{\gamma}_0,\theta_2,\theta_3)^\top$ denote the vector of cost parameters. Substituting the demand equation into the supply equation in (\ref{eq:systems}), and suppressing $(Z,W)$ for ease of exposition, gives
\begin{eqnarray}
P_{mt} = \frac{\theta_0 +\tilde{\gamma}_{mt}\alpha_0}{(1-\tilde{\gamma}_{mt}\alpha_1)} + \frac{\theta_3}{(1-\tilde{\gamma}_{mt}\alpha_1)} \times I_{mt} + \frac{\tilde{\gamma}_{mt}\varphi_{mt}+\vartheta_{mt}}{(1-\tilde{\gamma}_{mt}\alpha_1)}.\label{eq:price_reduced}
\end{eqnarray}

This reduced-form price equation provides the first part of the bridge. It gives an economic interpretation of the merger coefficient. Conditional on the post-merger conduct parameter, define $\beta^{\mathrm{RF}}=\frac{\theta_3}{(1-\tilde{\gamma}_{1}\alpha_1)}$. The direct marginal-cost effect of the merger is $\theta_3$, while the denominator determines how that cost change transmits to equilibrium prices, i.e., the cost pass-through. In particular, the coefficient is 
\begin{eqnarray}
\beta^{\mathrm{RF}} = \frac{\theta_3}{(1-\tilde{\gamma}_{1}\alpha_1)}=\frac{\theta_3}{1 - \frac{{\mathcal E}_{d}}{{\mathcal E}_{s}}} = \text{cost efficiency}\times\text{cost pass-through}, \label{eq:merger_passthrough}
\end{eqnarray}
where the second-to-last equality uses the fact that the demand elasticity ${\mathcal E}_{d}  = \alpha_1 \frac{P^*}{Q^*}$, the supply elasticity ${\mathcal E}_{s} = \frac{1}{\tilde{\gamma}_{mt}} \frac{P^*}{Q^*}$ for any $\frac{P^*}{Q^*}$, and their ratio is $\tilde{\gamma}_{1}\alpha_1$, and the last equality uses the definition of the cost pass-through rate \citep{WeylFabinger2013}.

The merger coefficient $\beta^{\mathrm{RF}}$ in Equation (\ref{eq:merger_passthrough}) holds post-merger conduct fixed and measures the price effect of the merger-induced change in marginal cost. Moreover, because $\tilde{\gamma}_1=\theta_1-\frac{\lambda_{\text{post}}}{\alpha_1}$, the pass-through of a given efficiency gain also depends on post-merger strategic conduct. Thus, for a given efficiency gain, the resulting price reduction depends on how the merger changes competitive conduct.

This decomposition clarifies a concern in merger evaluations and antitrust economics more broadly: besides affecting the costs, a merger can also affect strategic conduct and, therefore, the cost pass-through rate in \eqref{eq:merger_passthrough}. If the merger softens competition ($\lambda$ increases), supply becomes less elastic ($\mathcal{E}_s$ falls). A less elastic supply lowers the cost pass-through rate, so a smaller fraction of the cost efficiency $\theta_3$ reaches consumers as lower prices. Increasing conduct therefore hurts consumers by raising markups and by muting the effect of merger-induced efficiencies on prices.

We next determine the merger's total average price effect on treated markets. For ease of exposition, consider a $2\times2$ design, and let $P_{m,\mathrm{post}}(0)$ denote its post-merger price in the absence of the merger, $P_{m,\mathrm{post}}(1)$ its post-merger price under the merger, and $P_{m,\mathrm{pre}}(0)$ its pre-merger price. Suppose the primitive demand and cost shocks satisfy mean parallel trends across treated and untreated markets, i.e., $\mathbb{E}[\varphi_{m,\mathrm{post}}-\varphi_{m,\mathrm{pre}}\mid\mathrm{Treated}_m=1]=\mathbb{E}[\varphi_{m,\mathrm{post}}-\varphi_{m,\mathrm{pre}}\mid\mathrm{Treated}_m=0]$ and analogously for $\vartheta$.
Then the DiD estimand is
\begin{eqnarray}
\beta^{\mathrm{DiD}}&=&\mathbb{E}\!\left[P_{m,\mathrm{post}}(1)-P_{m,\mathrm{post}}(0)\mid\mathrm{Treated}_m=1\right]+\Big(\mathbb{E}\!\left[P_{m,\mathrm{post}}(0)-P_{m,\mathrm{pre}}(0)\mid\mathrm{Treated}_m=1\right]\nonumber\\
&&\qquad\qquad\qquad\qquad\qquad\qquad\qquad\qquad-\mathbb{E}\!\left[P_{m,\mathrm{post}}(0)-P_{m,\mathrm{pre}}(0)\mid\mathrm{Treated}_m=0\right]\Big). \qquad\label{eq:beta_did_def}
\end{eqnarray}
The second term in the parenthesis makes clear that parallel trends in primitive shocks alone are insufficient for DiD to identify the average treatment effect on the treated (ATT), which is the first term on the right-hand side of \eqref{eq:beta_did_def}. 

In particular, let $\lambda_T$ and $\lambda_C$ denote strategic conduct in treated and control markets absent the merger. If $\lambda_T\neq\lambda_C$, the two groups have different pricing slopes and thus different pass-through, so the same changes in demand and cost shocks can generate different changes in equilibrium prices. If $\lambda_T=\lambda_C$, the two groups have the same mapping from demand and cost shocks into equilibrium prices, so parallel trends in the primitive shocks imply parallel trends in prices and the second term in Equation \eqref{eq:beta_did_def} is zero, and $
\beta^{\mathrm{DiD}}=\mathbb{E}\!\left[P_{m,\mathrm{post}}(1)-P_{m,\mathrm{post}}(0)\mid\mathrm{Treated}_m=1\right].$ Thus, depending on the pre-merger conduct across markets, parallel trends in the primitive shocks need not imply parallel trends in prices, and DiD need not identify the ATT.

To characterize $\beta^{\mathrm{DiD}}$ in terms of model primitives under common pre-merger conduct, evaluate Equation \eqref{eq:price_reduced} at the three relevant states: the treated market after the merger under treatment, the same market after the merger period but without merger, and the pre-merger untreated markets. Substituting these equilibrium prices into Equation \eqref{eq:beta_did_def} and averaging over the treated markets gives a particularly simple decomposition. Define $\mathbb{S}^1_j:=\frac{\tilde{\gamma}_j}{1-\tilde{\gamma}_j\alpha_1}$ and $\mathbb{S}^2_j:=\frac{1}{1-\tilde{\gamma}_j\alpha_1}$ for $j\in\{0,1\}$ and let $\bar{\varphi}_{s}:=\mathbb{E}[\varphi_{m,s}\mid\mathrm{Treated}_m=1]$ and $\bar{\vartheta}_{s}:=\mathbb{E}[\vartheta_{m,s}$. When pre-merger conduct differs across treated and control markets, the coefficients mapping these shocks into equilibrium prices also differ. Consequently, prices may not satisfy the parallel assumption, as the following lemma establishes.

\begin{lemma}\label{lemma:did_bias}
In the $2\times2$ design with varying strategic conduct, suppose the primitive demand and cost shocks satisfy mean parallel trends across treated and untreated markets.
\begin{enumerate}[label=(\roman*)]
\item If $\lambda_T=\lambda_C$, then $\beta^{\mathrm{DiD}}$ identifies the average treatment effect on the treated and
\begin{eqnarray}
\beta^{\mathrm{DiD}} =\beta^{\mathrm{RF}} +
    \left(\frac{\theta_0+\tilde{\gamma}_1\alpha_0}{1-\tilde{\gamma}_1\alpha_1}
         -\frac{\theta_0+\tilde{\gamma}_0\alpha_0}{1-\tilde{\gamma}_0\alpha_1}\right)+
    (\mathbb{S}^1_1-\mathbb{S}^1_0)\bar{\varphi}_{\mathrm{post}}
   +(\mathbb{S}^2_1-\mathbb{S}^2_0)\bar{\vartheta}_{\mathrm{post}}.  \label{eq:did_decomp}
\end{eqnarray}
\item If $\lambda_T\neq\lambda_C$, parallel trends in the primitive shocks do not generally imply parallel trends in untreated prices, and $\beta^{\mathrm{DiD}}$ does not identify the ATT.
\end{enumerate}
\end{lemma}

The proof is in Appendix \ref{sec:proofs}. Lemma \ref{lemma:did_bias} highlights the distinction between imposing parallel trends directly on the equilibrium prices and deriving it from restrictions on the underlying demand and cost shocks. With common pre-merger conduct, the mapping from the primitive shocks into equilibrium prices is the same across treated and control markets, so parallel trends in the primitives imply parallel trends in prices. With different pre-merger conduct, this mapping differs across the two groups because conduct affects pass-through. Parallel trends in the primitives then need not imply parallel trends in prices. Thus, common pre-merger conduct and parallel trends in shocks are sufficient for parallel trends in prices.

So, although DiD that assumes parallel trends in prices directly can identify the ATT, it also entails the additional assumption that pre-merger conduct is the same across all markets.  

Under common pre-merger conduct, Lemma \ref{lemma:did_bias} shows that $\beta^{\mathrm{DiD}}$ identifies the average price effect of the merger on treated markets, whereas $\beta^{\mathrm{RF}}$ isolates the component arising from the merger-induced cost change. Consequently, if the merger does not improve cost (i.e., $\theta_3=0$), then $\beta^{\mathrm{RF}}=0$, but the merger can still affect prices through a change in conduct. Thus, the absence of an efficiency gain does not imply that the merger has no effect.

The structural approach can recover the total merger effect by comparing equilibrium prices under the merger and no-merger regimes. It relies on different identifying restrictions than parallel trends, i.e., the variation in the valid instruments. Moreover, because we can compare prices with and without the merger, the structural approach can evaluate the merger's price effect even in untreated markets, which DiD cannot identify. We return to this distinction and its implications for the merger evaluations in Section \ref{sec:implications}.

\subsubsection{Log-Linear Model\label{app:logspec}}
Next, we consider constant-elasticity demand and log-linear marginal cost.
For the ease of presentation, we use the same notations as those in the linear model, and define the demand and marginal cost functions as 
\begin{eqnarray}
\log Q_{mt}&=&\alpha_0+\alpha_1\log P_{mt}+\alpha_2Z_{mt}+\varphi_{mt},\label{eq:logdemand}\\
\log C_{mt}&=&\theta_0+\theta_1\log Q_{mt}+\theta_2W_{mt}+\theta_3I_{mt}+\vartheta_{mt}.\label{eq:logcost}
\end{eqnarray}
Here, the markup $h(Q_{mt}) = (\partial P_{mt}/\partial Q_{mt}) \times Q_{mt} = P_{mt}/\alpha_1 < 0$. Equating marginal revenue \eqref{eq:mr} and marginal cost (and continuing to suppress $Z$ and $W$ for ease of exposition) gives
\begin{eqnarray}
(1 - \alpha_1\theta_1)\log P_{mt} = -\log\!\left[1 + \frac{\lambda(I_{mt})}{\alpha_1}\right] + (\theta_0 + \theta_1\alpha_0) + \theta_3 \times I_{mt} + \theta_1\varphi_{mt} + \vartheta_{mt}.
\end{eqnarray}
Under common pre-merger conduct, i.e., $\lambda_T=\lambda_C=\lambda_{\text{pre}}$, define $\Lambda_{\text{pre}} = \log[1 + \lambda_{\text{pre}}/\alpha_1]$ and $\Lambda_{\text{post}} = \log[1 + \lambda_{\text{post}}/\alpha_1]$, then $\log[1 + \lambda(I_{mt})/\alpha_1] = \Lambda_{\text{pre}} + (\Lambda_{\text{post}} - \Lambda_{\text{pre}})\times I_{mt}$. Then the reduced form equation becomes
\begin{eqnarray}
\log P_{mt} = \Pi_0 + {\beta}^{\textrm{RF}}_{\log} \times I_{mt} + \Pi_Z Z_{mt} + \Pi_W W_{mt} +\varepsilon_{mt}^{\log},\label{eq:logprice_reduced}
\end{eqnarray}
where the parameter of interest ${\beta}^{\textrm{RF}}_{\log} =\frac{\theta_3 - (\Lambda_{\text{post}} - \Lambda_{\text{pre}})}{1-\alpha_1\theta_1}$, the intercept is $\Pi_0 = (\theta_0 + \theta_1\alpha_0 - \Lambda_{\text{pre}})/(1-\alpha_1\theta_1)$, $\Pi_Z = \theta_1\alpha_2/(1-\alpha_1\theta_1)$, and $\Pi_W = \theta_2/(1-\alpha_1\theta_1)$, and $\Lambda_{\text{post}} - \Lambda_{\text{pre}} = \log[(1+\lambda_{\text{post}}/\alpha_1)/(1+\lambda_{\text{pre}}/\alpha_1)]$ and the composite error is $\varepsilon_{mt}^{\log}=\frac{\theta_1}{1-\alpha_1\theta_1}\varphi_{mt}+\frac{1}{1-\alpha_1\theta_1}\vartheta_{mt}$.

Thus, Equation (\ref{eq:logprice_reduced}) is the reduced form of the log-linear DiD model in Equation (\ref{eq:did}). As before, the first question of interest is which economic object the merger coefficient represents. Under constant-elasticity demand, $\mathcal{E}_d=\alpha_1$, and under log-linear marginal cost, $\mathcal{E}_s=1/\theta_1$. The cost pass-through rate is $1/(1-\mathcal{E}_d/\mathcal{E}_s)=1/(1-\alpha_1\theta_1)$, so we have
\begin{equation}
\beta^{\mathrm{RF}}_{\log}=\frac{\theta_3-(\Lambda_{\text{post}}-\Lambda_{\text{pre}})}{1-\alpha_1\theta_1}
={\text{net-of-conduct cost efficiency}}\times{\text{cost pass-through rate}}.
\label{eq:log_passthrough}
\end{equation}

Thus, the DiD merger coefficient reflects both efficiency and strategic conduct. A negative coefficient need not by itself imply that the merger was pro-competitive, nor does a positive coefficient imply that the merger generated no efficiencies. Without the structural model, the two channels cannot be separated. However, because the conduct enters the reduced-form log-price equation as an additive term, it does not affect the coefficients on the demand and cost shocks. Consequently, even if pre-merger conduct in treated and control markets are different, parallel trends in the demand and cost shocks imply parallel trends in log prices. 

\begin{lemma}\label{lemma:did_bias_log}
Consider the $2\times2$ DiD design under the log-linear specification in Equations \eqref{eq:logdemand}--\eqref{eq:logcost}. Suppose $1+\lambda_{\textrm{pre}}/\alpha_1>0$ and $1+\lambda_{\textrm{post}}/\alpha_1>0$ and $1-\alpha_1\theta_1\neq0$. If the primitive demand and cost shocks satisfy parallel trends, then the DiD estimand identifies the average treatment effect on the treated whether pre-merger conduct is the same or different across treated and control markets, and
\begin{eqnarray}
\beta_{\log}^{\mathrm{DiD}}=\frac{\theta_3-(\Lambda_{\mathrm{post}}-\Lambda_{\mathrm{pre}})}{1-\alpha_1\theta_1}=\beta_{\log}^{\mathrm{RF}}.
\label{eq:log_did_att}
\end{eqnarray}
\end{lemma}

In the linear model, differences in pre-merger conduct change how demand and cost shocks pass through into prices, so parallel trends in the primitive shocks do not generally imply parallel trends in prices. In the log-linear model, conduct affects the level of log prices but not shock transmission, so the same problem does not arise. 

As we see next, we cannot identify the level of conduct, only the change in conduct, and only in the linear model. For the rest of the paper, we assume pre-merger conduct is the same across treated and control markets, i.e., $\lambda_T=\lambda_C=\lambda_{\mathrm{pre}}$. 

\subsection{Identification\label{sec:identification}}
Recovering the components of the merger's price effect requires identifying the demand, cost, and
conduct parameters that determine equilibrium prices. We develop the
identification argument using the linear model of
Section \ref{sec:linear_model}. The log-linear model has a different
equilibrium mapping, but most of the identification logic is
the same: excluded demand and cost shifters identify the demand and supply
relationships. The main difference is that, with a single demand slope, the
linear model identifies the merger-induced change in conduct,
$\Delta\lambda$, whereas the log-linear model identifies only efficiency augmented by some function of the conduct parameters.
In both cases, estimating two demand functions, say, for business and leisure passengers, is sufficient to separate conduct levels from the underlying cost parameters. We therefore present
the argument for the linear model, where the identification is especially
transparent, and return briefly to the log-linear model. We identify the demand-and-supply  slopes using standard exclusion
restrictions from the simultaneous-equations literature. In particular, the
cost shifters ${W}_{mt}$ affect prices but are excluded from demand,
while the demand shifters ${Z}_{mt}$ affect demand but are excluded from
supply. Under the usual relevance and orthogonality conditions, variation in
these instruments identifies the demand parameters
$\boldsymbol{\alpha}$ and the reduced-form supply parameters
$\tilde{\boldsymbol{\theta}}$ in Equation \eqref{eq:systems}, following the
classical argument dating back to \cite{Wright1928}.

The remaining issue is how much of the underlying supply structure we can recover from those reduced-form parameters. Recall that the pricing slope is $\tilde{\gamma}_{mt}$, and thus before and after the merger, they are $\tilde{\gamma}_0=\theta_1-\frac{\lambda_{\mathrm{pre}}}{\alpha_1}$ and $\tilde{\gamma}_1=\theta_1-\frac{\lambda_{\mathrm{post}}}{\alpha_1}$, respectively.
With a single demand slope $\alpha_1$, these two pricing slopes depend on three
underlying primitives: $\theta_1$, $\lambda_{\mathrm{pre}}$, and
$\lambda_{\mathrm{post}}$. The levels of marginal-cost curvature and conduct
therefore cannot be separated. Their difference, however, can
be identified as $\tilde{\gamma}_1-\tilde{\gamma}_0=-\frac{\lambda_{\mathrm{post}}-\lambda_{\mathrm{pre}}}{\alpha_1}$,
so that $\Delta\lambda=\lambda_{\mathrm{post}}-\lambda_{\mathrm{pre}}=-\alpha_1\left(\tilde{\gamma}_1-\tilde{\gamma}_0\right)$. We formalize these identification results below, including additional variation sufficient to recover the levels of strategic conduct.

\begin{proposition}\label{prop:conduct}
Suppose the demand and cost shocks satisfy $\mathbb{E}(\varphi_{mt})=\mathbb{E}(\vartheta_{mt})=0$, the exclusion restrictions hold, and the demand-and-supply instruments are orthogonal to their respective structural shocks. Suppose further that the corresponding population instrument--regressor matrices have full column rank. Then:

\begin{enumerate}[label=(\roman*)]

\item \label{prop:demand_cost}
The demand parameters ${\boldsymbol \alpha}:=(\alpha_0,\alpha_1,\alpha_2)^\top$ and the reduced-form supply parameters
$\tilde{\boldsymbol \theta}:= (\theta_0,\tilde{\gamma}_1,\tilde{\gamma}_0,\theta_2,\theta_3)^\top$ in Equation \eqref{eq:systems} are identified.

\item \label{prop:change}
The structural supply slope and conduct levels $(\theta_1,\lambda_{\mathrm{pre}},\lambda_{\mathrm{post}})$
are not separately identified when demand has a single price slope. However, the merger-induced change in strategic conduct is identified as $\Delta\lambda = -\alpha_1 \times (\tilde{\gamma}_1 - \tilde{\gamma}_0)$.

\item \label{prop:levels}
Suppose, in addition, that demand can be observed under two distinct price
slopes, $\alpha_1^1\neq\alpha_1^2$. Then $\theta_1$, $\lambda_{\mathrm{pre}}$, and
$\lambda_{\mathrm{post}}$ are separately identified.

\end{enumerate}
\end{proposition}
The proofs are in Appendix \ref{sec:proofs}.
Part \ref{prop:change} follows directly from the fact that marginal-cost
curvature and strategic conduct enter the pricing slope in the same way.
Indeed, if we change the strategic conduct and the scale parameter by $\chi/\alpha_1$ for any $\chi>0$, prices would be unaffected. In other words, the parameter vectors $(\theta_1,\lambda_{\mathrm{pre}},\lambda_{\mathrm{post}})$ and $\left(\theta_1+\frac{\chi}{\alpha_1}, \lambda_{\mathrm{pre}}+\chi,\lambda_{\mathrm{post}}+\chi\right)$
generate the same values of $\tilde{\gamma}_0$ and
$\tilde{\gamma}_1$, and are therefore observationally equivalent.
The common shift cancels when we difference the two pricing slopes, which identifies $\Delta\lambda$.

Identifying $\Delta\lambda$ does not by itself separate unilateral from
coordinated effects. With one fewer competitor, market power can increase even
if firms continue to compete according to the same underlying equilibrium
concept; any additional softening in behavior is a distinct coordinated
component. Separating these objects requires the pre-merger conduct level.

One route to identify $\lambda_{\mathrm{pre}}$ and $\lambda_{\mathrm{post}}$ is to use a richer demand system. Part \ref{prop:levels} shows that if we have two types of demand functions, e.g., one for business travelers and the other for leisure travelers, then we can identify the conduct and scale parameters. The idea is
closely related to the fact that conduct affects the pricing equation through the
interaction between perceived marginal revenue and the slope of demand. In contrast, the marginal-cost slope $\theta_1$ does not depend on the demand
slope. A change in the slope of demand therefore changes the way a given
conduct parameter appears in the pricing equation without changing
$\theta_1$.

In the empirical analysis, we do not impose either route, so we interpret
$\Delta\lambda$ as the merger-induced change in strategic conduct rather than
attempting to decompose it further. Accordingly, the estimates below report the control-market pricing slope $\tilde{\gamma}_C:=\theta_1-\lambda_C/\alpha_1$ rather than $\theta_1$, and the treated markets' conduct is identified relative to that of the control markets. The three questions require $\Delta\lambda$, pass-through, and the price effect, which depend only on the pricing slopes, not on the level of control-market conduct.\footnote{\cite{Corts1999} shows that static conduct models can give inconsistent estimates of conduct under efficient collusion because equilibrium conduct may vary with demand conditions. This critique does not apply to our setting because our objective is different: we identify the \emph{change} in conduct using an exogenous merger that can affect conduct, without assuming efficient collusion.}

In the log-linear model, the same exclusion restrictions identify the demand
and supply relationships, but with a single demand, the merger-induced cost effect and conduct change enter the intercept in the pricing Equation \eqref{eq:logprice_reduced} as $(\theta_3-(\Lambda_{\mathrm{post}}-\Lambda_{\mathrm{pre}}))$ and cannot be separately identified.

As in the linear model, however, observing the pricing relation under two distinct demand slopes identifies the conduct $(\lambda_{\textrm{pre}}, \lambda_{\textrm{post}})$. The intuition is that the common cost component is unchanged across the two regimes, whereas the conduct term $\log(1+\lambda/\alpha_1)$ changes with the demand
slope, identifying $\lambda_{\mathrm{pre}}$ and $\lambda_{\mathrm{post}}$ separately, and hence
also $\theta_3$ and $\Delta\lambda$.

\subsection{Implications for Merger Retrospectives \label{sec:implications}}
The preceding results help merger retrospectives by clarifying how to estimate the merger effect on prices using the DiD design and how to interpret the estimate in terms of efficiency and conduct. However, to use the estimates for retrospective evaluation, we must also specify the markets of interest. If the objective is to measure the effect in markets affected by the merger, the relevant object is the average treatment effect on treated markets. Lemmas \ref{lemma:did_bias} and \ref{lemma:did_bias_log} show that DiD identifies this effect when primitive demand and cost shocks satisfy parallel trends and, in the linear model, conduct is the same in treated and untreated markets.

If the objective is instead to assess what the estimated merger-induced changes in efficiency and conduct imply for a broader set of markets, the merger counterfactual must also be evaluated outside the treated sample, for which DiD is unsuitable. To see this formally, focus on the post-merger period, where $P_m(1)-P_m(0)$ denotes the market-specific merger effect on price. Averaging this effect over treated markets gives $\mathrm{ATT}=\mathbb{E}[P_m(1)-P_m(0)\mid\mathrm{Treated}_m=1]$, while averaging it over a set of markets $\mathcal M$ gives $\mathrm{ATE}_{\mathcal M}:=\mathbb{E}[P_m(1)-P_m(0)\mid m\in\mathcal M]$. The structural approach lets us recover both measures, provided the estimated merger-induced changes in costs and conduct apply to the markets being evaluated.

To understand when these measures differ, it helps to view them as different averages of the same underlying distribution of market-specific merger effects. Following the marginal treatment effect framework \citep{HeckmanVytlacil2005}, suppose that $\mathcal M$ includes the treated markets and let $U_m\in[0,1]$ be a uniform random variable that affects the likelihood that market $m$ is in the treated sample, with this likelihood decreasing in $U_m$. Furthermore, conditional on $U_m$, whether market $m$ is treated is independent of its merger effect $(P_m(1)-P_m(0))$. Then, $\mathrm{MTE}(u)=\mathbb{E}[P_m(1)-P_m(0)\mid U_m=u,m\in\mathcal M]$ is the average merger effect among markets at $U_m=u$. The value of using MTE is that \citet{HeckmanVytlacil2005} show that $\mathrm{ATT}=\int_0^1\mathrm{MTE}(u)\varrho(u)du$, where $\varrho(u)$ is the density of $U_m$ among treated markets, while $\mathrm{ATE}_{\mathcal M}=\int_0^1\mathrm{MTE}(u)du$, and therefore $\mathrm{ATE}_{\mathcal M}-\mathrm{ATT}=\int_0^1\mathrm{MTE}(u)(1-\varrho(u))du$. Thus, the ATT and $\mathrm{ATE}_{\mathcal M}$ can differ only when merger effects vary across markets, because the two estimands weight these effects differently.

To make this mechanism precise, consider first the linear model. After suppressing all covariates, the two pricing equations are $P_m(0)=\theta_0+\tilde{\gamma}_0Q_m(0)+\vartheta_m$ and $P_m(1)=\theta_0+\tilde{\gamma}_1Q_m(1)+\theta_3+\vartheta_m$, where $Q_m(1)$ and $Q_m(0)$ are, respectively, equilibrium quantities with and without merger. Because demand is unchanged by the merger, $Q_m(1)=Q_m(0)+\alpha_1[P_m(1)-P_m(0)]$. Subtracting the two pricing equations and using $\tilde{\gamma}_1-\tilde{\gamma}_0=-\Delta\lambda/\alpha_1$ gives 
\begin{eqnarray}
P_m(1)-P_m(0)=\frac{1}{1-\tilde{\gamma}_1\alpha_1}\left(\theta_3-\frac{\Delta\lambda}{\alpha_1}Q_m(0)\right).\label{eq:linear_counterfactual}
\end{eqnarray} 
The first component is the price effect of the merger-induced cost change, and the second is the price effect of the change in conduct, which depends on $Q_m(0)$. Then, from  \eqref{eq:linear_counterfactual}, we get $\mathrm{MTE}(u)=\left(\theta_3-\frac{\Delta\lambda}{\alpha_1}\mathbb{E}[Q_m(0)\mid U_m=u,m\in\mathcal M]\right)/\Big(1-\tilde{\gamma}_1\alpha_1\Big)$. If conduct does not change with the merger, i.e., $\Delta\lambda=0$, then $\mathrm{MTE}(u)=\theta_3/(1-\tilde{\gamma}_1\alpha_1)$ is constant and therefore $
\mathrm{ATE}_{\mathcal M}-\mathrm{ATT}=\theta_3/(1-\tilde{\gamma}_1\alpha_1)\int_0^1(1-\varrho(u))du=0,$ because $\int_0^1\varrho(u)du=1$. 
If conduct changes, $\Delta\lambda\neq0$, then $\mathrm{MTE}(u)$ varies with $u$ whenever $\mathbb{E}[Q_m(0)\mid U_m=u,m\in\mathcal M]$ varies with $u$, and because of the different weights placed on merger effects, we get $\mathrm{ATE}_{\mathcal M}\neq\mathrm{ATT}$.

The log-linear model gives a different result. Holding the market's demand and cost conditions fixed and comparing the merger and no-merger equilibria in Equation \eqref{eq:logprice_reduced} gives
\begin{eqnarray}
\log P_m(1)-\log P_m(0)=\frac{\theta_3-(\Lambda_{\mathrm{post}}-\Lambda_{\mathrm{pre}})}{1-\alpha_1\theta_1}=\beta_{\log}^{\mathrm{RF}}. \label{eq:log_counterfactual}
\end{eqnarray}
Thus, unlike in the linear model, the price effect is constant across all markets. Hence $\mathrm{MTE}_{\log}(u)=\beta_{\log}^{\mathrm{RF}}$ is a constant and $\mathrm{ATT}_{\log}=\mathrm{ATE}_{\mathcal M,\log}=\beta_{\log}^{\mathrm{RF}}$. Thus, even when conduct changes, the log-linear specification implies a constant treatment effect.
We summarize the implications for the linear and log-linear models in the following proposition.

\begin{proposition}\label{prop:first_bridge}
Let $\mathcal M$ be any set of markets over which we evaluate the merger effect.
\begin{enumerate}[label=(\alph*)]
\item In the linear model, the market-specific merger effect is given by Equation \eqref{eq:linear_counterfactual}. If $\Delta\lambda=0$ then $\mathrm{ATE}_{\mathcal M}=\mathrm{ATT}$.
\item In the log-linear model, the market-specific merger effect is given by Equation \eqref{eq:log_counterfactual} and $\mathrm{ATE}_{\mathcal M,\log}=\mathrm{ATT}_{\log}$ for every $\mathcal M$ and any conduct change $\Delta\lambda$ common across markets.
\end{enumerate}
\end{proposition}
Proposition \ref{prop:first_bridge} follows directly from the market-specific merger effects derived above. 
Its main insight is that the extrapolation of the ATT to other markets depends on merger-effect heterogeneity  \citep{HeckmanVytlacil2005}. In the linear model, Equation \eqref{eq:linear_counterfactual}, the merger effect depends on the counterfactual quantity $Q_m(0)$, which varies across markets. However, when $\Delta\lambda=0$, this heterogeniety is immaterial and the merger effect \emph{effectively} homogeneous, so $\mathrm{ATE}=\mathrm{ATE}_{\mathcal M}$. In the log-linear model, constant demand elasticity implies a homogeneous merger effect, Equation \eqref{eq:log_counterfactual}, so $\mathrm{ATT}_{\log}=\mathrm{ATE}_{\mathcal M,\log}$ even when $\Delta\lambda\neq0$.  

The homogeneity of merger effects in the log-linear model is an implication of constant demand elasticity. More generally, variation in conduct and cost pass-through across markets can generate heterogeneous price effects, creating a wedge between the two estimates. To see this, consider smooth demand and cost functions and suppose that the merger shifts marginal cost additively. Following the first-order approach of \citet{JeffeWeyl2013}, apply the pricing condition in Equation \eqref{eq:mr_mc} and let $\rho_m$ denote the pass-through of marginal cost into equilibrium prices, evaluated at the post-merger equilibrium in market $m$, which gives
\begin{eqnarray*}
\mathrm{MTE}(u)\simeq\theta_3 \mathbb{E}\left[\rho_m\mid U_m=u,m\in\mathcal{M}\right]-\Delta\lambda \mathbb{E}(\rho_m\frac{P_m(0)}{\mathcal{E}_{d,m}} \mid U_m=u,m\in\mathcal{M}).
\end{eqnarray*}
Thus, even when merger-induced changes in efficiency, $\theta_3$, and conduct, $\Delta\lambda$, are constant across markets, price effects can still be heterogeneous when cost pass-through varies. When $\Delta\lambda\neq0$, differences in demand conditions provide an additional source of heterogeneity. Importantly, even when $\Delta\lambda=0$, pass-through variation alone can generate heterogeneous merger effects.

The role of these functional form assumptions for conduct and merger analysis has been recognized in the literature \citep{Ellison1994, CrookeFroebTschantzWerden1999, FroebTschantzWerden2005, JeffeWeyl2013}. The functional form is also important for DiD design, not least because in general parallel trends in prices need not imply parallel trends in log prices and vice versa  \citep{RothSantanna2023}. 

Thus, our First Bridge connects the quasi-experimental and structural approaches to merger retrospectives. While DiD identifies the merger's price effect in treated markets under parallel trends, the structural approach can recover the same effect and distinguish its underlying mechanisms. Moreover, a structural model can evaluate counterfactual merger effects in other markets, whereas extrapolating the DiD estimate beyond treated markets generally requires additional assumptions about merger-effect heterogeneity.

However, the structural approach requires credible identification of the underlying demand and supply relationships. In merger retrospectives, this credibility depends in part on the choice of untreated markets and periods used to construct the moment conditions. In particular, control markets should provide valid moments and face demand and cost conditions comparable to those of the treated markets. The Second Bridge shows how a quasi-experimental method can inform this choice.

\section{Second Bridge: Structural Estimation with Quasi-experimental Weights \label{sec:synthetic_GMM}}
For our second methodological bridge, which we call
\emph{synthetic GMM}, we explore the possibility of using the quasi-experimental approach to improve the estimation of the structural approach. The motivation comes from recognizing that while it is difficult to pick appropriate 
control markets \citep{AshenfelterHoskenWeinberg2009}, in our merger setting, the problem is even more pronounced because there are only a handful of treated markets relative to a large number of untreated markets; see Section \ref{sec:data}.

Control markets must be similar to treated markets in that
they should face comparable costs and demand conditions, so that the
counterfactual evolution of treated markets in the absence of a merger can
be inferred from control markets. At the same time, they must be
independent in that they must not be affected by the merger. This
tradeoff has an important temporal dimension too that is often overlooked: Even if a control market
resembles a treated market at some time, the two may diverge. They may face the same industry-wide shocks but respond differently because they serve different passenger mixes, operate under different airport constraints, or have different underlying costs.

The standard GMM estimator assigns
equal weight to every market-period observation, meaning the moment conditions
used to identify efficiency gains and changes in conduct are almost entirely
determined by the untreated market-period observations.
As instruments are constructed to be valid for treated markets, invalid control markets may ``contaminate'' moment conditions. When those markets make up the overwhelming majority of the sample, as here, our estimates may also be contaminated.

We cannot test IV validity directly, but we can use the price dynamics of each
market in our estimation.
Price dynamics convey which untreated market-periods are better control markets, and we can use SDiD weights to capture that information. We can then reweight each market-period's contribution to the moment conditions using SDiD weights, relying more on observations with higher SDiD weights. The resulting estimator uses the balancing properties
of SDiD weights to construct valid moment conditions while retaining the economic interpretability
of the model parameters.

We establish two properties of synthetic GMM. First, when all moment conditions hold across all markets, Synthetic GMM is consistent. In this sense, reweighting does not introduce asymptotic bias, although it may affect inference. Second, when instruments are valid for treated markets
but invalid for some untreated markets, the standard GMM has bias, while the Synthetic GMM corrects for the contamination through SDiD weights, and its bias vanishes as the pre-merger window grows. Thus, Synthetic GMM may be useful precisely when we have a large, heterogeneous untreated pool with some potentially invalid IV.
We formalize this idea for the linear model using a linear factor
representation \citep{Abadie2021}.

\subsection{Synthetic GMM}

Our Synthetic GMM approach proceeds in two main steps (see Table \ref{tab:synthetic_gmm_algorithm} for a heuristic explanation). First, following \citet{ArkhangelskyAtheyHirshbergImbensWager2021}, we construct market-specific weights $\hat{\omega}$ that align pre-merger price changes between control and treated markets. Like synthetic control methods, we find weights such that the weighted average of the quarterly changes in residualized prices in the control markets closely matches that of the treated markets across all pre-merger periods. For each pre-merger and post-merger window, the time weights are equal. We discuss this issue in Appendices \ref{section:sgmm_bias_details} and \ref{sec:few_treated_inference}.

Second, we use these weights to estimate the model parameters by reweighting the supply-side moment conditions, which we call Synthetic GMM (S-GMM). Instead of the standard sample analog of the supply moment conditions $\frac{1}{MT}\sum_{m,t}g^S(\cdot;\cdot)$, we compute synthetic-weighted sample moments $\sum_{m,t}\omega_m r_t g^S(\cdot;\cdot)$, where $\omega_m$ applies market weights and $r_t$ applies time weights within each window. 
This reweighting is a principled way to select control markets, as it weights supply moments more heavily toward control markets whose pre-merger price dynamics more closely resemble those of the treated markets.

\begin{table}[ht!!]
\centering
\caption{Algorithm for Synthetic GMM\label{tab:synthetic_gmm_algorithm}}
\begin{threeparttable}
\begin{algorithm}[H]
\begin{algorithmic}[1]
\State {Data:} $\mathcal{D}_{mt}:=\{P_{mt},(\boldsymbol{W}_{mt},\boldsymbol{Z}_{mt},
  Q_{mt})\}$ with treated markets $\mathcal{T}$ and control markets
  $\mathcal{C}$ and $\fTpre$ and $\fTpost$ as pre
  and post-merger. Let $\bk:=(\boldsymbol{\theta},\boldsymbol{\alpha},\boldsymbol{\lambda})$ be the parameter vector.
\vspace{0.2em}
\State Construct residualized prices $\widetilde{P}_{mt}$ and their pre-merger first differences $\Delta\widetilde{P}_{mt}$, and determine market weights $\hat{\omega}$ as the solution to \eqref{eq:w_opt}.
\vspace{0.2em}
\State Define weights:
  $\omega_{m}=\hat{\omega}_m$ if $m\in\mathcal{C}$,
  and $\omega_{m}=\frac{1}{|\mathcal{T}|}$ if $m\in\mathcal{T}$;
  $r_{t}=\hat{\tau}_t$ if $t\in\fTpre$,
  and $r_{t}=\frac{1}{T_1}$ if
  $t\in\fTpost$,
  where $\sum_{m\in\mathcal{C}}\hat{\omega}_{m}=1$ and $\sum_{ t\in \fTpre}\hat{\tau}_{t}=1$ by construction.
\State Construct moments: $\gtil(\bk)=\Bigl(\tfrac{1}{MT}\sum_{m,t} g^D(\mathcal{D}_{mt};\bk^D)^{\top},\
  \sum_{m,t}\omega_{m}  r_{t}  g^S(\mathcal{D}_{mt};\bk^S)^{\top}\Bigr)^{\top}$.
\vspace{0.2em}
\State Synthetic GMM:
  $\hat{\bk}_{S\text{-GMM}}:=\arg\min_{\bk}
   \gtil(\bk)^{\top}\What\gtil(\bk)$.
\end{algorithmic}
\end{algorithm}
\end{threeparttable}
\end{table}

\medskip\noindent\emph{Weights.}\quad
Let $\widetilde{P}_{mt}$ be the residualized price for market $m$ at time $t$, obtained by removing the effects of the demand and cost shifters. We construct the weights using first differences of residualized prices, $\Delta\widetilde P_{mt}:=\widetilde P_{mt}-\widetilde P_{m,t-1}$. For the first pre-merger quarter $t_1$, $\widetilde P_{m,t_0}$ is the residualized price in the quarter immediately preceding the pre-merger window, so that we have $T_0$ pre-merger price changes for each market.
Following \citet{ArkhangelskyAtheyHirshbergImbensWager2021}, we define the market and time weights, respectively, as the solutions to
\begin{align}
  (\hat{\omega}_0,\hat{\boldsymbol{\omega}})
  &:= \arg\min_{\substack{\omega_0\in\R, \boldsymbol{\omega}\geq\mathbf{0}\\ \sum_{m\in\cC}\omega_m=1}}
    \sum_{t\in\fTpre}
    \biggl(\omega_0+\sum_{m\in\cC}\omega_m\Delta\widetilde{P}_{mt}
    -\frac{1}{M_T}\sum_{m\in\cT}\Delta\widetilde{P}_{mt}\biggr)^{\!2}
    +\zeta^2T_0\|\boldsymbol{\omega}\|^2,
  \label{eq:w_opt}\\
  (\hat{\tau}_0,\hat{\boldsymbol{\tau}})
  &:= \arg\min_{\substack{\tau_0\in\R, \boldsymbol{\tau}\geq\mathbf{0}\\ \sum_{t\in\fTpre}\tau_t=1}}
    \sum_{m\in\cC}
    \biggl(\tau_0+\sum_{t\in\fTpre}\tau_t\Delta\widetilde{P}_{mt}
    -\frac{1}{T_1}\sum_{t'\in\fTpost}\Delta\widetilde{P}_{mt'}\biggr)^{\!2}.
  \label{eq:r_opt}
\end{align}
The regularization parameter in Equation (\ref{eq:w_opt}) is $\zeta=(M_TT_1)^{1/4}\widehat\sigma$, where $\widehat\sigma$ is the standard deviation, pooled over control markets and pre-merger quarters, of the within-window first differences of residualized prices, that is, of the matched series excluding its first row.

The first term in Equation (\ref{eq:w_opt}) rewards weights that make the pre-merger price changes of the weighted controls resemble those of the treated markets. The second term regularizes the weights, penalizing solutions that place excessive weight on a small number of control markets. The intercept $\omega_0$ allows the weighted controls and treated markets to differ by a constant in their quarterly price changes, corresponding to a linear trend in the gap between their price levels. Equation (\ref{eq:r_opt}) analogously chooses pre-merger time weights by matching the post-merger price changes of the control markets.

The corresponding market and time weights are
\begin{eqnarray}
  \omega_m = \begin{cases}
    \hat{\omega}_m & m\in\cC,\\[2pt]
    1/M_T          & m\in\cT,
  \end{cases}\qquad
  r_t = \begin{cases}
    \hat{\tau}_t & t\in\fTpre,\\[2pt]
    1/T_1        & t\in\fTpost.
  \end{cases}
  \label{eq:combined_weights}
\end{eqnarray}
Thus, the treated markets and the control pool each receive total market weight one, while the pre- and post-merger periods each receive total time weight one.\footnote{In our empirical analysis, the time weights that solve Equation (\ref{eq:r_opt}) using the first differences of residualized prices are close to being uniform. Therefore, in Synthetic GMM estimation we set $\hat{\tau}_t=1/T_0, \forall t\in\fTpre$.} 

For notational ease, let $\mathcal{D}_{mt} := \{P_{mt}, Q_{mt}, \boldsymbol{Z}_{mt}, \boldsymbol{W}_{mt}, I_{mt}\} \in \mathcal{D} \subset \mathbb{R}^{d_{\mathcal{D}}}$ denote the data for market $m$ at time $t$, and let $\boldsymbol{\kappa} := (\boldsymbol{\theta}, \boldsymbol{\alpha}, \boldsymbol{\lambda}) \in \Sigma \subset \mathbb{R}^{d_{\kappa}}$ denote the full parameter vector, with $\boldsymbol{\kappa}^0$ denoting the true value. The moment function is $g: \mathcal{D} \times \Sigma \to \mathbb{R}^{d_g}$, where $d_g \geq d_{\kappa}$.

Having constructed the market and time weights, we now use them to reweight the moment conditions that would otherwise enter standard GMM. Our concern is specific to the supply side: the merger is modeled as affecting costs and conduct rather than demand, and the large pool of control markets is particularly important for the supply-side moments. We therefore keep the demand moments unchanged and apply the synthetic weights only to the supply moments. The factor model below makes precise how differences across control markets can generate supply-side instrument invalidity and why pre-merger price dynamics can be informative about this heterogeneity.

\begin{definition}[Synthetic GMM]\label{def:sgmm}
Partition the moment function into its demand-and-supply  blocks,
$g(\mathcal{D}_{mt};\bk)=\bigl(g^D(\mathcal{D}_{mt};\bk^D)^{\top}, g^S(\mathcal{D}_{mt};\bk^S)^{\top}\bigr)^{\top}$, as in Appendix \ref{app:affine}, such that $\bk^D\cap \bk^S=\emptyset$. The Synthetic GMM estimator is
$\hat{\bk}_{S\text{-GMM}}=\arg\min_{\bk\in\Sigma}\gtil(\bk)^{\top}\What\gtil(\bk)$, with the weighted sample moment $\gtil(\bk) = \left(\frac{1}{MT}\sum_{m=1}^{M}\sum_{t=1}^{T} g^D(\mathcal{D}_{mt};\bk^D), \sum_{m=1}^{M}\sum_{t=1}^{T}\omega_m r_t  g^S(\mathcal{D}_{mt};\bk^S)\right)^\top$, 
where $\omega_m$ and $r_t$ are defined in Equation (\ref{eq:combined_weights}), and the weighting matrix is block diagonal, $\What=\operatorname{diag}(\What_D,\What_S)$. The demand moments are the unweighted sample average; the SDiD weights enter only the supply moments. Finally, suppose there exist population weights $\omega^*=(\omega_1^*,\ldots,\omega_M^*)^{\top}$ and $r^*=(r_1^*,\ldots,r_T^*)^{\top}$ with $\omega_m^*\geq0$, $r_t^*\geq0$, and bounded total mass, such that $\muw(\bk)=\Big(\frac{1}{MT}\sum_{m,t}\E[g^D(\mathcal{D}_{mt};\bk^D)]^{\top},\ \sum_{m,t}\omega_m^*r_t^*\E[g^S(\mathcal{D}_{mt};\bk^S)]^{\top}\Big)^{\top}$, which is the population moment, 
 is well defined.
\end{definition}

We begin with the baseline case in which all moment conditions are correctly specified, so the instruments are valid for every market in the sample. In this case, reweighting valid supply moments does not change their population target, and Synthetic GMM is consistent.

\begin{theorem}[Baseline Consistency]\label{thm:sgmm_consistency}
Under Assumptions \ref{ass:standard_gmm_start}--\ref{ass:weight_bounds} defined in Appendix \ref{section:sgmm_formal}, Synthetic GMM is consistent, i.e., $\hat{\bk}_{S\text{-GMM}}\xrightarrow{p}\boldsymbol{\kappa}^0$, as $M,T\to\infty$.
\end{theorem}

The proof is in Appendix \ref{proof:sgmm_consistency}. Assumption \ref{ass:standard_gmm_start} requires that the moment conditions hold for every market, treated and control alike, i.e., $\mathbb{E}[g(\mathcal{D}_{mt};\boldsymbol{\kappa}^0)]=\mathbf{0}$ for all $m,t$. Under this condition, reweighting changes individual observations' contributions but does not distort the population moments. Consistency then follows from the standard extremum-estimator argument in \citet{Hansen1982} and \citet{NeweyMcFadden1994}, provided the synthetic DiD market weights $\hat{\omega}_m$ converge uniformly to well-defined population limits, satisfy the required bounds, and the weighted moment conditions identify the parameters. The time weights are fixed constants. We next consider the case in which the supply-side moment conditions may fail for some control markets.

\subsubsection{Bias Analysis\label{sec:biasanalysis}}

We now characterize and compare the asymptotic bias of Synthetic GMM and standard GMM when the supply-side moment conditions may fail for some control markets. We focus only on the linear demand-supply model, present the assumptions and main results here, and leave the intermediate steps, regularity conditions, and proofs to Appendix \ref{section:sgmm_formal}.

\paragraph{Factor representation and instrument validity.}

To connect this misspecification to observable market characteristics, we model the unobservable cost shock using a linear factor structure \citep[e.g.,][]{AbadieDiamondHainmueller2010}. Because the validity of the supply-side moment conditions may vary across control markets, we focus on the cost shock.

\begin{assumption}[Factor Model]
\label{ass:factor}
The cost shock satisfies $\vartheta_{mt}=\mumt^{\top}\mathbf F_t+\epsilon_{mt}$, where $\mathbf F_t\in\R^K$ is a vector of common time-varying factors, and the pre-merger factor matrix $\mathbf F_{\mathrm{pre}}=(\mathbf F_t^\top)_{t\in\fTpre}$
has full column rank. The vector $\mumt\in\R^K$ is a vector of
market-specific factor loadings fixed over time, and $\epsilon_{mt}$ is a
market-time idiosyncratic shock with
$\E[\epsilon_{mt}]=0$, $\mathbb V(\epsilon_{mt})=\sigma^2$, and
$\E[\mathbf F_t\epsilon_{mt}]=\mathbf 0$. Furthermore, letting
$\mathbf A_{mt}^S\in\R^{d_S}$ denote the vector of supply-side instruments,
the idiosyncratic shock is orthogonal to ${\mathbf A_{mt}}^S$, i.e., $\E[\mathbf A_{mt}^S\epsilon_{mt}]=\mathbf 0.$ 
\end{assumption}

The rank condition requires sufficient independent variation in the pre-merger changes of the common factors to identify differences in their market-specific loadings. In particular, pre-merger price changes cannot reveal the loading on a combination of factors whose change is constant across all pre-merger quarters.

The common factors capture aggregate cost conditions shared across markets, such as jet fuel prices, airport congestion, or industry-wide capacity conditions. The market-specific loading $\mumt$ determines how strongly each route is exposed to these factors. For example, a hub-to-hub route
may load heavily on business-travel factors, while a leisure route may load
heavily on vacation-demand factors. 
So, two similar routes have similar loadings. Markets with similar loadings therefore respond similarly to changes in common cost conditions.

Under the factor structure, supply-side IV validity can be written as $\E\bigl[\mathbf{A}_{mt}^S\vartheta_{mt}\bigr]
= \underbrace{\E\bigl[\mathbf{A}_{mt}^S\mathbf{F}_t^{\top}\bigr]}_{=\Gamt}\mumt
+ \underbrace{\E\bigl[\mathbf{A}_{mt}^S\epsilon_{mt}\bigr]}_{=\mathbf{0}}
= \Gamt\mumt=: \delt.$ 
Thus, the instrument is valid for market $m$ at time $t$ when $\delt=\mathbf 0$. For the merger evaluation to be interpretable, we require the instruments to be valid in the treated markets.

\begin{assumption}[IV Validity for Treated Markets]\label{ass:iv_treated}
The supply instruments are valid for every treated market in every period, so $\Gamt\mumt=\mathbf{0}$ for all $m\in\mathcal{T}$ and all $t$.
\end{assumption}
Assumption \ref{ass:iv_treated} provides a benchmark for evaluating the consequences of IV invalidity in the control markets. If the instruments are also invalid in the treated markets, the parameters relevant for merger evaluation cannot be identified, regardless of how the untreated markets are weighted. We therefore maintain IV validity in the treated markets and focus on the possible bias from invalid instruments in the untreated markets.

\paragraph{Bias of Synthetic GMM.}
Using the linearity of the demand-and-supply equations in the parameters $\bk$ (Appendix \ref{app:affine}), and Assumptions \ref{ass:factor}--\ref{ass:iv_treated}, the two blocks of the Synthetic GMM objective can therefore be analyzed separately because the Jacobian and weighting matrix are block diagonal, and the demand and supply moments depend on distinct parameter vectors. Let $\mathbf{G}^S_w=\sum_{m,t}\omega_m^*r_t^*\E[\mathbf{B}^S_{mt}]$, and let $\mathcal{W}_S$ be the probability limit of $\What_S$, and define $\mathbf{L}_{S,w}\equiv\bigl(\mathbf{G}_w^{S\top}\mathcal{W}_S\mathbf{G}^S_w\bigr)^{-1}\mathbf{G}_w^{S\top}\mathcal{W}_S$. Then the bias of Synthetic GMM is then 
\begin{equation}
\bksw-\bkz=\begin{pmatrix}\mathbf{0}\\ \mathbf{L}_{S,w}\dtil\end{pmatrix},
\label{eq:bias_syn}
\end{equation}
where the first block contains the demand parameters $\bk^D$, and $
\dtil:=\sum_{m\in\cC}\sum_t\omega_m^*r_t^*\Gamt\mumt$ is the \emph{weighted average invalidity} of the control pool. Thus, the supply-side bias is determined by the invalidity that remains after reweighting, and it is zero whenever $\dtil=\mathbf 0$.

The SDiD weights are constructed from price dynamics; see Equation \eqref{eq:price_reduced} for the linear model and Equation \eqref{eq:logprice_reduced} for the log-linear model. They may assign greater weight to untreated markets with cost and demand conditions similar to those in treated markets. Since these conditions affect equilibrium prices, price dynamics likely contain information about the underlying supply-side IV invalidity.\footnote{We construct the market weights using residualized prices alone because the reduced-form price equation incorporates the quantity channel through the demand equation, including the demand shock $\varphi_{mt}$.} If price dynamics are uninformative about this heterogeneity, however, Synthetic GMM offers no advantage relative to standard GMM that uses uniform weights, and in fact it may be less precise.

The linear model makes this link explicit. In the pre-merger period, strategic conduct is common across markets. Using Equation \eqref{eq:price_reduced} and the factor representation in Assumption \ref{ass:factor}, the residualized price can be written as $
\widetilde P_{mt}=\frac{\tilde\gamma_0\varphi_{mt}+\vartheta_{mt}}{1-\tilde\gamma_0\alpha_1}
=\frac{\mumt^{\top}\mathbf F_t+\tilde\gamma_0\varphi_{mt}+\epsilon_{mt}}{1-\tilde\gamma_0\alpha_1}.$
The same market-specific loading $\mumt$ therefore affects the pre-merger price dynamics used to construct the weights and, through $\delt=\Gamt\mumt$, the supply-side IV invalidity. Pre-merger price histories can therefore inform which control markets have factor loadings similar to those of the treated markets. The following assumption formalizes the condition under which balancing these price dynamics also balances the heterogeneity relevant for IV validity.

\begin{assumption}[Pre-Merger Price Balance]
\label{ass:balance}
The market weights in \eqref{eq:w_opt} achieve pre-treatment balance in the first differences of residualized prices, $\Delta\widetilde P_{mt}$, at rate $O_p(T_0^{-1/2})$, up to the intercept $\hat\omega_0$. Moreover, this balance implies the same rate for the market-specific factor loadings $\mumt$ that determine supply-side IV invalidity through $\delt=\Gamt\mumt$.
\end{assumption}

Assumption \ref{ass:balance} is a high-level condition linking the weights to the structural moments. We estimate the weights from price dynamics, while the bias in \eqref{eq:bias_syn} depends on $\mumt$.
The IV invalidity is $\delt=\Gamt\mumt$, so the assumption ensures that better balance in price dynamics improves balance in $\mumt$, reducing $\dtil$ and therefore the S-GMM bias. The quality of the pre-merger price balance provides empirical evidence for this condition.
We impose the $O_p(T_0^{-1/2})$ rate to capture the improvement in balance with the pre-merger periods. For more on pre-treatment fit and synthetic-control estimators see \cite{FermanPinto2021}.

\begin{theorem}[Bias of Synthetic GMM]
\label{thm:syn_bias}
Suppose Assumptions \ref{ass:factor}, \ref{ass:iv_treated}, and \ref{ass:balance} hold, together with the regularity conditions required for consistency and identification. Then $\hat{\bk}_{S-GMM}\convp\bksw$, where $\bksw-\bkz$ is given by \eqref{eq:bias_syn} and $\|\bksw-\bkz\|=O_p(T_0^{-1/2})$.
\end{theorem}

The proof is in Appendix \ref{section:sgmm_bias_details}. Under Assumption \ref{ass:balance}, the weighted invalidity $\dtil$ shrinks as the pre-merger window grows, so the bias in \eqref{eq:bias_syn} converges to zero. In our empirical application, $T_0$ is finite and equals eight quarters, so the result provides a formal link between improved pre-merger balance and lower bias.\footnote{For notational simplicity, we present the results for the level moment conditions. The precise conditions are stated in Appendix \ref{section:sgmm_formal}. Our estimating equations are in first differences; the argument applies after replacing $g(\mathcal{D}_{mt};\kappa)$ by $\Delta g(\mathcal{D}_{mt};\kappa)$ and adjusting the other covariates, under the conditions stated in Appendix \ref{section:sgmm_bias_details}. The market weights are solved on the pre-merger first differences of residualized prices (Equation \ref{eq:w_opt}), the same transformation used for the moments. Because the factor loadings are fixed over time, balance in price changes and balance in price levels target the same loadings, subject to the rank condition in Assumption \ref{ass:factor}.}

\paragraph{Bias of Standard GMM.}

Standard GMM is the special case with uniform weights. Following the same argument as above, the relevant misspecification term is the \emph{average instrument invalidity in the control pool}, $\dbar$, multiplied by the control-market share $M_C/M$. Let $
\mathbf L=(\mathbf G^{\top}\Wmat\mathbf G)^{-1}\mathbf G^{\top}\Wmat\in\mathbb R^{d_\kappa\times d_g},$ 
where $\mathbf G$ is the Jacobian of the population moment conditions and $\Wmat$ is the GMM weighting matrix. Partition $\mathbf L$ conformably with the demand-and-supply  moments as $\mathbf L=(\mathbf L_D,\mathbf L_S)$, where $\mathbf L_D\in\mathbb R^{d_\kappa\times d_D}$ and $\mathbf L_S\in\mathbb R^{d_\kappa\times d_S}$, with $d_g=d_D+d_S$. Because the demand moments remain valid and misspecification arises only through the supply-side instruments, the bias is
\begin{eqnarray}
\bks-\bkz
&=&\frac{M_C}{M}\mathbf L
\begin{pmatrix}
\mathbf 0\\
\dbar
\end{pmatrix}
=\frac{M_C}{M}\mathbf L_S\dbar.
\label{eq:bias_std}
\end{eqnarray}
Thus, supply-side moment violations bias the structural parameters whenever $\mathbf L_S\dbar\neq\mathbf 0$.

In our estimator, the weighting matrix is block diagonal across the demand-and-supply moments, $\What=\operatorname{diag}(\What_D,\What_S)$. Because the Jacobian is also block diagonal and no parameter enters both blocks (Appendix \ref{app:affine}), the demand-and-supply parameters are estimated separately. The demand rows of $\mathbf L_S$ in Equation \eqref{eq:bias_std} are therefore zero, consistent with the zero demand block in Equation \eqref{eq:bias_syn}. Consequently, supply-side instrument invalidity does not bias the demand parameters in our estimator. This separation relies on the block-diagonal weighting matrix; with nonzero off-diagonal blocks, such as an efficient weighting matrix constructed from the joint covariance of the demand-and-supply moments, the demand rows of $\mathbf L_S$ need not be zero.

\begin{theorem}[Bias of GMM]
\label{thm:std_bias}
Suppose Assumptions \ref{ass:factor} and \ref{ass:iv_treated} hold, together with the regularity conditions required for consistency and identification (as stated in Appendix \ref{section:sgmm_formal}). If $\mathbf L_S\dbar\neq\mathbf 0$, then $\hat{\bk}_{\mathrm{GMM}}\convp\bks$, where $\bks-\bkz$ is given by Equation \eqref{eq:bias_std}, and $\bks\neq\bkz$. Furthermore, if $\mathbf L_S\dbar$ does not converge to $\mathbf 0$ as $M,T\rightarrow\infty$, then the asymptotic bias does not vanish.
\end{theorem}

The proofs are in Appendix \ref{section:sgmm_bias_details}. Theorems \ref{thm:syn_bias} and \ref{thm:std_bias} differ in how untreated-market invalidity enters the bias. Standard GMM depends on the unweighted average invalidity $\dbar$, which need not shrink as the pre-merger window grows. Synthetic GMM instead depends on the weighted invalidity $\dtil$, which is $O_p(T_0^{-1/2})$ under Assumption \ref{ass:balance}. Thus, the advantage of Synthetic GMM depends on the information contained in pre-merger prices. Under the factor structure and balance condition, SDiD weighting reduces the contribution of invalid untreated-market moments as the pre-merger window grows. Without this relationship, matching price dynamics do not reduce the bias, which we show in the Monte Carlo exercise.

\subsection{Monte Carlo Experiments}\label{sec:montecarlo}

We use Monte Carlo experiments to evaluate S-GMM's finite-sample performance when some untreated markets violate the supply-side moment conditions. This exercise is designed to illustrate the distinction between Theorems \ref{thm:syn_bias} and \ref{thm:std_bias}. When the supply-side IVs are valid in every market, GMM and S-GMM should recover the same parameters. When the IVs remain valid in treated markets but fail in some control markets, those invalid moments contaminate GMM. Synthetic GMM can reduce this contamination when the market heterogeneity generating instrument invalidity is reflected in the pre-merger price dynamics used to construct the SDiD weights. In this exercise, we vary the fraction of invalid markets and compare the bias and precision of the two estimators. 

\paragraph{Data Generating Process.}
We simulate data from the linear demand-and-supply model from Section \ref{sec:structural_retrospective}. Demand in market $m$ and period $t$ is $
Q_{mt}=\alpha_0+\alpha_1P_{mt}+\alpha_2Z_{mt}+\alpha_3Z^{(2)}_{mt}+\varphi_{mt},$ 
where $(\alpha_0,\alpha_1,\alpha_2,\alpha_3)=(10,-2,1,1)$, $Z_{mt}$ and $Z^{(2)}_{mt}$ are independent standard normal variables, i.e., $Z_{mt}, Z^{(2)}_{mt})\sim {\mathcal N}(0, 1)$, and $\varphi_{mt}\sim {\mathcal N}(0,0.5^2)$. Marginal cost is $C_{mt}=\theta_0+\theta_1Q_{mt}+\theta_2W_{mt}+\theta_3I_{mt}+\vartheta_{mt},$ 
where $(\theta_0,\theta_1,\theta_2,\theta_3)=(2,0.3,0.5,-0.5)$ and $W_{mt}\sim {\mathcal N}(0,1)$. Pre-merger conduct is $\lambda_{\mathrm{pre}}=0.3$ and the merger increases conduct by $\Delta\lambda=0.15$. Thus, $\tilde\gamma_0=\theta_1-\lambda_{\mathrm{pre}}/\alpha_1=0.45$ before the merger and $\tilde\gamma_1=0.525$ after the merger. The demand and pricing equations determine prices and quantities.

The cost shock has the factor structure $
\vartheta_{mt}=\varpi_{m1}F_{1t}+\varpi_{m2}F_{2t}+\varepsilon_{mt},$ 
where $F_{1t}$ and $F_{2t}$ are independent AR(1) processes with autoregressive coefficient $0.6$ and innovation variance $0.64$, and $\varepsilon_{mt}\sim {\mathcal N}(0,0.5^2)$. The factor loadings vary across markets. For the second factor, control-market loadings satisfy $\varpi_{m2}\sim {\mathcal N}(1,0.3^2)$ in the baseline design, while treated-market loadings are drawn from the same distribution. Additional designs vary the location and dispersion of the treated-market loadings relative to the controls.

The first factor generates the supply-side moment violation. By construction, treated markets satisfy the supply moment conditions. Among control markets, we designate a fraction $\pi$ as invalid. The supply instrument is $A^S_{mt}=\frac{\nu_{mt}+dZ^{(2)}_{mt}+cF_{1t}}{\operatorname{sd}(A^S)},$
where $d=1$. The $Z^{(2)}_{mt}$ component provides relevance, while the $F_{1t}$ component generates correlation between the instrument and the cost shock when $\varpi_{m1}\neq0$. We set $\varpi_{m1}=0$ for treated markets and valid control markets. For invalid control markets, $\varpi_{m1}$ is nonzero, so that $\E[A^S_{mt}\vartheta_{mt}]\neq0$. 
The coefficient $c$ controls the magnitude of this violation. 

The design also links instrument invalidity to the information used by S-GMM. The second factor loading $\varpi_{m2}$ selects invalid control markets by their distance from the treated markets, while the magnitude of their first loading $\varpi_{m1}$ depends on that distance. As a result, the same market heterogeneity that makes some controls less comparable to the treated markets is reflected in their pre-merger price dynamics. Synthetic GMM can therefore use those dynamics to reduce the weight placed on untreated markets that contribute more strongly to the invalid moments. This construction is the finite-sample analog of the link imposed by Assumption \ref{ass:balance}. Appendix \ref{app:mc_design_appendix} gives the exact construction and also considers designs in which instrument invalidity is less informative, or uninformative, about pre-merger price similarity.

Our baseline sample contains $M_T=10$ treated markets, $M_C=400$ control markets, and $T_0=T_1=8$ pre- and post-merger periods. The main experiment varies the fraction of invalid control markets over $\pi\in\{0,0.3,0.5\}$. For each $\pi$, we simulate $R=500$ times. Appendix \ref{app:mc_design_appendix} reports additional experiments varying the magnitude and detectability of instrument invalidity, the distribution of factor loadings, the number of treated and control markets, and the length of the pre-merger window.

For a scalar parameter $\psi$ with true value $\psi_0$ and estimator $j\in\{\mathrm{GMM},\mathrm{S-GMM}\}$, let $\hat\psi_{jr}$ denote its estimate in Monte Carlo replication $r$. We report the absolute value of the median bias, i.e., $|\operatorname{median}_r(\hat\psi_{jr}-\psi_0)|,$ and the root mean squared error, i.e., $\operatorname{RMSE}(\hat\psi_j)=\sqrt{\frac{1}{R}\sum_{r=1}^{R}(\hat\psi_{jr}-\psi_0)^2}$. 
For brevity and ease of presentation, we focus only on the cost-efficiency parameter, $\theta_3$, and the change in strategic conduct, $\Delta\lambda$.

\begin{table}[t!!]
\centering
\caption{Monte Carlo Performance for the Cost Efficiency Parameter\label{tab:mc_results}}
\begin{threeparttable}
\begin{tabular}{lcccc}
\toprule
& \multicolumn{2}{c}{Absolute Median Bias} & \multicolumn{2}{c}{RMSE} \\
\cmidrule(lr){2-3}\cmidrule(lr){4-5}
Invalid Markets & GMM & S-GMM & GMM & S-GMM \\
 &  & (as \% of GMM) &  & (as \% of GMM) \\
\midrule
0\%   & 0.100 & 73.9 & 2.31 & 42.8 \\
30\%  & 0.293 & 34.3 & 4.05 & 29.1 \\
50\%  & 0.525 & 14.1 & 3.20 & 33.8 \\
\bottomrule
\end{tabular}
\begin{figurenotes}The table reports Monte Carlo performance of GMM and S-GMM for the merger-induced cost efficiency parameter, $\theta_3$. Each design uses $R=500$ Monte Carlo replications. The fraction of invalid markets is the share of untreated markets for which the supply-side instrument violates the moment condition. Demand moments and supply moments for treated markets are always valid. 
\end{figurenotes}
\end{threeparttable}
\end{table}

Table \ref{tab:mc_results} reports the main comparison for the cost-efficiency parameter, $\theta_3$. When all control-market moments are valid, $\pi=0$, both standard and Synthetic GMM estimate the parameters around their true values. The bias of standard GMM across the relevant designs increases from $0.100$ when no controls are invalid to $0.293$ when 30 percent are invalid and $0.525$ when 50 percent are invalid. The corresponding S-GMM biases are $0.074$, $0.101$, and $0.074$. Thus, increasing the share of invalid controls progressively shifts standard GMM away from the true parameter, whereas the S-GMM is substantially less sensitive to the additional invalid moments. The same pattern appears in RMSE: across these designs, S-GMM's RMSE is between 29.1 and 42.8 percent of standard GMM's RMSE.

\begin{figure}[t!]
\centering
\caption{Histograms of the Estimates\label{fig:mc_hist}}
\includegraphics[width=\textwidth]{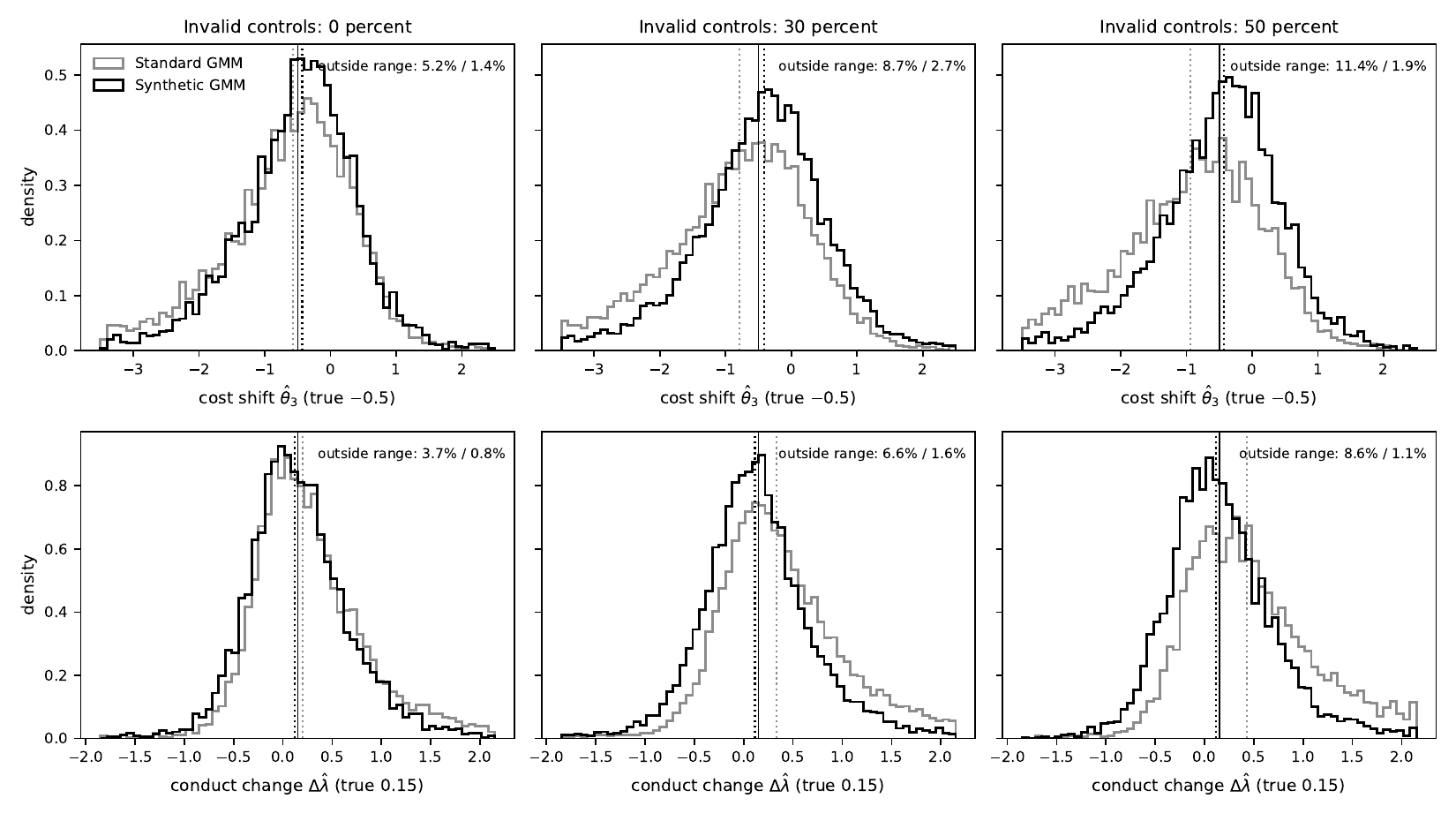}
\begin{figurenotes}
 Each panel shows histograms of the estimates from 500 Monte Carlo replications for standard GMM and Synthetic GMM. The solid vertical line indicates the true parameter value, and the dashed vertical lines indicate the median estimates for the two estimators. The percentages reported in each panel
indicate the share of estimates outside the plotted range, with standard GMM reported first.
\end{figurenotes}
\end{figure}

Figure \ref{fig:mc_hist} presents the Monte Carlo density (histogram) of the
estimated efficiency and conduct effects.\footnote{For numerical convenience, the simulations apply synthetic weights to both demand- and supply-side moments. Weighting only the supply-side
moments yields materially similar results.} 
As the fraction of invalid controls increases, the standard GMM estimate histogram shifts away from the true parameter and develops a longer tail. The density of S-GMM estimates is more stable. These results support the claim that S-GMM performs better than GMM because, in this design, the SDiD weights contain information about which control markets contribute more strongly to the misspecified moments.

The gains from Synthetic GMM in these designs depend on pre-merger price dynamics being informative about which control markets have valid supply moments. 
In Appendix \ref{app:mc_design_appendix} we verify this mechanism by varying how informative pre-merger prices are about the heterogeneity generating instrument invalidity. 
As expected, S-GMM performs better when this relationship is stronger and when a longer pre-merger window provides more information for constructing the weights. In Appendix \ref{app:mc_design_appendix}, we examine this mechanism by varying how informative pre-merger prices are about instrument invalidity. When this relationship is
stronger, Synthetic GMM places less weight on invalid
controls and vice versa.  This exercise shows that Synthetic GMM can use price dynamics to reduce the influence of invalid control markets when those dynamics are
informative about moment violations. The gains
therefore come from the information used to construct
the weights (Assumption \ref{ass:balance}), rather than from reweighting itself.

\section{Empirical Findings \label{sec:structural_estimates}}

We now present the empirical findings for each of the four mergers. 
We consider both the linear and log-linear specifications and, for the linear model, report estimates from standard GMM and Synthetic GMM. Our main specification allows the merger-induced change in conduct to vary with pre-merger market structure, measured by the number of competitors.

\subsection{Instrumental Variables}

Identification relies on demand and cost shifters that generate variation in prices and quantities, but are excluded from the demand and cost equations. Demand shifters affect consumers' willingness to travel without directly affecting airlines' marginal costs, while cost shifters affect airlines' costs or competitive conditions without directly affecting consumer utility. Because price and quantity are jointly determined, shifters on one side of the market can be used to instrument for the other side.

\paragraph{Demand Shifters.}
We use three demand shifters. The first two are the number of markets served from the origin, and the number of markets served from the destination, excluding the markets served by the merging airlines. These variables capture the value of access to the air-travel network because serving more destinations makes air travel more attractive relative to other modes of transportation. Although an individual carrier's network can affect its marginal costs \citep{CilibertoMurryTamer2021}, our measures aggregate destinations served by other (non-merging) carriers and so they likely affect demand without directly changing the merging airlines' costs. The third shifter is population-based market size, which captures changes in potential travel demand in the market.

The demand equation is estimated in first differences, with the change in price as the endogenous variable. Its excluded instruments combine changes in the cost shifters described below, interactions of the demand and cost shifters with the treated-market indicator, and lagged pre-merger levels of the demand shifters. This gives ten excluded instruments each for DL--NW, UA--CO, and AA--US and eight for WN--FL, where the Southwest potential-entrant instrument is omitted because Southwest is a merging carrier.\footnote{We construct the treatment interactions by multiplying changes in the three demand shifters and two cost shifters by the treated-market indicator. This gives five interactions for DL--NW, UA--CO, and AA--US and four for WN--FL, where the Southwest shifter is omitted because Southwest is a merging carrier. The lagged instruments use origin and destination network size in the last pre-merger quarter and market size four quarters earlier, each entering at the first post-merger transition. Market size is measured by the annual change in population and standardized within each merger sample to have mean zero and standard deviation one. Although the legacy-carrier count affects costs, we exclude it from the demand instruments.}

\paragraph{Cost Shifters.}
We use three types of cost shifters. The first measures the number of airlines present at both the origin and destination airports, capturing the pool of potential entrants. Following the literature \citep[e.g.,][]{GoolsbeeSyverson2008}, presence at both endpoints increases the probability of entering a route and affects competitive conditions. We distinguish legacy carriers from other potential entrants because their costs and competitive effects may differ. The second shifter indicates whether Southwest is a potential entrant and captures the competitive pressure associated with its presence at both endpoints.

The third cost shifter is the number of distinct regional carriers an airline uses to serve the market. Regional-carrier use is related to operating costs through aircraft utilization, fuel consumption, crew requirements, and network organization \citep{ForbesLederman2009, AryalCampbellCilibertoKhmelnitskaya2026}. Conditional on the observed product characteristics, we assume that regional-carrier use does not directly affect the demand. Together, these variables provide excluded supply-side variation for identifying the pricing equation.

\subsection{Empirical Specification}

Before presenting the results, we present the first-difference pricing equation we estimate. To maintain consistency with the DiD reasoning that led to Equation \eqref{eq:DiD_fd}, we allow the merger-induced cost effect to evolve over the eight quarters since the merger. 
For the linear model, the pricing equation in \eqref{eq:systems} becomes
\begin{equation}
P_{mt}=\theta_0+\left(\theta_1-\frac{\lambda_{mt}}{\alpha_1}\right)Q_{mt}+\theta_2W_{mt}+\sum_{k=1}^{8}\theta_{3,k} {\mathcal I}_{m,t,k}+\tilde\digamma_m+\vartheta_{mt},
\label{eq:structural_price_results}
\end{equation}
where, recall that $ {\mathcal I}_{m,t,k}=1$ if market $m$ has been at least $k$ quarters post-merger at time $t$, and zero otherwise. This is the same cumulative definition used in Section \ref{sec:retrospective}, so that $\Delta  {\mathcal I}_{m,t,k}=\text{Transition}_{m,t,k}$. As in the DiD specification, we exclude the merger quarter and difference the first post-merger quarter against the last pre-merger quarter.

Let $D^{j}_{mt}$, $j\in\{\text{control},\text{treated-pre},\text{treated-post}\}$, indicate control (i.e., untreated) and treated markets and, within treated markets, before and after the merger, and let $\text{Transition}_{mt}$ denote the quarter in which a treated market moves from pre- to post-merger. First differencing Equation \eqref{eq:structural_price_results} gives us the estimating equation:
\begin{eqnarray*}
\Delta P_{mt} &= &\left[\sum_{j} \left(\theta_1 - \frac{\lambda_j}{\alpha_1}\right) D^j_{mt}\right] \Delta Q_{mt}
- \frac{\Delta\lambda_m}{\alpha_1}  Q_{m,t-1}  \text{Transition}_{mt} \\
&&+ \theta_2 \Delta W_{mt} + \sum_{k=1}^{8} \theta_{3,k}  \text{Transition}_{m,t,k} + \Delta \vartheta_{mt}.
\end{eqnarray*}
The first-differenced pricing equation therefore separates the merger-induced change in conduct, $\Delta\lambda_m$, from the sequence of merger-induced cost shifts $\{\theta_{3,k}\}$. We summarize the latter by $\bar{\theta}_3:=\sum_{k=1}^{8}\frac{9-k}{8}\theta_{3,k},$
the average cost effect over the eight post-merger quarters.

\paragraph{Conduct and market structure.}
The theoretical model treats $\Delta\lambda$ as common across treated markets. 
In the estimation of the linear model, we allow it to vary with pre-merger market structure to capture the fact that coordination can be easier to sustain among fewer firms \citep{Harrington2017}. 
To implement this idea, we let the merger-induced change in conduct depend on the number of airlines $N_m$ serving the market before the merger: $\lambda_m^j=\lambda_0^j+\lambda_N^j \log N_m$ for $j\in\{\mathrm{pre},\mathrm{post}\}$, so that $\Delta\lambda_m=\Delta\lambda_0+\Delta\lambda_N\log N_m$. A larger $\lambda$ denotes less competition, so a negative $\Delta\lambda_N$ means a softening that is larger in markets with fewer competitors. Substituting into the pricing equation, the coefficients on $Q_{mt}$ and on $Q_{mt}\log N_m$ are, respectively, $\varsigma_0^j=\theta_1-\lambda_0^j/\alpha_1$ and $\varsigma_N^j=-\lambda_N^j/\alpha_1$, identified from variation in $Q_{mt}$ and $N_m$ before and after the merger. Since $\alpha_1$ is identified from the demand equation, $\Delta\lambda_0=-\alpha_1(\varsigma_0^{\mathrm{post}}-\varsigma_0^{\mathrm{pre}})$ and $\Delta\lambda_N=-\alpha_1(\varsigma_N^{\mathrm{post}}-\varsigma_N^{\mathrm{pre}})$. The same holds in first differences, where the transition row carries $-(\Delta\lambda_0+\Delta\lambda_N\log N_m)Q_{m,t-1}/\alpha_1$ and the pricing instruments are interacted with $\log N_m$ for the quantity interactions. We maintain that pre-merger $N_m$ is exogenous to the structural shocks conditional on the shifters and instruments, and interpret $\Delta\lambda_N$ as heterogeneity in the conduct effect by market structure, not as the effect of adding or removing a competitor.

In the log-linear model, conduct enters through the intercept, as $-\Lambda$ with $\Lambda=\log(1+\lambda/\alpha_1)$, so the level of the change, $\Delta\Lambda_0$, enters only together with the first-quarter cost shift and is not identified; its gradient in $\log N_m$, $\psi_N=-\Delta\Lambda_N$, is identified under a cost shift common across $N_m$. We maintain that pre-merger $N_m$ is exogenous to the structural shocks conditional on the shifters and instruments, and read $\Delta\lambda_N$ and $\psi_N$ as heterogeneity of the conduct effect by market structure, not as the effect of adding or removing a competitor.

\subsection{Inference with Few Treated Markets.} Some of the merger parameters are identified from only a small number of treated markets (Section \ref{sec:data}). For inference on these parameters, we follow \citet{ConleyTaber2011}, while maintaining the assumption that the merger is exogenous: we assign the merger to placebo sets of untreated markets, as many as were treated and with the same merger date, re-solve the market weights and re-estimate the model in every draw, $B=100$ times, and use the placebo estimates, centered at their mean, as the reference distribution under the null of no merger effect, for each merger effect, for the two jointly, and for the price effect. Appendix \ref{sec:few_treated_inference} gives the procedure, the $p$-values, and the 95 percent placebo intervals.

\subsection{Findings}
Table \ref{tab:merger_effects} reports the estimates for the four mergers. In the linear model, we separately estimate changes in marginal cost and conduct and their contributions to the overall price effect. The log-linear model identifies the overall proportional price effect and the conduct gradient. We report standard GMM and Synthetic GMM estimates with 95\% placebo-based confidence intervals, alongside the corresponding DiD estimates for comparison.\footnote{Demand estimates, elasticities, and first-stage statistics are reported in Appendix \ref{sec:appendix_gmm_full}, and standard errors in Appendix Tables \ref{tab:conductN_main} and \ref{tab:weighted_system}. The estimated demand coefficient is negative in all cases. As explained above, we also use placebo inference due to the small number of treated markets (Appendix \ref{sec:few_treated_inference}).}

Overall, we find no evidence of merger-induced cost savings. Estimated price effects are generally small relative to their uncertainty, except for WN--FL, for which the structural estimates indicate an increase of approximately \$10 per one-way trip.

\paragraph{Efficiencies.}
For DL--NW and AA--US, the estimated changes in marginal cost are close to zero under both standard and Synthetic GMM. For UA--CO, standard GMM estimates a marginal-cost increase of \$6.2, but its 95\% placebo interval is $[-\$21.5,\$33.4]$. Thus, we find no evidence of either cost savings or cost increases for these three mergers. For WN--FL, by contrast, Synthetic GMM estimates a marginal-cost increase of approximately \$27, with a placebo interval that excludes zero, suggesting that the merger increased costs.

Estimated pass-through rates range from 0.78 to 0.94 across specifications, implying that most changes in marginal cost are reflected in fares. Consequently, differences in estimated cost and conduct changes, and not the differences in cost pass--through, account for most of the variation in price effects across mergers.

\paragraph{Market Power.}
We next examine whether the mergers changed competitive conduct. Recall that a higher $\lambda$ indicates less competitive conduct and that the conduct effect may

\begin{landscape}
\thispagestyle{empty}
\begin{table}[htp!]
\centering
\caption{Merger Effects: Efficiencies, Market Power and Prices\label{tab:merger_effects}}
\scalebox{0.865}{\hspace{-1.2in}
\begin{tabular}{lccccccc|ccc}
\toprule
& \multicolumn{7}{c|}{Linear model (\$ per one-way trip)}
& \multicolumn{3}{c}{Log-linear model} \\
\cmidrule(lr){2-8}\cmidrule(lr){9-11}
& \multicolumn{2}{c}{Conduct $\Delta\lambda(N)$}
& Cost
& Efficiency
& Market-power
& \multicolumn{2}{c|}{Price effect}
& Conduct
& \multicolumn{2}{c}{Price effect} \\
\cmidrule(lr){2-3}\cmidrule(lr){7-8}\cmidrule(lr){10-11}
Merger
& $N=3$ & $N=5$
& effect
& component
& component
& Model & DiD
& gradient
& Model & DiD \\
\midrule

\multicolumn{11}{l}{DL--NW} \\

$-$\emph{GMM}
& $0.121$ & $0.058$ & $-3.7$ & $-3.2$ & $3.7$ & $0.5$ & $-1.4$

& $-0.118$ & $0.007$ & $-0.004$ \\

&

$[0.015, 0.181]$ & $[0.000, 0.091]$ & $[-9.7, 10.5]$ & $[-8.9, 10.3]$

& $[-7.8, 9.9]$ & $[-6.0, 9.6]$ & $[-12.3, 8.1]$

& $[-0.108, 0.012]$ & $[-0.026, 0.042]$ & $[-0.055, 0.040]$ \\

$-$\emph{S-GMM}

& $0.098$ & $0.038$ & $-2.4$ & $-2.1$ & $2.6$ & $0.4$ & $3.5$

& $-0.107$ & $0.000$ & $0.012$ \\

&

$[-0.013, 0.165]$ & $[-0.020, 0.072]$ & $[-7.9, 11.7]$ & $[-8.1, 12.0]$

& $[-8.7, 8.8]$ & $[-6.0, 10.0]$ & $[-2.6, 10.8]$

& $[-0.097, 0.024]$ & $[-0.028, 0.033]$ & $[-0.020, 0.049]$ \\[0.4em]

\multicolumn{11}{l}{UA--CO} \\

$-$\emph{GMM}

& $0.247^{\dagger}$ & $0.103$ & $6.2$ & $5.1$ & $1.1$ & $6.2$ & $0.4$

& $-0.085$ & $0.015$ & $-0.009$ \\

&

$[-0.598, 1.144]$ & $[-0.249, 0.481]$ & $[-21.5, 33.4]$ & $[-23.0, 45.9]$

& $[-36.1, 34.1]$ & $[-16.2, 37.1]$ & $[-15.6, 14.8]$

& $[-0.271, 0.063]$ & $[-0.045, 0.077]$ & $[-0.087, 0.062]$ \\

$-$\emph{S-GMM}$^{\spadesuit}$

& $n/a$ & $n/a$ & $n/a$ & $n/a$ & $n/a$ & $n/a$ & $5.5$

& $-0.088$ & $0.010$ & $0.015$ \\

&

& & & & & & $[-5.9, 17.7]$

& $[-0.275, 0.058]$ & $[-0.048, 0.071]$ & $[-0.037, 0.073]$ \\[0.4em]

\multicolumn{11}{l}{WN--FL} \\

$-$\emph{GMM}

& $-0.121$ & $-0.065$ & $38.0$ & $37.1$ & $-26.8$ & $10.3$ & $6.4$

& $-0.181$ & $0.078$ & $0.041$ \\

&

$[-0.220, -0.004]$ & $[-0.111, -0.011]$ & $[24.9, 48.1]$ & $[23.2, 47.9]$

& $[-34.6, -17.2]$ & $[2.9, 16.5]$ & $[-2.7, 14.3]$

& $[-0.272, -0.116]$ & $[0.047, 0.104]$ & $[-0.003, 0.079]$ \\

$-$\emph{S-GMM}$^{\ddagger}$

& $-0.072$ & $-0.038$ & $27.2$ & $26.3$ & $-15.4$ & $10.9$ & $8.4$

& $-0.129$ & $0.065$ & $0.076$ \\

&

$[-0.175, 0.049]$ & $[-0.085, 0.016]$ & $[14.5, 37.2]$ & $[13.4, 36.4]$

& $[-23.6, -5.3]$ & $[3.6, 17.2]$ & $[1.4, 13.4]$

& $[-0.220, -0.064]$ & $[0.035, 0.090]$ & $[0.046, 0.098]$ \\[0.4em]

\multicolumn{11}{l}{AA--US} \\

$-$\emph{GMM}

& $0.132$ & $0.036$ & $-0.2$ & $-0.2$ & $3.1$ & $2.9$ & $-0.2$

& $0.006$ & $0.012$ & $0.000$ \\

&

$[-0.130, 0.420]$ & $[-0.084, 0.163]$ & $[-14.6, 11.7]$ & $[-18.8, 13.1]$

& $[-16.8, 27.4]$ & $[-7.8, 12.6]$ & $[-11.4, 9.6]$

& $[-0.078, 0.108]$ & $[-0.022, 0.043]$ & $[-0.054, 0.047]$ \\

$-$\emph{S-GMM}

& $0.145$ & $0.040$ & $-0.4$ & $-0.3$ & $3.5$ & $3.2$ & $4.8$

& $0.000$ & $0.012$ & $0.010$ \\

&

$[-0.087, 0.409]$ & $[-0.080, 0.169]$ & $[-13.9, 10.6]$ & $[-17.0, 12.7]$

& $[-13.7, 23.9]$ & $[-8.0, 13.8]$ & $[-3.6, 12.6]$

& $[-0.077, 0.097]$ & $[-0.021, 0.043]$ & $[-0.026, 0.043]$ \\
\bottomrule
\end{tabular}}

\begin{figurenotes}
The table reports merger effects estimated separately for each merger using the combined sample of nonstop and connecting markets. For each merger, we report standard GMM and Synthetic GMM (S-GMM) estimates, together with the corresponding first-difference DiD estimate of the price effect. The left panel reports estimates from the linear model in dollars per one-way trip. The right panel reports estimates from the log-linear model in log points. In the linear model, the change in conduct, $\Delta\lambda(N)=\Delta\lambda_0+\Delta\lambda_N\log N$, is reported at $N=3$ and $N=5$. The cost effect, $\bar{\theta}_3$, is the average merger-induced change in marginal cost over the eight post-merger quarters. The efficiency component is the resulting price effect after post-merger conduct pass-through, and the market-power component is the price effect of the conduct change. Their sum is the structural price effect. In the log-linear model, the conduct gradient is $\psi_N/(1-\alpha_1\theta_1)$. Under the restriction that the cost shift is common across $N$, the conduct change is not separately identified. Brackets report 95 percent placebo intervals based on $B=100$ placebo assignments.\\
$^{\dagger}$ For UA--CO, $N=3$ lies below the range of pre-merger competitor counts in treated markets.\\
$^{\ddagger}$ The WN--FL S-GMM pricing first stage is $4.01$, so placebo inference can over-reject and under-cover.\\
$^{\spadesuit}$ UA--CO linear S-GMM estimates are omitted (and marked n/a) because the estimated pass-through denominator changes sign across observed market structures.\end{figurenotes}
\end{table}
\end{landscape}
\noindent depend on the pre-merger number of competitors. For DL--NW, UA--CO, and AA--US, the estimated conduct changes are positive and larger in markets with fewer firms. By contrast, for WN--FL, conduct decreases, suggesting more competitive conduct after the merger.

The evidence is mixed. Under standard GMM, the placebo intervals for the conduct changes exclude zero for DL--NW and WN--FL, but not for UA--CO or AA--US. Under Synthetic GMM, the estimated changes are generally smaller, and their placebo intervals include zero. The log-linear estimates provide additional evidence of a more competitive conduct for WN--FL, while the conduct gradients for the other mergers are imprecise.Conduct is identified through changes in the relationship between prices and quantities, and this requires sufficient variation in the instruments. The first-stage statistics in the airline data can be weak. Weak first stages make it difficult to separately identify cost and conduct effects, even when the overall price effect can be estimated more precisely.

\paragraph{Price Effects.}
We combine the estimated cost and conduct changes to obtain the overall effect on one-way fares, averaged across treated markets. The clearest result is for WN--FL, where both standard and Synthetic GMM estimate a fare increase of about \$10 per one-way trip, and the effect is statistically significant. The log-linear estimates imply price increases of approximately 6.5\% to 7.8\%, consistent with a positive price effect under both functional forms.

The different components of the price effects help explain this result. Under standard GMM, the increase in marginal cost raises fares by an estimated \$37.1, while the change toward more competitive conduct lowers them by \$26.8. The resulting net increase is \$10.3. Under Synthetic GMM, the corresponding components are \$26.3 and $-\$15.4$, yielding a net increase of \$10.9. These estimates suggest that higher marginal costs resulted in the fare increase, which was partly offset by a shift toward more competitive conduct.\footnote{One qualification for WN--FL is that, as discussed in Footnote \ref{footnote:WN_bag}, AirTran charged separately for baggage and reservation changes, whereas Southwest included these services in the fare. Before the merger, AirTran collected up to \$12.70 per one-way passenger trip from these fees, an amount comparable to the estimated \$10 fare increase. Part of the measured increase may therefore reflect the shift from AirTran's unbundled pricing to Southwest's bundled pricing rather than an increase in the price of an unchanged bundle of services.}

For the other three mergers, we find price effects are statistically indistinguishable from zero. Standard GMM estimates fare changes of \$0.5 for DL--NW, \$6.2 for UA--CO, and \$2.9 for AA--US, all with placebo intervals containing zero. The corresponding Synthetic GMM estimates for DL--NW and AA--US are also small, and the log-linear estimates lead to the same conclusion. Although some conduct estimates suggest less competitive behavior, the evidence is too imprecise to establish an increase in fares.

The structural estimates can also be used to evaluate price effects outside the treated markets. These estimates are close to the treated-market effects; see Appendix Table \ref{tab:weighted_system}.

\paragraph{Comparing with DiD estimates.}
We can compare the structural estimates with the DiD estimates. For standard GMM, we use the first-difference DiD estimates with controls from Table \ref{tab:did_main}. For Synthetic GMM, we use synthetic DiD with the same market weights as the structural estimator. Thus, within each row, the structural and reduced-form estimates use the same comparison markets and weights.

The DiD estimates broadly corroborate the structural approach. Neither approach finds evidence of a fare reduction. For WN--FL, the first-difference DiD estimate is \$6.4, and the synthetic-DiD estimate is \$8.4, compared with structural estimates of approximately \$10. The first-difference placebo interval includes zero, whereas the synthetic-DiD interval excludes zero. For the other mergers, the DiD estimates are statistically indistinguishable from zero. Thus, the two approaches yield a similar assessment of the overall price effects. In contrast, the structural approach also shows how these effects depend on the changes in marginal cost and competitive conduct.

\paragraph{Synthetic versus Standard GMM.}
Synthetic GMM uses data on price dynamics to assign greater weight to untreated markets that resemble the treated markets. To the extent that these dynamics are informative about the heterogeneity underlying supply-side IV invalidity, we have shown that reweighting may reduce bias relative to standard GMM.

In our empirical application, standard and Synthetic GMM yield similar estimates for DL--NW and AA--US. For WN--FL, Synthetic GMM produces smaller estimates of both the marginal-cost increase and the change in conduct. However, because the two components offset one another, the estimated net fare effect remains nearly unchanged, increasing from \$10.3 under standard GMM to \$10.9 under Synthetic GMM.

Reweighting can also reduce the identifying variation available to estimate the structural parameters by concentrating on fewer untreated markets. For WN--FL, the effective number of control markets decreases from 1,132 to 179, while the pricing-equation first-stage statistic declines from 10.1 under standard GMM to 4.01 under Synthetic GMM. At this level of instrument strength, placebo inference may over-reject and produce intervals with insufficient coverage. 

This identification problem is more pronounced for the UA--CO merger, which has only 11 treated markets. Although SDiD weighting selects control markets with similar price dynamics, the resulting pricing-equation estimates are unstable. In particular, estimated pass-through changes sign across the number of firms, implying negative pass-through in some markets. We therefore omit the corresponding linear-model estimates from Table \ref{tab:did_main}, and interpret it as reflecting the identifying variation available in our data rather than an inherent limitation of Synthetic GMM.

Taken together, the estimates show the gains of bridging the two approaches. While both approaches are consistent and suggest these mergers did not lower prices, the structural estimates explain why. In the WN–FL merger, the estimates suggest that costs and conduct
can move in opposite directions and in the AA–US merger, they show how changes in conduct can vary with pre-merger market structure.

\section{Conclusion}

This paper develops a two-way bridge between quasi-experimental and structural approaches to merger retrospectives. The first bridge derives the DiD price effect from a demand-and-supply model and shows how merger-induced changes in marginal cost and conduct determine that effect. It also makes explicit the conditions under which DiD identifies the effect in treated markets and when that effect can be extrapolated to other markets. 
The second bridge uses synthetic-DiD weights based on price dynamics to choose comparison markets for structural estimation, yielding Synthetic GMM. It shows when this reweighting can reduce bias from invalid comparison markets.

We apply the framework to four major U.S. airline mergers. We find no evidence that any of the four mergers reduced fares and no evidence of merger-induced efficiency gains. For WN--FL, the estimates indicate higher marginal cost together with more competitive conduct, so the conduct change partly offsets the effect of higher costs on fares. For American--US Airways, the estimated conduct change is less competitive in markets with fewer pre-merger competitors, consistent with concerns about coordinated effects, but it is imprecise. 

The two approaches complement each other in this application. The
quasi-experimental analysis identifies the net price effect, while the
structural analysis shows whether it reflects changes in costs, conduct,
or both. These channels can offset one another, as in the Southwest--AirTran
merger, so similar price effects need not imply similar changes in
efficiency or market power. More broadly, our framework is not specific to the four mergers we study here. Applying it to the full history of domestic airline mergers since deregulation in 1978 would let us examine how these different merger effects have evolved across the industry. We leave this question for future research.

A related literature evaluates merger simulations by comparing predictions based on pre-merger data with realized post-merger prices \citep[e.g.,][]{Peters2006, BjornerstedtVerboven2016}. 
\cite{Peters2006} finds that allowing for merger-induced
changes in conduct improves these predictions. By linking the two
approaches within a common framework, the First Bridge instead allows
the estimates of the merger price effects from the two approaches to be compared directly with each other, providing an internal consistency check.
Additionally, the First Bridge also makes explicit what the data
identify, what additional assumptions are needed to interpret the merger
effect, and whether that effect applies beyond the treated markets.
The Second Bridge illustrates that the credibility of the structural estimates 
depends on the choice of comparison markets, and makes that choice transparent.

For applied researchers, the framework suggests a sequence of choices when
conducting a merger retrospective. The starting point is to assess which
parallel-trends assumption is credible, including whether it should hold
for fares or log fares. A DiD estimate then provides the net price effect
in treated markets, but understanding the mechanism requires
a demand-and-supply model, credible instruments, and assumptions about how
costs and conduct vary with the merger. The choice of comparison markets
also matters: when untreated markets differ in their pre-merger price
dynamics, synthetic DiD weighting offers a way to reduce potential bias in
the structural estimates. Finally, researchers should distinguish effects
identified in treated markets from those extrapolated to other markets,
and assess the precision of each estimate. 
In summary, we hope to propose a more disciplined and robust way to evaluate merger retrospectives.

The two bridges were developed for airline mergers, but neither depends
on the industry. The same logic may apply to other interventions in which
quasi-experimental variation identifies an effect and an economic model
is needed to recover its mechanisms or evaluate counterfactuals beyond
the observed setting. One important agenda for future research is to extend our approach to differentiated-product
demand and supply, building on \cite{BerryHaile2014}. Merger effects can
still be mapped into changes in costs and conduct, and untreated markets
can still be reweighted using pre-merger outcomes. However, identifying
changes in conduct with differentiated products \citep{Nevo1998} and
extending the bias analysis to nonlinear systems present additional
challenges. 

\bibliographystyle{aea}
\bibliography{merger}

\singlespacing
\begin{center}
\noindent\textbf{\huge{Appendix}}
\end{center}
\setcounter{section}{0}
\setcounter{equation}{0}
\setcounter{figure}{0}
\setcounter{table}{0}
\renewcommand{\thesection}{\Alph{section}}
\renewcommand{\theHsection}{appendix.\Alph{section}}   
\numberwithin{equation}{section}
\numberwithin{figure}{section}
\numberwithin{table}{section}
\renewcommand{\theequation}{\thesection.\arabic{equation}}
\renewcommand\thefigure{\thesection.\arabic{figure}}
\renewcommand\thetable{\thesection.\arabic{table}}

\section{Data Construction}
\subsection{Origin and Destination Survey and T-100 Domestic Market}
Our primary data source is the Domestic Origin and Destination Survey (DB1B), a 10\% sample of airline tickets from all reporting carriers. The U.S. DOT's National Transportation Library provides this dataset.
 We clean the data by removing the following:
\begin{itemize}
\item Tickets with more than six coupons or more than three coupons in either direction.
\item Tickets sold by foreign carriers flying between two U.S. airports.
\item Tickets that are part of international travel.
\item Tickets involving non-contiguous domestic travel (Hawaii, Alaska, and Territories).
\item Tickets with questionable fare credibility or bulk fare indicators.
\item Tickets that are neither one-way nor round-trip.
\item Interline tickets (travel on more than one airline in a directional trip).
\end{itemize}
We treat round-trip tickets as two separate directional trips, with each direction priced at half the (round-trip) price on the ticket.
 Similarly, we count a passenger on a round-trip as one passenger in both directions.
 We exclude tickets cheaper than \$25, those in the top and bottom 1\% of the year-quarter fare distribution, and airlines serving fewer than 90 passengers per quarter. All the fares and values are expressed in 2012Q2 dollars.\footnote{We also use baggage and reservation-change fee revenue from the U.S. DOT's Form 41 Schedule P-1.2 to construct AirTran's average ancillary fees per passenger, which is useful to understand the WN-FL merger.} The T-100 dataset contains domestic nonstop segment data reported by U.S. air carriers, including carrier, origin, destination, and available capacity. We use this dataset to gather information on airlines' capacity along their nonstop routes.

\begin{landscape}
\thispagestyle{empty}
\begin{table}
\caption{Summary Statistics for Covariates and Instruments\label{tab:iv_summary}}
\hspace*{-0.8in}\begin{threeparttable}
\scalebox{0.98}{\begin{tabular}{l*{4}{ccc}}
\toprule
 & \multicolumn{3}{c}{DL--NW}
 & \multicolumn{3}{c}{UA--CO}
 & \multicolumn{3}{c}{WN--FL}
 & \multicolumn{3}{c}{AA--US} \\
\cmidrule(lr){2-4}
\cmidrule(lr){5-7}
\cmidrule(lr){8-10}
\cmidrule(lr){11-13}
Variable
& Mean & SD & (c)
& Mean & SD & (c)
& Mean & SD & (c)
& Mean & SD & (c) \\
\midrule
\multicolumn{13}{l}{\emph{Demand and Cost Shifters}} \\
Markets served from origin 
& $1.320$ & $0.492$ & $0.070$
& $1.476$ & $0.442$ & $0.055$
& $1.566$ & $0.479$ & $0.052$
& $1.430$ & $0.444$ & $0.055$ \\
Markets served from destination 
& $1.366$ & $0.463$ & $0.064$
& $1.497$ & $0.380$ & $0.055$
& $1.512$ & $0.419$ & $0.051$
& $1.448$ & $0.417$ & $0.055$ \\
Market size
& $0.644$ & $1.122$ & $0.046$
& $0.815$ & $1.278$ & $0.034$
& $0.677$ & $1.060$ & $0.039$
& $0.704$ & $1.089$ & $0.014$ \\
Southwest potential entrant
& $0.582$ & $0.493$ & $0.073$
& $0.730$ & $0.444$ & $0.133$
&n/a &n/a &n/a 
& $0.791$ & $0.407$ & $0.062$ \\
Legacy carriers at endpoints
& $3.610$ & $0.708$ & $0.349$
& $2.916$ & $0.454$ & $0.327$
& $3.958$ & $0.630$ & $0.381$
& $1.961$ & $0.201$ & $0.082$ \\
Number of regional carriers used
& $0.399$ & $0.311$ & $0.116$
& $0.386$ & $0.271$ & $0.108$
& $0.378$ & $0.297$ & $0.113$
& $0.393$ & $0.284$ & $0.112$ \\
\addlinespace
\multicolumn{13}{l}{\emph{Interaction instruments}} \\
Change in origin markets $\times$ treated
& $0.000$ & $0.014$ & $0.962$
& $0.000$ & $0.008$ & $0.979$
& $0.000$ & $0.010$ & $0.961$
& $0.000$ & $0.009$ & $0.967$ \\
Change in destination markets $\times$ treated
& $0.000$ & $0.014$ & $0.961$
& $0.000$ & $0.009$ & $0.978$
& $0.000$ & $0.013$ & $0.961$
& $0.000$ & $0.011$ & $0.967$ \\
Change in market size $\times$ treated
& $0.001$ & $0.013$ & $0.959$
& $0.000$ & $0.005$ & $0.978$
& $0.001$ & $0.005$ & $0.958$
& $0.001$ & $0.005$ & $0.963$ \\
Change in Southwest potential entry $\times$ treated
& $0.000$ & $0.020$ & $1.000$
& $0.000$ & $0.020$ & $1.000$
& n/a&n/a &n/a& $0.000$ & $0.012$ & $1.000$ \\
Change in legacy-carrier count $\times$ treated
& $-0.002$ & $0.094$ & $0.991$
& $-0.001$ & $0.044$ & $0.998$
& $-0.004$ & $0.076$ & $0.994$
& $0.000$ & $0.028$ & $0.999$ \\
Change in regional carriers used $\times$ treated
& $0.000$ & $0.032$ & $0.959$
& $0.000$ & $0.018$ & $0.978$
& $0.000$ & $0.028$ & $0.958$
& $0.001$ & $0.025$ & $0.963$ \\
Lagged origin markets $\times$ first post-merger quarter
& $0.003$ & $0.064$ & $0.997$
& $0.002$ & $0.054$ & $0.999$
& $0.004$ & $0.088$ & $0.997$
& $0.003$ & $0.069$ & $0.998$ \\
Lagged destination markets $\times$ first post-merger quarter
& $0.003$ & $0.063$ & $0.997$
& $0.002$ & $0.058$ & $0.999$
& $0.003$ & $0.069$ & $0.997$
& $0.003$ & $0.067$ & $0.998$ \\
Lagged market size $\times$ first post-merger quarter
& $0.001$ & $0.035$ & $0.997$
& $0.001$ & $0.061$ & $0.999$
& $0.001$ & $0.045$ & $0.997$
& $0.002$ & $0.052$ & $0.998$ \\
\bottomrule
\end{tabular}}

\begin{figurenotes}
Statistics are computed using treated and control markets in each
merger's estimation sample (the base specification in Appendix
Table \ref{tab:conduct_all_markets}). The numbers of markets served from the origin and destination exclude the merging airlines. For demand and cost shifters, column (c)
reports the standard deviation of the quarterly first difference,
except for market size, for which it reports the standard deviation
of the four-quarter change. 
For the interaction instruments, column (c)
reports the fraction of observations equal to zero. All interaction
instruments are zero in control markets. Market size is standardized.
The legacy-carrier-count interaction instruments are included in the
pricing equation but not price in the demand equation. The Southwest
potential-entry indicator and its interaction are excluded from
the WN--FL specification.
\end{figurenotes}
\end{threeparttable}
\end{table}
\end{landscape}

\subsection{Descriptive Statistics for the Structural Instruments}
\label{sec:appendix_iv_stats}

Table \ref{tab:iv_summary} reports descriptive statistics for the
shifters and instruments used in the structural estimation of each
merger. The statistics are computed using both treated and control
markets in the estimation sample. For each shifter, we report its mean,
standard deviation, and the standard deviation of its change over
time, which is relevant because the model is estimated in first
differences. For the interaction instruments, we report the mean,
standard deviation, and fraction of observations equal to zero.

The demand shifters are the number of markets served from the origin
and destination airports, excluding the merging carriers, and market
size. They enter the demand equation and also instrument
quantity in the pricing equation. The cost shifters are an indicator
for Southwest as a potential entrant, the number of legacy carriers
present at the market's endpoints but not serving the market
(excluding the merging carriers), and the number of distinct regional
carriers used by a main carrier to serve the market. These variables
enter the pricing equation directly. Except for the legacy-carrier
count, they also instrument price in the demand equation. The
Southwest potential-entry indicator is excluded from the WN--FL
specification because Southwest is one of the merging carriers.

We also use instruments that interact the shifters with treatment
status. These include changes in the demand and cost shifters
interacted with the treated-market indicator, and lagged
levels of the demand shifters interacted with an indicator for the
first post-merger quarter in treated markets. The lagged values are
taken from the preceding quarter for the origin and destination
market counts and from four quarters earlier for market size.
All these interactions instrument quantity in the pricing equation.
They also instrument price in the demand equation, except for the
interaction involving the legacy-carrier count. 
Even after differencing, the exogenous shifters retain variation across all four mergers, while the interaction instruments are zero in control markets. The latter are informative only in a relatively small number of treated markets, ranging from 11 to 50 across mergers.

\section{Proofs for First Bridge \label{sec:proofs}}
\begin{proof}[Proof of Lemma \ref{lemma:did_bias}]
\label{app:proof_did_bias}
(i) Write the reduced-form error of \eqref{eq:price_reduced} as
\begin{equation}
  \varepsilon_{mt}=\frac{\tilde{\gamma}_{mt}}{(1-\tilde{\gamma}_{mt}\alpha_1)}\varphi_{mt} + \frac{1}{(1-\tilde{\gamma}_{mt}\alpha_1)}\vartheta_{mt} := \mathbb{S}^1_{mt} \varphi_{mt} + \mathbb{S}^2_{mt} \vartheta_{mt}.
  \label{eq:composite_error}
\end{equation}
We use the reduced form \eqref{eq:price_reduced} and the composite
error decomposition \eqref{eq:composite_error} to determine all four terms in the definition of the DiD \eqref{eq:did_definition}. Recall that
$I_{mt}=1$ only in the treated post-period cell; the pricing slope
is $\tilde{\gamma}_1$ in that cell and
$\tilde{\gamma}_0$ in the other three cells.
As in the text, $\bar{\varphi}_s$ and $\bar{\vartheta}_s$ are the treated markets' mean shocks in period $s\in\{\mathrm{pre},\mathrm{post}\}$; let
$\bar{\varphi}^{u}_s:=\mathbb{E}[\varphi_{m,s}\mid\mathrm{Treated}_m=0]$ and $\bar{\vartheta}^{u}_s:=\mathbb{E}[\vartheta_{m,s}\mid\mathrm{Treated}_m=0]$ be the untreated markets'.
Mean parallel trends is $\bar{\varphi}_{\mathrm{post}}-\bar{\varphi}_{\mathrm{pre}}=\bar{\varphi}^{u}_{\mathrm{post}}-\bar{\varphi}^{u}_{\mathrm{pre}}$, and likewise for $\vartheta$.

Taking expectations of \eqref{eq:price_reduced} in each of the four cells:
\begin{align}
\mathbb{E}[P_{m,\mathrm{post}}\mid\mathrm{Treated}_m=1]
  &= \frac{\theta_0+\tilde{\gamma}_1\alpha_0+\theta_3}{1-\tilde{\gamma}_1\alpha_1}
   + \mathbb{S}^1_1\bar{\varphi}_{\mathrm{post}} + \mathbb{S}^2_1\bar{\vartheta}_{\mathrm{post}},
  \label{eq:cell_tp}\\[6pt]
\mathbb{E}[P_{m,\mathrm{pre}}\mid\mathrm{Treated}_m=1]
  &= \frac{\theta_0+\tilde{\gamma}_0\alpha_0}{1-\tilde{\gamma}_0\alpha_1}
   + \mathbb{S}^1_0\bar{\varphi}_{\mathrm{pre}}  + \mathbb{S}^2_0\bar{\vartheta}_{\mathrm{pre}},
  \label{eq:cell_tpre}\\[6pt]
\mathbb{E}[P_{m,\mathrm{post}}\mid\mathrm{Treated}_m=0]
  &= \frac{\theta_0+\tilde{\gamma}_0\alpha_0}{1-\tilde{\gamma}_0\alpha_1}
   + \mathbb{S}^1_0\bar{\varphi}^{u}_{\mathrm{post}} + \mathbb{S}^2_0\bar{\vartheta}^{u}_{\mathrm{post}},
  \label{eq:cell_cp}\\[6pt]
\mathbb{E}[P_{m,\mathrm{pre}}\mid\mathrm{Treated}_m=0]
  &= \frac{\theta_0+\tilde{\gamma}_0\alpha_0}{1-\tilde{\gamma}_0\alpha_1}
   + \mathbb{S}^1_0\bar{\varphi}^{u}_{\mathrm{pre}}  + \mathbb{S}^2_0\bar{\vartheta}^{u}_{\mathrm{pre}}.
  \label{eq:cell_cpre}
\end{align}

Subtracting \eqref{eq:cell_cpre} from \eqref{eq:cell_cp}, the untreated markets' mean change is
$\mathbb{S}^1_0(\bar{\varphi}^{u}_{\mathrm{post}}-\bar{\varphi}^{u}_{\mathrm{pre}})+\mathbb{S}^2_0(\bar{\vartheta}^{u}_{\mathrm{post}}-\bar{\vartheta}^{u}_{\mathrm{pre}})
=\mathbb{S}^1_0(\bar{\varphi}_{\mathrm{post}}-\bar{\varphi}_{\mathrm{pre}})+\mathbb{S}^2_0(\bar{\vartheta}_{\mathrm{post}}-\bar{\vartheta}_{\mathrm{pre}})$,
the equality by mean parallel trends. Subtracting it from the treated markets' mean change, \eqref{eq:cell_tp} minus \eqref{eq:cell_tpre}, the terms in $\bar{\varphi}_{\mathrm{pre}}$ and $\bar{\vartheta}_{\mathrm{pre}}$ cancel, and \eqref{eq:did_definition} gives the desired expression \eqref{eq:did_decomp}
\begin{equation*}
{\beta}^{\mathrm{DiD}}
=\frac{\theta_3}{1-\tilde{\gamma}_1\alpha_1} + \left(\left(
\frac{\theta_0+\tilde{\gamma}_1\alpha_0}{1-\tilde{\gamma}_1\alpha_1}
-\frac{\theta_0+\tilde{\gamma}_0\alpha_0}{1-\tilde{\gamma}_0\alpha_1}\right)
+(\mathbb{S}^1_1-\mathbb{S}^1_0)\bar{\varphi}_{\mathrm{post}}
+(\mathbb{S}^2_1-\mathbb{S}^2_0)\bar{\vartheta}_{\mathrm{post}}\right).
\end{equation*}
The right-hand side is also the average treatment effect on the treated: the treated markets' mean post-merger price under the merger is \eqref{eq:cell_tp}, and without the merger it is the equilibrium at $\tilde{\gamma}_0$ with the same shocks, $\frac{\theta_0+\tilde{\gamma}_0\alpha_0}{1-\tilde{\gamma}_0\alpha_1}+\mathbb{S}^1_0\bar{\varphi}_{\mathrm{post}}+\mathbb{S}^2_0\bar{\vartheta}_{\mathrm{post}}$, and the difference of the two is the expression above. So $\beta^{\mathrm{DiD}}$ identifies the average treatment effect on the treated.
It then follows that if $\Delta\lambda=0$, we have $\tilde{\gamma}_1=\tilde{\gamma}_0$ by
Proposition \ref{prop:conduct}-\ref{prop:change}, hence $\mathbb{S}^1_1=\mathbb{S}^1_0$, and $\mathbb{S}^2_1=\mathbb{S}^2_0$, and ${\beta}^{\mathrm{DiD}}={\beta}^{\mathrm{RF}}$.
When $\Delta\lambda\neq 0$, $\tilde{\gamma}_1\neq\tilde{\gamma}_0$, so
$\mathbb{S}^1_1\neq \mathbb{S}^1_0$, and $\mathbb{S}^2_1\neq \mathbb{S}^2_0$, and except in the knife-edge cases, the term in the parentheses above is
non-zero and ${\beta}^{\mathrm{DiD}}\neq{\beta}^{\mathrm{RF}}$.

(ii) From \eqref{eq:price_reduced} and
\eqref{eq:composite_error}, the changes in untreated
equilibrium prices are
\begin{align*}
\mathbb{E}[P_{m,\mathrm{post}}(0)-P_{m,\mathrm{pre}}(0)
\mid \mathrm{Treated}_m=1]
&=
\mathbb{S}^1_0
(\bar{\varphi}_{\mathrm{post}}-\bar{\varphi}_{\mathrm{pre}})
+
\mathbb{S}^2_0
(\bar{\vartheta}_{\mathrm{post}}-\bar{\vartheta}_{\mathrm{pre}}),\\
\mathbb{E}[P_{m,\mathrm{post}}(0)-P_{m,\mathrm{pre}}(0)
\mid \mathrm{Treated}_m=0]
&=
\frac{\theta_1-\lambda_C/\alpha_1}
{1-(\theta_1-\lambda_C/\alpha_1)\alpha_1}
(\bar{\varphi}^{u}_{\mathrm{post}}-\bar{\varphi}^{u}_{\mathrm{pre}})\\
&\quad+
\frac{1}{1-(\theta_1-\lambda_C/\alpha_1)\alpha_1}
(\bar{\vartheta}^{u}_{\mathrm{post}}-\bar{\vartheta}^{u}_{\mathrm{pre}}).
\end{align*}
As in part (i), mean parallel trends in the primitive shocks is
$\bar{\varphi}_{\mathrm{post}}-\bar{\varphi}_{\mathrm{pre}}
=\bar{\varphi}^{u}_{\mathrm{post}}-\bar{\varphi}^{u}_{\mathrm{pre}}$,
and likewise for $\vartheta$. Using these equalities,
the difference in untreated price trends becomes 
\begin{eqnarray*}
&&\mathbb{E}[P_{m,\mathrm{post}}(0)-P_{m,\mathrm{pre}}(0)
\mid\mathrm{Treated}_m=1]
-\mathbb{E}[P_{m,\mathrm{post}}(0)-P_{m,\mathrm{pre}}(0)
\mid\mathrm{Treated}_m=0]\\
&=&
\left(\mathbb{S}^1_0-\frac{\theta_1-\lambda_C/\alpha_1}
{1-(\theta_1-\lambda_C/\alpha_1)\alpha_1}\right)
(\bar{\varphi}_{\mathrm{post}}-\bar{\varphi}_{\mathrm{pre}})
+\left(\mathbb{S}^2_0-\frac{1}
{1-(\theta_1-\lambda_C/\alpha_1)\alpha_1}\right)
(\bar{\vartheta}_{\mathrm{post}}-\bar{\vartheta}_{\mathrm{pre}}).
\end{eqnarray*}
If $\lambda_T\neq\lambda_C$, this expression is
generally nonzero, so the parallel trends in the
primitive shocks do not generally imply parallel trends
in untreated prices. Hence, the DiD design does not 
identify the average treatment effect on the treated
\citep{Abadie2005}. 

To see the difference in the present setting, let $\widetilde{\beta}^{\mathrm{DiD}}$ denote the DiD estimate when $\lambda_T\neq\lambda_C$. Then using the definition
of DiD in \eqref{eq:did_definition}, we get 
\begin{eqnarray*}
\widetilde{\beta}^{\mathrm{DiD}}
&=&
\frac{\theta_3}{1-\tilde{\gamma}_1\alpha_1}
+\left(\frac{\theta_0+\tilde{\gamma}_1\alpha_0}{1-\tilde{\gamma}_1\alpha_1}
-\frac{\theta_0+\tilde{\gamma}_0\alpha_0}{1-\tilde{\gamma}_0\alpha_1}\right)
+(\mathbb{S}^1_1-\mathbb{S}^1_0)\bar{\varphi}_{\mathrm{post}}
+(\mathbb{S}^2_1-\mathbb{S}^2_0)\bar{\vartheta}_{\mathrm{post}}\\
&&+\left(\mathbb{S}^1_0-\frac{\theta_1-\lambda_C/\alpha_1}
{1-(\theta_1-\lambda_C/\alpha_1)\alpha_1}\right)
(\bar{\varphi}_{\mathrm{post}}-\bar{\varphi}_{\mathrm{pre}})
+\left(\mathbb{S}^2_0-\frac{1}
{1-(\theta_1-\lambda_C/\alpha_1)\alpha_1}\right)
(\bar{\vartheta}_{\mathrm{post}}-\bar{\vartheta}_{\mathrm{pre}})\\
&=&
\beta^{\mathrm{DiD}}
+\left(\mathbb{S}^1_0-\frac{\theta_1-\lambda_C/\alpha_1}
{1-(\theta_1-\lambda_C/\alpha_1)\alpha_1}\right)
(\bar{\varphi}_{\mathrm{post}}-\bar{\varphi}_{\mathrm{pre}})
+\left(\mathbb{S}^2_0-\frac{1}
{1-(\theta_1-\lambda_C/\alpha_1)\alpha_1}\right)
(\bar{\vartheta}_{\mathrm{post}}-\bar{\vartheta}_{\mathrm{pre}})
\end{eqnarray*}
where the second equality uses $\beta^{\mathrm{DiD}}$ defined in \eqref{eq:did_decomp}. 
Consequently, $\widetilde{\beta}^{\mathrm{DiD}}
\neq\beta^{\mathrm{DiD}}$, so the DiD does not identify the average effect of merger on prices in the treated markets.
\end{proof}

\begin{proof}[Proof of Lemma \ref{lemma:did_bias_log}]
Suppress the observed demand and cost shifters $Z_{mt}$ and $W_{mt}$, since the argument is conditional on them. Equation \eqref{eq:logprice_reduced} then becomes
\begin{equation}
\log P_{mt}=\Pi_0+\beta_{\log}^{\mathrm{RF}}\times I_{mt}+\varepsilon_{mt}^{\log}. \label{eq:proof_log_rf}
\end{equation}
Let $A_{\log}=\frac{\theta_1}{1-\alpha_1\theta_1}$ and $B_{\log}=\frac{1}{1-\alpha_1\theta_1}$, so that $\varepsilon_{mt}^{\log}=A_{\log}\varphi_{mt}+B_{\log}\vartheta_{mt}$. Importantly, $A_{\log}$ and $B_{\log}$ are invariant to treatment status and strategic conduct. Taking first differences gives $\Delta\varepsilon_m^{\log}=A_{\log}\Delta\varphi_m+B_{\log}\Delta\vartheta_m$, and therefore
\begin{eqnarray*}
\mathbb{E}\left[\Delta\varepsilon_m^{\log}\mid\mathrm{Treated}_m=1\right]&-&\mathbb{E}\left[\Delta\varepsilon_m^{\log}\mid\mathrm{Treated}_m=0\right]\\
&=&A_{\log}\times\left\{\mathbb{E}\left[\Delta\varphi_m\mid\mathrm{Treated}_m=1\right]-\mathbb{E}\left[\Delta\varphi_m\mid\mathrm{Treated}_m=0\right]\right\}\\
&&+B_{\log}\times\left\{\mathbb{E}\left[\Delta\vartheta_m\mid\mathrm{Treated}_m=1\right]-\mathbb{E}\left[\Delta\vartheta_m\mid\mathrm{Treated}_m=0\right]\right\}=0,\end{eqnarray*}
where the last equality follows from parallel trends in the primitive demand and cost shocks. Thus, the composite error also satisfies parallel trends. Applying the $2\times2$ DiD estimand to Equation \eqref{eq:proof_log_rf} therefore gives
\begin{eqnarray*}
\beta_{\log}^{\mathrm{DiD}}=\beta_{\log}^{\mathrm{RF}}+\mathbb{E}\left[\Delta\varepsilon_m^{\log}\mid\mathrm{Treated}_m=1\right]-\mathbb{E}\left[\Delta\varepsilon_m^{\log}\mid\mathrm{Treated}_m=0\right]=\beta_{\log}^{\mathrm{RF}}=\frac{\theta_3-\left(\Lambda_{\mathrm{post}}-\Lambda_{\mathrm{pre}}\right)}{1-\alpha_1\theta_1}.
\end{eqnarray*}
\end{proof}

\begin{proof}[Proof of Proposition \ref{prop:conduct}--\ref{prop:demand_cost}]
Consider the demand-and-supply  functions governing equilibrium outcomes in market $m$ during period $t$,
\begin{eqnarray*}
Q_{mt} &=& \alpha_0 + \alpha_1 P_{mt} + \alpha_2^\top \mathbf{Z}_{mt} + \varphi_{mt}; \quad \text{(Demand)} \label{eq:demand-GA}\\
P_{mt} &=& \theta_0 + \tilde{\gamma}_{mt} Q_{mt} + \theta_2^\top \mathbf{W}_{mt} + \theta_3 I_{mt} + \vartheta_{mt},\text{(Supply)}
\end{eqnarray*}
where $\mathbf{Z}_{mt}$ is a $d_z \times 1$ vector of demand shifters, $\mathbf{W}_{mt}$ is a $d_w \times 1$ vector of cost shifters, $I_{mt}$ is a scalar merger indicator, and $\varphi_{mt}$ and $\vartheta_{mt}$ represent unobserved demand and supply shocks, respectively.
Recall that the pricing slope is $\tilde{\gamma}_0$ when $I_{mt}=0$ and $\tilde{\gamma}_1$ when $I_{mt}=1$. We can therefore rewrite the supply equation as
\begin{equation}
P_{mt}=\theta_0+\tilde{\gamma}_0Q_{mt}+(\tilde{\gamma}_1-\tilde{\gamma}_0)I_{mt}Q_{mt} +\theta_2^\top\mathbf W_{mt}+\theta_3I_{mt}+\vartheta_{mt}.\text{(Supply)}
\label{eq:supply-GA}
\end{equation}
The parameters $\boldsymbol{\alpha}=(\alpha_0,\alpha_1,\alpha_2^\top)^\top$ and $\boldsymbol{\theta}=(\theta_0,\tilde{\gamma}_0,\tilde{\gamma}_1-\tilde{\gamma}_0,\theta_2^\top,\theta_3)^\top$ are $(2+d_z)\times1$ and $(4+d_w)\times1$ demand-and-supply  parameter vectors, respectively. Identification of $\boldsymbol{\theta}$ is equivalent to
identification of $\tilde{\boldsymbol{\theta}}=(\theta_0,\tilde{\gamma}_0,\tilde{\gamma}_1,\theta_2^\top,\theta_3)^\top$, since $\tilde{\gamma}_1$ is recovered from
$\tilde{\gamma}_0$ and $\tilde{\gamma}_1-\tilde{\gamma}_0$. Because the pricing slope may change after the merger, the supply equation contains two endogenous regressors, $Q_{mt}$ and $I_{mt}Q_{mt}$. Identifying their coefficients requires excluded
instruments that provide independent variation in both regressors.
We therefore use the demand shifters and their interactions with
the merger indicator as instruments.

The orthogonality conditions that ensure our instrumental variables provide exogenous variation are as follows:
\begin{eqnarray}
\mathbb{E}[\mathbf{W}_{mt} \varphi_{mt}] &=& \mathbf{0}_{d_w \times 1}; \quad \text{(Cost shifters orthogonal to demand shocks)} \label{eq:orth1-GA}\\
\mathbb{E}[\mathbf{Z}_{mt} \vartheta_{mt}] &=& \mathbf{0}_{d_z \times 1}; \quad \text{(Demand shifters orthogonal to supply shocks)} \label{eq:orth2-GA}\\
\mathbb{E}[I_{mt} \vartheta_{mt}] &=& 0; \quad \text{(Merger indicator orthogonal to supply shocks)} \label{eq:orth3-GA}\\
\mathbb{E}[\mathbf{Z}_{mt} \varphi_{mt}] &=& \mathbf{0}_{d_z \times 1}; \quad \text{(Demand shifters orthogonal to demand shocks)} \label{eq:orth4-GA}\\
\mathbb{E}[\mathbf{W}_{mt} \vartheta_{mt}] &=& \mathbf{0}_{d_w \times 1}; \quad \text{(Cost shifters orthogonal to supply shocks)} \label{eq:orth5-GA}\\
\mathbb{E}[\varphi_{mt}] &=& 0,\quad \mathbb{E}[\vartheta_{mt}] = 0. \quad \text{(Mean zero demand-and-supply  shocks)} \label{eq:orth6-GA}\\
\mathbb{E}[I_{mt}\mathbf{Z}_{mt}\vartheta_{mt}] &=& \mathbf{0}_{d_z\times1}. \quad \text{(Interactions orthogonal to supply shocks)} \label{eq:orth7-GA}
\end{eqnarray}
The required orthogonality conditions include those for the interacted demand shifters, since their validity does not follow from the separate orthogonality of the demand shifters and the
merger indicator to the supply shock.
From the orthogonality conditions \eqref{eq:orth1-GA} and \eqref{eq:orth4-GA}, we can construct moment conditions for demand identification. The orthogonality of cost shifters to demand shocks yields
\begin{equation}
\mathbb{E}[\mathbf{W}_{mt} (Q_{mt} - \alpha_0 - \alpha_1 P_{mt} - \alpha_2^\top\mathbf{Z}_{mt})] = \mathbf{0}_{d_w \times 1}. \label{eq:mom_demand1-GA}
\end{equation}
Additionally, the orthogonality of demand shifters to demand shocks is
\begin{equation}
\mathbb{E}[\mathbf{Z}_{mt} (Q_{mt} - \alpha_0 - \alpha_1 P_{mt} - \alpha_2^\top\mathbf{Z}_{mt})] = \mathbf{0}_{d_z \times 1}. \label{eq:mom_demand2-GA}
\end{equation}

For supply identification, the orthogonality conditions \eqref{eq:orth2-GA}, \eqref{eq:orth3-GA}, \eqref{eq:orth5-GA}, and \eqref{eq:orth7-GA} allow us to construct the required moment conditions. The orthogonality of demand shifters to supply shocks gives
\begin{equation}
\mathbb{E}[\mathbf{Z}_{mt} (P_{mt} - \theta_0 - \tilde{\gamma}_{mt} Q_{mt} - \theta_2^\top \mathbf{W}_{mt} - \theta_3 I_{mt})] = \mathbf{0}_{d_z \times 1}. \label{eq:mom_supply1-GA}
\end{equation}
Then the orthogonality of the merger to supply shocks gives
\begin{equation}
\mathbb{E}[I_{mt} (P_{mt} - \theta_0 - \tilde{\gamma}_{mt} Q_{mt} - \theta_2^\top\mathbf{W}_{mt} - \theta_3 I_{mt})] = 0. \label{eq:mom_supply2-GA}
\end{equation}
Furthermore, the orthogonality of cost shifters to supply shocks yields
\begin{equation}
\mathbb{E}[\mathbf{W}_{mt} (P_{mt} - \theta_0 - \tilde{\gamma}_{mt} Q_{mt} - \theta_2^\top \mathbf{W}_{mt} - \theta_3 I_{mt})] = \mathbf{0}_{d_w \times 1}, \label{eq:mom_supply3-GA}
\end{equation}
and the orthogonality of the interacted demand shifters to supply shocks additionally gives
\begin{equation}
\mathbb{E}[I_{mt}\mathbf{Z}_{mt}(P_{mt}-\theta_0-\tilde{\gamma}_0Q_{mt}-\left(\tilde{\gamma}_1-\tilde{\gamma}_0\right)I_{mt}Q_{mt}-\theta_2^\top\mathbf{W}_{mt}-\theta_3I_{mt})]=\mathbf{0}_{d_z\times1}. \label{eq:mom_supply4-GA}
\end{equation}

For notational simplicity, we define the following regressors in matrix form:
\begin{align*}
\mathbf{X}_{D,mt} &= (1,P_{mt},\mathbf{Z}_{mt}^\top)^\top;\quad\!\!
\mathbf{X}_{S,mt}=(1,Q_{mt},I_{mt}Q_{mt},\mathbf{W}_{mt}^\top,I_{mt})^\top\\
\mathbf{Z}_{D,mt} &= (1,\mathbf{W}_{mt}^\top,\mathbf{Z}_{mt}^\top)^\top;\quad\!\!
\mathbf{Z}_{S,mt}=(1,\mathbf{Z}_{mt}^\top,(I_{mt}\mathbf{Z}_{mt})^\top,I_{mt},\mathbf{W}_{mt}^\top)^\top.
\end{align*}
We can then express the expanded set of moment conditions in matrix form. For demand parameters, combining equations \eqref{eq:mom_demand1-GA} and \eqref{eq:mom_demand2-GA} gives $\mathbb{E}[\mathbf{Z}_{D,mt}(Q_{mt}-\mathbf{X}_{D,mt}^\top\boldsymbol{\alpha})]=\mathbf{0}_{(1+d_w+d_z)\times1}$, which implies the following system $
\mathbb{E}[\mathbf{Z}_{D,mt}\mathbf{X}_{D,mt}^\top]\boldsymbol{\alpha}=\mathbb{E}[\mathbf{Z}_{D,mt}Q_{mt}].$
Similarly, for supply parameters, combining equations \eqref{eq:mom_supply1-GA}--\eqref{eq:mom_supply4-GA} gives $\mathbb{E}[\mathbf{Z}_{S,mt}(P_{mt}-\mathbf{X}_{S,mt}^\top\boldsymbol{\theta})]=\mathbf{0}_{(2+2d_z+d_w)\times1}$, which corresponds to the system
$\mathbb{E}[\mathbf{Z}_{S,mt}\mathbf{X}_{S,mt}^\top]\boldsymbol{\theta}=\mathbb{E}[\mathbf{Z}_{S,mt}P_{mt}].
$

For the demand parameters $\boldsymbol{\alpha}$ to be identified, the matrix $\mathbb{E}[\mathbf{Z}_{D,mt}\mathbf{X}_{D,mt}']$ must be of full column rank. This requires:
\begin{equation}
\text{rank}(\mathbb{E}[\mathbf{Z}_{D,mt}\mathbf{X}_{D,mt}^\top])=2+d_z. \label{eq:rank_demand-GA}
\end{equation}
Explicitly, this condition becomes
\begin{equation*}
\text{rank}\begin{pmatrix}
1 & \mathbb{E}[P_{mt}] & \mathbb{E}[\mathbf{Z}_{mt}]^\top \\
\mathbb{E}[\mathbf{W}_{mt}] & \mathbb{E}[\mathbf{W}_{mt}P_{mt}] & \mathbb{E}[\mathbf{W}_{mt}\mathbf{Z}_{mt}^\top] \\
\mathbb{E}[\mathbf{Z}_{mt}] & \mathbb{E}[\mathbf{Z}_{mt}P_{mt}] & \mathbb{E}[\mathbf{Z}_{mt}\mathbf{Z}_{mt}^\top]
\end{pmatrix}=2+d_z. \label{eq:rank_demand_explicit-GA}
\end{equation*}

Similarly, for the supply parameters $\boldsymbol{\theta}$ to be identified, we require
\begin{equation}
\text{rank}(\mathbb{E}[\mathbf{Z}_{S,mt}\mathbf{X}_{S,mt}^\top])=4+d_w, \label{eq:rank_supply-GA}
\end{equation}
which translates to the following rank condition
\begin{equation*}
\text{rank}\begin{pmatrix}
1 & \mathbb{E}[Q_{mt}] & \mathbb{E}[I_{mt}Q_{mt}] & \mathbb{E}[\mathbf{W}_{mt}]^\top & \mathbb{E}[I_{mt}] \\
\mathbb{E}[\mathbf{Z}_{mt}] & \mathbb{E}[\mathbf{Z}_{mt}Q_{mt}] & \mathbb{E}[I_{mt}\mathbf{Z}_{mt}Q_{mt}] & \mathbb{E}[\mathbf{Z}_{mt}\mathbf{W}_{mt}^\top] & \mathbb{E}[\mathbf{Z}_{mt}I_{mt}] \\
\mathbb{E}[I_{mt}\mathbf{Z}_{mt}] & \mathbb{E}[I_{mt}\mathbf{Z}_{mt}Q_{mt}] & \mathbb{E}[I_{mt}^2\mathbf{Z}_{mt}Q_{mt}] & \mathbb{E}[I_{mt}\mathbf{Z}_{mt}\mathbf{W}_{mt}^\top] & \mathbb{E}[I_{mt}^2\mathbf{Z}_{mt}] \\
\mathbb{E}[I_{mt}] & \mathbb{E}[I_{mt}Q_{mt}] & \mathbb{E}[I_{mt}^2Q_{mt}] & \mathbb{E}[I_{mt}\mathbf{W}_{mt}]^\top & \mathbb{E}[I_{mt}^2] \\
\mathbb{E}[\mathbf{W}_{mt}] & \mathbb{E}[\mathbf{W}_{mt}Q_{mt}] & \mathbb{E}[I_{mt}\mathbf{W}_{mt}Q_{mt}] & \mathbb{E}[\mathbf{W}_{mt}\mathbf{W}_{mt}^\top] & \mathbb{E}[\mathbf{W}_{mt}I_{mt}]
\end{pmatrix}=4+d_w. \label{eq:rank_supply_explicit-GA}
\end{equation*}
The order conditions require at least one excluded instrument
for the demand equation and sufficient excluded instruments
for the two endogenous regressors in the supply equation.
For demand, there are $1+d_w+d_z$ instruments and $2+d_z$
parameters, so we need $d_w\geq1$. Similarly, for supply, there
are $2+2d_z+d_w$ instruments and $4+d_w$ parameters,
so we need $d_z\geq1$. These order conditions and the full-column-rank
conditions above ensure the parameters are identified.

Under the rank conditions \eqref{eq:rank_demand-GA} and \eqref{eq:rank_supply-GA}, when the system is exactly identified, the parameter vectors have unique solutions given by $\boldsymbol{\alpha} = \left(\mathbb{E}[\mathbf{Z}_{D,mt}\mathbf{X}_{D,mt}^\top]\right)^{-1}\mathbb{E}[\mathbf{Z}_{D,mt}Q_{mt}]$ and $\boldsymbol{\theta} = \left(\mathbb{E}[\mathbf{Z}_{S,mt}\mathbf{X}_{S,mt}^\top]\right)^{-1}\mathbb{E}[\mathbf{Z}_{S,mt}P_{mt}]$. 
When the system is over-identified ($d_w>1$ for demand or $d_z>1$ for supply), the demand parameter $\boldsymbol{\alpha}$ is defined as the GMM estimator $\boldsymbol{\alpha}=\arg\min_{\boldsymbol{\alpha}}\mathbf{g}_D(\boldsymbol{\alpha})^\top\mathcal{W}_D\mathbf{g}_D(\boldsymbol{\alpha})$, with $\mathbf{g}_D(\boldsymbol{\alpha})\equiv\mathbb{E}\!\left[\mathbf{Z}_{D,mt}\bigl(Q_{mt}-\mathbf{X}_{D,mt}^\top\boldsymbol{\alpha}\bigr)\right]$ and $\mathcal{W}_D$ a positive definite weight matrix. The solution is unique under the rank condition \eqref{eq:rank_demand-GA}, and an analogous argument applies to $\boldsymbol{\theta}$ under \eqref{eq:rank_supply-GA}.
\end{proof}

\begin{proof}[Proof of Proposition \ref{prop:conduct}-\ref{prop:change}]
Two parameter sets $(\theta_1, \lambda_\text{pre}, \lambda_\text{post})$ and $(\theta_1^*, \lambda_\text{pre}^*, \lambda_\text{post}^*)$ are observationally equivalent if they generate the same values for the observables:
\begin{eqnarray*}
\tilde{\gamma}_0 = \theta_1 - \frac{\lambda_\text{pre}}{\alpha_1}=\theta_1^* - \frac{\lambda_\text{pre}^*}{\alpha_1}= \tilde{\gamma}_0^*; \quad 
\tilde{\gamma}_1 = \theta_1 - \frac{\lambda_\text{post}}{\alpha_1}=\theta_1^* - \frac{\lambda_\text{post}^*}{\alpha_1}= \tilde{\gamma}_1^*.
\end{eqnarray*}
We show non-identification by constructing at least two observationally equivalent parameter sets. Let $(\theta_1, \lambda_\text{pre}, \lambda_\text{post})$ be any parameter set that satisfies the two equations above. For any arbitrary constant $\chi \in \mathbb{R}$, define the alternative parameter set $(\theta_1^* = \theta_1 + \frac{\chi}{\alpha_1}; \lambda_\text{pre}^* = \lambda_\text{pre} +\chi; \lambda_\text{post}^* = \lambda_\text{post} + \chi)$.
We can verify that this alternative parameter set generates the same observables. Indeed, note that $\tilde{\gamma}_0^* =
\theta_1^* - \frac{\lambda_\text{pre}^*}{\alpha_1}
= \theta_1 + \frac{\chi}{\alpha_1} - \frac{\lambda_\text{pre} + \chi}{\alpha_1}
= \theta_1 - \frac{\lambda_\text{pre}}{\alpha_1} = \tilde{\gamma}_0,$
and similarly, we get $\tilde\gamma_1^*=\theta_1^* - \lambda_\text{post}^*/\alpha_1
= \tilde\gamma_1$.
However, for any $\chi \neq 0$, we have $\theta_1^* \neq \theta_1$, $\lambda_\text{pre}^* \neq \lambda_\text{pre}$, and $\lambda_\text{post}^* \neq \lambda_\text{post}$. As $\chi$ was arbitrary, the parameters $(\theta_1, \lambda_\text{pre},\lambda_\text{post})$ are not identified.
Next, we show that the change in conduct $\Delta\lambda:= \lambda_\text{post} - \lambda_\text{pre}$ is identified.
To this end, we show that $\Delta\lambda$ takes the same value across all observationally equivalent parameter sets. Consider any two observationally equivalent parameter sets $(\theta_1, \lambda_\text{pre}, \lambda_\text{post})$ and $(\theta_1^*, \lambda_\text{pre}^*, \lambda_\text{post}^*)$ that  generate the same observables $\tilde{\gamma}_0$ and $\tilde{\gamma}_1$. It is easy to check that for both these sets,
$
\Delta\lambda = \lambda_\text{post} - \lambda_\text{pre}=-\alpha_1 \times (\tilde{\gamma}_1- \tilde{\gamma}_0) = -\alpha_1 \times (\tilde{\gamma}_1^* - \tilde{\gamma}_0^*) = \lambda_\text{post}^* - \lambda_\text{pre}^*=\Delta\lambda.$
Therefore, $\Delta\lambda$ takes the same value across all observationally equivalent parameter sets, so $\Delta\lambda$ is identified.
\end{proof}

\begin{proof}[Proof of Proposition \ref{prop:conduct}-\ref{prop:levels}]
Two parameter sets $(\theta_1, \lambda_\text{pre}, \lambda_\text{post})$
and $(\theta_1^*, \lambda_\text{pre}^*, \lambda_\text{post}^*)$ are observationally equivalent if $(\tilde{\gamma}_0^{1}, \tilde{\gamma}_1^{1}, \tilde{\gamma}_0^{2}, \tilde{\gamma}_1^{2})=(\tilde{\gamma}_0^{1*}, \tilde{\gamma}_1^{1*}, \tilde{\gamma}_0^{2*}, \tilde{\gamma}_1^{2*})$. The parameters $(\theta_1, \lambda_\text{pre}, \lambda_\text{post})$ are  identified if for any other parameter set $(\theta_1^*, \lambda_\text{pre}^*, \lambda_\text{post}^*)$ that is observationally equivalent, we have $(\theta_1, \lambda_\text{pre}, \lambda_\text{post}) = (\theta_1^*, \lambda_\text{pre}^*, \lambda_\text{post}^*)$.
Suppose there are two parameter sets $(\theta_1, \lambda_\text{pre}, \lambda_\text{post})$ and $(\theta_1^*, \lambda_\text{pre}^*, \lambda_\text{post}^*)$ that are observationally equivalent.
Then
\begin{align}
\tilde{\gamma}_0^{1}=\theta_1 - \frac{\lambda_\text{pre}}{\alpha_1^{1}} &= \theta_1^* - \frac{\lambda_\text{pre}^*}{\alpha_1^{1}}=\tilde{\gamma}_0^{1*} \label{eq:alpha01},\\
\tilde{\gamma}_1^{1}=\theta_1 - \frac{\lambda_\text{post}}{\alpha_1^{1}} &= \theta_1^* - \frac{\lambda_\text{post}^*}{\alpha_1^{1}}=\tilde{\gamma}_1^{1*},\label{eq:alpha11}\\
\tilde{\gamma}_0^{2}=\theta_1 - \frac{\lambda_\text{pre}}{\alpha_1^{2}} &= \theta_1^* - \frac{\lambda_\text{pre}^*}{\alpha_1^{2}}=\tilde{\gamma}_0^{2*},\label{eq:alpha02}\\
\tilde{\gamma}_1^{2}=\theta_1 - \frac{\lambda_\text{post}}{\alpha_1^{2}} &= \theta_1^* - \frac{\lambda_\text{post}^*}{\alpha_1^{2}}=\tilde{\gamma}_1^{2*}.\label{eq:alpha12}
\end{align}
Rearranging Equations (\ref{eq:alpha01}) and (\ref{eq:alpha02}) gives $(\theta_1 - \theta_1^*)=\frac{\lambda_\text{pre}}{\alpha_1^{1}} - \frac{\lambda_\text{pre}^*}{\alpha_1^{1}} $ and $(\theta_1 - \theta_1^*)=\frac{\lambda_\text{pre}}{\alpha_1^{2}} - \frac{\lambda_\text{pre}^*}{\alpha_1^{2}}$, respectively. Equating the two and simplifying the expressions further gives $
\lambda_\text{pre}\times\left(\frac{\alpha_1^{2} - \alpha_1^{1}}{\alpha_1^{1}\alpha_1^{2}}\right) = \lambda_\text{pre}^*\times\left(\frac{\alpha_1^{2} - \alpha_1^{1}}{\alpha_1^{1}\alpha_1^{2}}\right).
$
Since $\alpha_1^{1}\neq\alpha_1^{2}$, the two terms in the parentheses in the equation above are non-zero; therefore, $\lambda_\text{pre}=\lambda_\text{pre}^*$.
Applying the same logic as above to Equations (\ref{eq:alpha11}) and (\ref{eq:alpha12}), it follows that  we get $\lambda_\text{post}=\lambda_\text{post}^*$.
Finally, we note that because $\lambda_\text{pre}=\lambda_\text{pre}^*$ and from (\ref{eq:alpha01}) $(\theta_1 - \theta_1^*)=\frac{\lambda_\text{pre}}{\alpha_1^{1}} - \frac{\lambda_\text{pre}^*}{\alpha_1^{1}}=0$, as shown above, we get $\theta_1 = \theta_1^*$.
Now that we have established $\lambda_\text{pre} = \lambda_\text{pre}^*$ and $\lambda_\text{post} = \lambda_\text{post}^*$.
This proves that the parameters $(\theta_1, \lambda_\text{pre}, \lambda_\text{post})$ are identified, and are given by   $\theta_1 = \frac{\alpha_1^{1}\tilde{\gamma}_0^{1} - \alpha_1^{2}\tilde{\gamma}_0^{2}}{\alpha_1^{1} - \alpha_1^{2}}; \lambda_\text{pre} =\alpha_1^{1}\times(\theta_1 - \tilde{\gamma}_0^{1}); \lambda_\text{post} = \alpha_1^{1}\times(\theta_1 - \tilde{\gamma}_1^{1}).$
\end{proof}

\section{Second Bridge: Asymptotics of Synthetic GMM \label{section:sgmm_formal}}
The appendix proceeds in three steps. First, we show that the moment conditions are linear in the parameter vector. This linear representation lets us express the bias in closed form. Second, we prove the consistency of the Synthetic GMM under the assumption that all moments are valid. Third, we allow supply-side moment violations in some control markets and compare the resulting biases of standard and Synthetic GMM.

\subsection{Linear Structure of the Moment Function \label{app:affine}}

\medskip\noindent\emph{Notation and structural model.}\quad
Let $m\in\{1,\ldots,M\}$ index city-pair markets and
$t\in\{1,\ldots,T\}$ index quarterly time periods. Treated markets
$\cT$ (with $M_T=|\cT|$ elements) and control markets
$\cC=\{1,\ldots,M\}\setminus\cT$ (with $M_C=|\cC|$ elements) partition
the sample, where $M=M_T+M_C$. Time periods are partitioned into pre-merger
periods $\fTpre$ ($T_0=|\fTpre|$ elements) and
post-merger periods $\fTpost$ ($T_1=|\fTpost|$
elements). The treatment indicator is
$I_{mt}=\mathbf{1}\{m\in\cT, t\in\fTpost\}$.

The parameter vector is
$\bk=(\alpha_0,\alpha_1,\boldsymbol{\alpha}_2^{\top},\theta_0,
\tilde{\gamma}_0,\tilde{\gamma}_1,\boldsymbol{\theta}_2^{\top},
\theta_3)^{\top}\in\Sigma\subset\R^{d_{\kappa}}$,
with true value $\bkz$. The observed data are
$\mathcal{D}_{mt}=(P_{mt},Q_{mt},\mathbf{Z}_{mt},\mathbf{W}_{mt},
I_{mt})$. Let $\mathbf{A}_{mt}^D\in\R^{d_D}$ and
$\mathbf{A}_{mt}^S\in\R^{d_S}$ collect the demand and supply
instruments. The stacked moment function is
\begin{equation*}
  g(\mathcal{D}_{mt};\bk)
  =
  \begin{pmatrix}
    \mathbf{A}_{mt}^D\bigl(Q_{mt} - \alpha_0
      - \alpha_1 P_{mt}
      - \boldsymbol{\alpha}_2^{\top}\mathbf{Z}_{mt}\bigr)\\[4pt]
    \mathbf{A}_{mt}^S\bigl(P_{mt} - \theta_0
      - \tilde{\gamma}_0 G_{mt}^{0}Q_{mt}
      - \tilde{\gamma}_1 G_{mt}^{1}Q_{mt}
      - \boldsymbol{\theta}_2^{\top}\mathbf{W}_{mt}
      - \theta_3 I_{mt}\bigr)
  \end{pmatrix}\in\R^{d_g},
\end{equation*}
where $G_{mt}^{0}=1-I_{mt}$ and
$G_{mt}^{1}=I_{mt}$ partition the sample into pre- and
post-merger regimes. The exclusion restrictions require
$\E[\mathbf{A}_{mt}^S\vartheta_{mt}]=\mathbf{0}$ and
$\E[\mathbf{A}_{mt}^D\varphi_{mt}]=\mathbf{0}$. We maintain the demand
condition for all markets and periods throughout; the supply condition,
whose validity may vary across control markets, is the focus of this
section.

\medskip\noindent\emph{Linear structure.}\quad
Since $I_{mt}=1$ only for treated markets in post-merger periods,
$\tilde{\gamma}_{mt}$ equals $\tilde{\gamma}_0$ when $I_{mt}=0$ and
$\tilde{\gamma}_1$ when $I_{mt}=1$, which can be written as
$$
 \tilde{\gamma}_{mt}(\bk)= \tilde{\gamma}_0 G_{mt}^{0}
    + \tilde{\gamma}_1 G_{mt}^{1}.
$$ 
The key point is that $\mumt$ determines both the pre-merger price dynamics used to construct the SDiD weights and the supply-side IV invalidity through $\delt=\Gamt\mumt$.
This enters the supply equation linearly in $\tilde{\gamma}_0$ and
$\tilde{\gamma}_1$. Define the demand-and-supply  regressor vectors
\begin{equation*}
  \mathbf{x}_{mt}^D
  = \begin{pmatrix}1 \\ P_{mt} \\ \mathbf{Z}_{mt}\end{pmatrix},
  \qquad
  \mathbf{x}_{mt}^S
  = \begin{pmatrix}1 \\ G_{mt}^{0}Q_{mt} \\
      G_{mt}^{1}Q_{mt} \\ \mathbf{W}_{mt} \\ I_{mt}
    \end{pmatrix},
\end{equation*}
and the parameter sub-vectors
$\bk^D=(\alpha_0,\alpha_1,\boldsymbol{\alpha}_2^{\top})^{\top}$
and
$\bk^S=(\theta_0,\tilde{\gamma}_0,\tilde{\gamma}_1,
\boldsymbol{\theta}_2^{\top},\theta_3)^{\top}$.
The residuals
$e_{mt}^D(\bk)=Q_{mt}-(\bk^D)^{\top}\mathbf{x}_{mt}^D$
and
$e_{mt}^S(\bk)=P_{mt}-(\bk^S)^{\top}\mathbf{x}_{mt}^S$
are linear in $\bk^D$ and $\bk^S$ respectively.
The moment function, therefore, separates into a data term and a linear term in $\bk$:
\begin{equation*}
  g(\mathcal{D}_{mt};\bk)
  =
  \begin{pmatrix}
    \mathbf{A}_{mt}^D Q_{mt} \\
    \mathbf{A}_{mt}^S P_{mt}
  \end{pmatrix}
  -
  \underbrace{
    \begin{pmatrix}
      \mathbf{A}_{mt}^D(\mathbf{x}_{mt}^D)^{\top} & \mathbf{0} \\
      \mathbf{0} & \mathbf{A}_{mt}^S(\mathbf{x}_{mt}^S)^{\top}
    \end{pmatrix}
  }_{=: \mathbf{B}_{mt}}
  \bk.
  \label{eq:app_g_matrix}
\end{equation*}
The matrix $\mathbf{B}_{mt}=-\nabla_{\bk}g(\mathcal{D}_{mt};\bk)$ is
block-diagonal, depending on the data $\mathcal{D}_{mt}$ but not on
$\bk$. Evaluating at $\bkz$ and differencing gives
\begin{equation}
  g(\mathcal{D}_{mt};\bk) - g(\mathcal{D}_{mt};\bkz)
  = \mathbf{B}_{mt}(\bkz-\bk),
  \label{eq:affine_diff}
\end{equation}
so $g$ is affine in $\bk$ with Jacobian $-\mathbf{B}_{mt}$. The demand-and-supply  blocks
are
\begin{eqnarray*}
  \mathbf{B}_{mt}^D
  &=& \mathbf{A}_{mt}^D(\mathbf{x}_{mt}^D)^{\top}
  = \mathbf{A}_{mt}^D
    \begin{pmatrix}1 & P_{mt} & \mathbf{Z}_{mt}^{\top}\end{pmatrix}
  \in\R^{d_D\times(2+d_Z)},\\[4pt]
  \mathbf{B}_{mt}^S
  &=& \mathbf{A}_{mt}^S(\mathbf{x}_{mt}^S)^{\top}
  = \mathbf{A}_{mt}^S
    \begin{pmatrix}
      1 & G_{mt}^{0}Q_{mt} & G_{mt}^{1}Q_{mt}
      & \mathbf{W}_{mt}^{\top} & I_{mt}
    \end{pmatrix}
  \in\R^{d_S\times(4+d_W)}.
\end{eqnarray*} 
The block-diagonal structure reflects the exclusion restrictions: demand
instruments enter only the demand block, and supply instruments only the
supply block, so supply-side misspecification propagates exclusively
through the supply block.

\subsection{Consistency\label{section:consistency}}

Next, we show that Synthetic GMM is consistent when the demand- and supply-side moment conditions hold for all markets and periods. Throughout this subsection, we maintain $\E[\mathbf A_{mt}^D\varphi_{mt}]=\mathbf0$ and $\E[\mathbf A_{mt}^S\vartheta_{mt}]=\mathbf0$ for all $m$ and $t$. We assume that the data are independent across markets and time. We begin with some assumptions.

\begin{assumption}[Moment Condition]\label{ass:standard_gmm_start}
The parameter space $\Sigma$ is compact with $\bkz\in\mathrm{int}(\Sigma)$ and $\mathbb{E}[g(\mathcal{D}_{mt};  \boldsymbol{\kappa}^0)]=\boldsymbol{0}$ for all $m$ and $t$.
\end{assumption}
\begin{assumption}[Sampling]\label{ass:sampling}
The data are independent across markets and time.
\end{assumption}

\begin{assumption}[Identification]\label{ass:identification}
The weighted population moment $\mu_w(\bk)$ is uniquely zero at $\bkz$. Equivalently, for every $\varepsilon>0,
\inf_{\bk\in\Sigma:\ \|\bk-\bkz\|\geq \varepsilon}
\|\mu_w(\bk)\|>0$.
\end{assumption}

\begin{assumption}[Measurability and Continuity]\label{ass:continuity}
$(i)$ $g(\mathcal{D}_{mt};  \boldsymbol{\kappa})$ is measurable in $\mathcal{D}_{mt}, \forall\boldsymbol{\kappa} \in \Sigma$ and continuous in $\boldsymbol{\kappa}$ for each $\mathcal{D}_{mt}$. $(ii)$ The map $\boldsymbol{\kappa} \mapsto \mathbb{E}[g(\mathcal{D}_{mt};  \boldsymbol{\kappa})]$ is continuous on $\Sigma$.
\end{assumption}

\begin{assumption}[Differentiability]\label{ass:differentiability}
$g(\mathcal{D}_{mt};  \boldsymbol{\kappa})$ is continuously differentiable in $\boldsymbol{\kappa}$ on $\Sigma$ for almost every $\mathcal{D}_{mt}$.
\end{assumption}

\begin{assumption}[Boundedness]\label{ass:moment_bounds}
Uniformly over $m$ and $t$, $(i)$ $\mathbb{E}[\sup_{\boldsymbol{\kappa} \in \Sigma} \|g(\mathcal{D}_{mt};  \boldsymbol{\kappa})\|^2] \leq
\bar{g}^2< \infty$; and the Jacobian $(ii)$ $\mathbb{E}[\sup_{\boldsymbol{\kappa} \in \Sigma} \|\nabla_{\boldsymbol{\kappa}} g(\mathcal{D}_{mt};  \boldsymbol{\kappa})\|^2] < \infty$.
\end{assumption}

\begin{assumption}[Full Rank]\label{ass:rank}
Let  $G_w:=\sum_{m=1}^{M}\sum_{t=1}^{T}\omega_m^*r_t^* \mathbb{E}\!\left[\nabla_{\kappa}g(\mathcal D_{mt};\kappa^0)\right]
$ be the weighted population Jacobian and has
full column rank, $\operatorname{rank}(G_w)=d_\kappa$.
\end{assumption}

\begin{assumption}[Weighting Matrix]\label{ass:standard_gmm_end}
$\What=\operatorname{diag}(\What_D,\What_S) \xrightarrow{p} \mathcal{W}=\operatorname{diag}(\mathcal{W}_D,\mathcal{W}_S)$, where $\mathcal{W}_D$ and $\mathcal{W}_S$ are positive definite.
\end{assumption}

\begin{assumption}[Weight Consistency]\label{ass:weight_consistency}
The  population weights $\omega^*$ and $r^*$ are such that $
\max_{1 \leq m \leq M}  |{\omega}_m - \omega_m^*| = o_p(1/M)$, and  $\max_{1 \leq t \leq T} |{r}_t - r_t^*| = o_p(1/T).$
\end{assumption}

\begin{assumption}[Asymptotic]\label{ass:asymptotic}
As $M,T\to\infty$, $M/T\to\rho\in(0,\infty)$, $M_C/M\to\iota\in(0,1)$, and $T_0/T\to\pi_0\in(0,1)$.
\end{assumption}
\begin{assumption}[Bounded Weights]\label{ass:weight_bounds}
There exist constants $0  < C_\omega < \infty$ and $0  < C_r < \infty$ such that w.p.1, $0 \leq M {\omega}_m \leq C_\omega$ for all $m$ and $0 \leq  T {r}_t \leq C_r$ for all $t$, and the same bounds hold for the population weights, i.e., $0 \leq M\omega_m^* \leq C_\omega$ and $0 \leq T r_t^* \leq C_r$. In addition, the total market and time weight masses are bounded, i.e., $\sum_{m=1}^{M}\omega_m=O_p(1),
\sum_{t=1}^{T}r_t=O_p(1),$ and $\sum_{m=1}^{M}\omega_m^*=O(1), \sum_{t=1}^{T}r_t^*=O(1)$.
\end{assumption}

Assumptions \ref{ass:standard_gmm_start}--\ref{ass:standard_gmm_end} are standard GMM conditions \citep{Hansen1982,NeweyMcFadden1994}, while Assumptions \ref{ass:weight_consistency}--\ref{ass:weight_bounds} govern the behavior of the synthetic DiD weights \citep{ArkhangelskyAtheyHirshbergImbensWager2021}. Assumption \ref{ass:identification} requires that the weighted population moment $\mu_w(\bk)$ be uniquely zero at $\bkz$. Under the linear demand-and-supply system, this is equivalent to a weighted rank condition. Recall from Appendix \ref{app:affine} that the moment function is affine in the parameter vector, so that it can be written as $
g(\mathcal{D}_{mt};\bk)= \mathbf{Z}_{mt}(Y_{mt} -
\mathbf{X}_{mt}^\top\boldsymbol{\kappa}) =a_{mt}-B_{mt}\bk.$ Therefore, $ \mu_w(\bk)= \mathbf{A}_w-\mathbf{B}_w\bk,$ where
$\mathbf{A}_w:=\sum_{m=1}^{M}\sum_{t=1}^{T}\omega_m^*r_t^*\E[a_{mt}]$ and $\mathbf{B}_w:=\sum_{m=1}^{M}\sum_{t=1}^{T}
\omega_m^*r_t^*
\E[B_{mt}].$
Hence $\mu_w(\bk)=\mathbf{0}$ has a unique solution if  $\operatorname{rank}(\mathbf{B}_w)=d_\kappa$. Some SDiD weights may be zero; identification only requires that the observations receiving positive weight retain sufficient variation to identify $\bkz$. Under uniform weights, $\omega_m^*=1/M$ and $r_t^*=1/T$, this reduces to the standard GMM population identification condition.

Assumption \ref{ass:rank} with Definition \ref{def:sgmm} implies that the demand moments are unweighted, so $\mathbf{B}_w=\operatorname{diag}(\mathbf{B}^D,\mathbf{G}^S_w)$ with $\mathbf{B}^D=\frac{1}{MT}\sum_{m,t}\E[\mathbf{B}^D_{mt}]$ and $\mathbf{G}^S_w=\sum_{m,t}\omega_m^*r_t^*\E[\mathbf{B}^S_{mt}]$, and the rank condition holds if and only if $\operatorname{rank}(\mathbf{B}^D)=d_{\kappa^D}$, the standard GMM condition for the demand equation, and $\operatorname{rank}(\mathbf{G}^S_w)=d_{\kappa^S}$, where $d_{\kappa^D}$ and $d_{\kappa^S}$ are the numbers of demand-and-supply  parameters. Assumptions \ref{ass:weight_consistency} and \ref{ass:weight_bounds} below concern the weights of the supply moments; the demand moments' weights, $1/M$ and $1/T$, equal their population values and satisfy the bounds with $C_\omega=C_r=1$.

Assumption \ref{ass:weight_consistency} requires the estimated market weights to converge to well-defined population counterparts (the time weights, $1/T_0$ and $1/T_1$, are constants), while Assumption \ref{ass:weight_bounds} prevents individual observations from receiving asymptotically excessive weight. The condition $T_0/T\to\pi_0\in(0,1)$ in Assumption \ref{ass:asymptotic} ensures that both the pre- and post-merger windows grow with $T$. Since $r_t=1/T_1$ in post-merger periods, it follows that $T r_t=T/T_1\to(1-\pi_0)^{-1}$, so the post-merger time weights satisfy Assumption \ref{ass:weight_bounds}. To establish the consistency of Synthetic GMM, we first need to show that the weighted sample moment converges uniformly to its population counterpart despite using estimated market and time weights. Lemma \ref{lem:uniform_lln} establishes this uniform convergence. We then use the lemma to show that the Synthetic GMM objective converges uniformly to a population objective uniquely minimized at $\bkz$, which yields Theorem \ref{thm:sgmm_consistency} by the standard extremum-estimator argument.

\begin{lemma}\label{lem:uniform_lln}
Under Assumptions \ref{ass:standard_gmm_start}, \ref{ass:sampling}, \ref{ass:moment_bounds}, \ref{ass:weight_consistency}, \ref{ass:asymptotic}, and \ref{ass:weight_bounds},
$
\sup_{\boldsymbol{\kappa}\in\Sigma}\left\|\gtil(\boldsymbol{\kappa})-\mu_w(\boldsymbol{\kappa})\right\|\xrightarrow{p}0.
$
\end{lemma}
\begin{proof}
Decompose
\begin{align*}
\gtil(\boldsymbol{\kappa}) - \mu_w(\boldsymbol{\kappa})
&= \underbrace{\sum_{m,t} {\omega}_m {r}_t \bigl[g(\mathcal{D}_{mt}; \boldsymbol{\kappa}) - \mathbb{E}[g(\mathcal{D}_{mt}; \boldsymbol{\kappa})]\bigr]}_{A_M(\boldsymbol{\kappa})}
+ \underbrace{\sum_{m,t} ({\omega}_m {r}_t - \omega_m^* r_t^*) \mathbb{E}[g(\mathcal{D}_{mt}; \boldsymbol{\kappa})]}_{B_M(\boldsymbol{\kappa})}.
\end{align*}

Consider first $A_M(\boldsymbol{\kappa})$. By the affine structure of the moment function (Appendix \ref{app:affine}), $
g(\mathcal D_{mt};\bk)-\E[g(\mathcal D_{mt};\bk)]=g(\mathcal D_{mt};\bkz)-\E[g(\mathcal D_{mt};\bkz)]+\left(\mathbf B_{mt}-\E[\mathbf B_{mt}]\right)(\bkz-\bk).$
Therefore, $
A_M(\bk)=\sum_{m,t}\omega_mr_t\left[g(\mathcal D_{mt};\bkz)-\E[g(\mathcal D_{mt};\bkz)]\right]+\left\{\sum_{m,t}\omega_mr_t\left(\mathbf B_{mt}-\E[\mathbf B_{mt}]\right)\right\}(\bkz-\bk)$, and since $\Sigma$ is compact,
\begin{eqnarray*}
\sup_{\bk\in\Sigma}\|A_M(\bk)\|\leq\left\|\sum_{m,t}\omega_mr_t\left[g(\mathcal D_{mt};\bkz)-\E[g(\mathcal D_{mt};\bkz)]\right]\right\|+\operatorname{diam}(\Sigma)\left\|\sum_{m,t}\omega_mr_t\left(\mathbf B_{mt}-\E[\mathbf B_{mt}]\right)\right\|.
\end{eqnarray*}
We show that both terms on the right-hand side are $o_p(1)$. For the first term, add and subtract the population weights:
\begin{eqnarray}
&&\sum_{m,t}\omega_mr_t\left[g(\mathcal D_{mt};\bkz)-\E[g(\mathcal D_{mt};\bkz)]\right]=\sum_{m,t}\omega_m^*r_t^*\left[g(\mathcal D_{mt};\bkz)-\E[g(\mathcal D_{mt};\bkz)]\right]\notag\\
&&+\sum_{m,t}\left(\omega_mr_t-\omega_m^*r_t^*\right)\left[g(\mathcal D_{mt};\bkz)-\E[g(\mathcal D_{mt};\bkz)]\right].\label{eq:A}
\end{eqnarray}
Using independence across markets, Assumptions \ref{ass:moment_bounds} and \ref{ass:weight_bounds},
\begin{eqnarray*}
\E\left\|\sum_{m,t}\omega_m^*r_t^*\left[g(\mathcal D_{mt};\bkz)-\E[g(\mathcal D_{mt};\bkz)]\right]\right\|^2=\sum_m(\omega_m^*)^2\E\left\|\sum_t r_t^*\left[g(\mathcal D_{mt};\bkz)-\E[g(\mathcal D_{mt};\bkz)]\right]\right\|^2\\
\leq\left(\max_m\omega_m^*\right)\left(\sum_m\omega_m^*\right)\max_m\E\left\|\sum_t r_t^*\left[g(\mathcal D_{mt};\bkz)-\E[g(\mathcal D_{mt};\bkz)]\right]\right\|^2\\=O(M^{-1}),
\end{eqnarray*}
where the last equality follows because, by the Cauchy--Schwarz inequality and Assumption \ref{ass:moment_bounds}, $
\E\left\|\sum_t r_t^*\left[g(\mathcal D_{mt};\bkz)-\E[g(\mathcal D_{mt};\bkz)]\right]\right\|^2\leq\left(\sum_t r_t^*\right)\sum_t r_t^*\E\left\|g(\mathcal D_{mt};\bkz)-\E[g(\mathcal D_{mt};\bkz)]\right\|^2=O(1).$
Hence the population-weight term is $O_p(M^{-1/2})=o_p(1)$.
For the second term on the right-hand-side of Equation \eqref{eq:A}, using $\omega_mr_t-\omega_m^*r_t^*=({\omega}_m-\omega_m^*)r_t+\omega_m^*(r_t-r_t^*)$, we get
\begin{eqnarray*}
&&\left\|\sum_{m,t}(\omega_mr_t-\omega_m^*r_t^*)\left[g(\mathcal D_{mt};\bkz)-\E[g(\mathcal D_{mt};\bkz)]\right]\right\|\\
&&\leq\max_m|\omega_m-\omega_m^*|\sum_{m,t}r_t\left\|g(\mathcal D_{mt};\bkz)-\E[g(\mathcal D_{mt};\bkz)]\right\|\\
&&+\max_t|r_t-r_t^*|\sum_{m,t}\omega_m^*\left\|g(\mathcal D_{mt};\bkz)-\E[g(\mathcal D_{mt};\bkz)]\right\|.
\end{eqnarray*}
By Assumptions \ref{ass:moment_bounds} and \ref{ass:weight_bounds}, the two sums on the right-hand side are $O_p(M)$ and $O_p(T)$, respectively. Assumption \ref{ass:weight_consistency} therefore implies that the first term is $o_p(M^{-1})O_p(M)=o_p(1)$ and the second is $o_p(T^{-1})O_p(T)=o_p(1)$. Thus, the second term on the right-hand side of Equation \eqref{eq:A} is $o_p(1)$.

The same argument, using Assumption \ref{ass:moment_bounds} (ii), gives
$
\left\|\sum_{m,t}\omega_mr_t\left(\mathbf B_{mt}-\E[\mathbf B_{mt}]\right)\right\|=o_p(1),
$
and because $\operatorname{diam}(\Sigma)<\infty$, it follows that $\sup_{\bk\in\Sigma}\|A_M(\bk)\|=o_p(1)$.

Next, consider the second term, $B_M(\boldsymbol{\kappa})$, and write ${\omega}_m {r}_t - \omega_m^* r_t^* = ({\omega}_m - \omega_m^*){r}_t + \omega_m^*({r}_t - r_t^*)$. Let $\bar{M}_g := \sup_{\boldsymbol{\kappa},m,t} \|\mathbb{E}[g(\mathcal{D}_{mt}; \boldsymbol{\kappa})]\|$, which is finite, by Assumption  \ref{ass:moment_bounds}.
Then
\begin{align*}
\sup_{\boldsymbol{\kappa}} \|B_M(\boldsymbol{\kappa})\|
&\leq \bar{M}_g \sum_{m,t} |{\omega}_m - \omega_m^*|  {r}_t + \bar{M}_g \sum_{m,t} \omega_m^*  |{r}_t - r_t^*| \\
&\leq \bar{M}_g \times \Bigl(M\max_m |{\omega}_m - \omega_m^*| \Bigr) \times \Bigl(\sum_t {r}_t\Bigr)
    + \bar{M}_g \times \Bigl(T \max_t |{r}_t - r_t^*| \Bigr) \times \Bigl(\sum_m \omega_m^*\Bigr).
\end{align*}
By Assumption \ref{ass:weight_consistency}, $M\max_m|{\omega}_m-\omega_m^*|=o_p(1)$ and $T\max_t|{r}_t-r_t^*|=o_p(1),$ and by Assumption \ref{ass:weight_bounds}, the sums $\sum_t r_t$ and $\sum_m\omega_m^*$ are bounded, therefore, $\sup_{\bk\in\Sigma}\|B_M(\bk)\|=o_p(1)$. Hence, we get $\sup_{\boldsymbol{\kappa} \in \Sigma} \|\tilde{g}(\boldsymbol{\kappa}) - \mu_w(\boldsymbol{\kappa})\| \xrightarrow{p} 0$.
\end{proof}

\begin{proof}[Proof of Theorem \ref{thm:sgmm_consistency}]\label{proof:sgmm_consistency}
Define the sample objective $Q_M(\boldsymbol{\kappa}) := \gtil(\boldsymbol{\kappa})^\top \What \gtil(\boldsymbol{\kappa})$ and the population objective $Q(\boldsymbol{\kappa}) := \mu_w(\boldsymbol{\kappa})^\top \mathcal{W} \mu_w(\boldsymbol{\kappa})$. By Assumption \ref{ass:identification}, $\mu_w(\bkz)=\mathbf0$ and $\bkz$ is the unique zero of $\mu_w(\bk)$ on $\Sigma$. Since $\mathcal{W}$ is positive definite (Assumption  \ref{ass:standard_gmm_end}), $Q(\boldsymbol{\kappa}^0) = 0 < \mu_w(\boldsymbol{\kappa})^\top \mathcal{W} \mu_w(\boldsymbol{\kappa}) = Q(\boldsymbol{\kappa})$ for all $\boldsymbol{\kappa} \neq \boldsymbol{\kappa}^0$. Therefore, $\boldsymbol{\kappa}^0$ is the unique minimizer of $Q(\boldsymbol{\kappa})$.
\color{black}{\color{violet!75!black}\textbf{[end of verified port] }}

Next, we show that the sample objective function $Q_M$ converges uniformly to the population $Q$.
To this end, we begin by noting the following relationship,
\begin{eqnarray*}
|Q_M(\boldsymbol{\kappa}) - Q(\boldsymbol{\kappa})|&=& \bigl|\gtil(\boldsymbol{\kappa})^\top \What \gtil(\boldsymbol{\kappa}) - \mu_w(\boldsymbol{\kappa})^\top \mathcal{W} \mu_w(\boldsymbol{\kappa})\bigr|\\
&=&\bigl|\gtil(\boldsymbol{\kappa})^\top \What \gtil(\boldsymbol{\kappa}) - \mu_w(\boldsymbol{\kappa})^\top \mathcal{W} \mu_w(\boldsymbol{\kappa})+\mu_w(\boldsymbol{\kappa})^\top \What \mu_w(\boldsymbol{\kappa})-\mu_w(\boldsymbol{\kappa})^\top \What \mu_w(\boldsymbol{\kappa})\bigr|\\
&\leq& \underbrace{\bigl|\gtil(\boldsymbol{\kappa})^\top \What \gtil(\boldsymbol{\kappa}) - \mu_w(\boldsymbol{\kappa})^\top \What \mu_w(\boldsymbol{\kappa})\bigr|}_{\text{term1}}
+ \underbrace{\bigl|\mu_w(\boldsymbol{\kappa})^\top (\What - \mathcal{W}) \mu_w(\boldsymbol{\kappa})\bigr|}_{\text{term 2}}
\end{eqnarray*}
where the second equality comes from adding and subtracting $\mu_w(\boldsymbol{\kappa})^\top \What \mu_w(\boldsymbol{\kappa})$ and the inequality is the triangle inequality.
Then, we note by adding and subtracting $\gtil(\boldsymbol{\kappa})^\top \What \mu_w(\boldsymbol{\kappa})$ from \text{term 1} and applying triangle inequality, we get
\begin{eqnarray*}
\text{term1}
&=& \bigl|\gtil(\boldsymbol{\kappa})^\top \What [\tilde{g}(\boldsymbol{\kappa}) - \mu_w(\boldsymbol{\kappa})]
+ [\gtil(\boldsymbol{\kappa}) - \mu_w(\boldsymbol{\kappa})]^\top \What \mu_w(\boldsymbol{\kappa})\bigr|\\
&\leq& \|\gtil(\boldsymbol{\kappa})\| \times \|\What\| \times \|\gtil(\boldsymbol{\kappa}) - \mu_w(\boldsymbol{\kappa})\|
+ \|\gtil(\boldsymbol{\kappa}) - \mu_w(\boldsymbol{\kappa})\|\times \|\What\| \times \|\mu_w(\boldsymbol{\kappa})\| \nonumber \\
&=& \|\What\| \times \|\tilde{g}(\boldsymbol{\kappa}) - \mu_w(\boldsymbol{\kappa})\| \times \bigl[\|\gtil(\boldsymbol{\kappa})\| + \|\mu_w(\boldsymbol{\kappa})\|\bigr]. \label{eq:T1_bound}
\end{eqnarray*}
Next, the submultiplicative property of the operator norm, $|a^\top B a| \leq \|a\|^2 \|B\|$ for any vector $a$ and matrix $B$, we can bound $\text{term2}\leq \|\mu_w(\boldsymbol{\kappa})\|^2 \cdot \|\What - \mathcal{W}\|$.

Combining the bounds of the two terms and taking the supremum over $\boldsymbol{\kappa} \in \Sigma$, we get
\begin{eqnarray*}
\sup_{\boldsymbol{\kappa}} |Q_M(\boldsymbol{\kappa}) - Q(\boldsymbol{\kappa})|
&\leq& \underbrace{\|\What\|}_{=O_p(1),   as   \hat{\mathcal{W}} \xrightarrow{p} \mathcal{W}\text{  Assumption  \ref{ass:standard_gmm_end}}} \times \underbrace{\sup_{\boldsymbol{\kappa}} \|\gtil(\boldsymbol{\kappa}) - \mu_w(\boldsymbol{\kappa})\|}_{=o_p(1), \text{  Lemma  \ref{lem:uniform_lln}}} \times \sup_{\boldsymbol{\kappa}}\bigl[\|\gtil(\boldsymbol{\kappa})\| \\&& + \|\mu_w(\boldsymbol{\kappa})\|\bigr] + \sup_{\boldsymbol{\kappa}} \|\mu_w(\boldsymbol{\kappa})\|^2 \times \underbrace{\|\What - \mathcal{W}\|}_{=o_p(1), \text{  Assumption  \ref{ass:standard_gmm_end}}}. \label{eq:uniform_bound}
\end{eqnarray*}
Then, we note that $\sup_{\boldsymbol{\kappa}} \|\mu_w(\boldsymbol{\kappa})\|<\infty$ because $\mu_w$ is continuous (Assumption  \ref{ass:continuity}) and $\Sigma$ is compact. Furthermore, $\sup_{\boldsymbol{\kappa}} \|\gtil(\boldsymbol{\kappa})\| \leq \sup_{\boldsymbol{\kappa}} \|\gtil(\boldsymbol{\kappa}) - \mu_w(\boldsymbol{\kappa})\| + \sup_{\boldsymbol{\kappa}} \|\mu_w(\boldsymbol{\kappa})\| = O_p(1)$.
Hence, $
\sup_{\boldsymbol{\kappa}} |Q_M(\boldsymbol{\kappa}) - Q(\boldsymbol{\kappa})| \leq O_p(1) \times o_p(1) \times O_p(1) + O(1) \times o_p(1) = o_p(1).$
Finally, because $\Sigma$ is compact, $Q(\boldsymbol{\kappa})$ is continuous by Assumption \ref{ass:continuity}, $\bkz$ is its unique minimizer, and $\sup_{\boldsymbol{\kappa}}|Q_M(\boldsymbol{\kappa})-Q(\boldsymbol{\kappa})|\xrightarrow{p}0$, the extremum estimator theorem \citep{NeweyMcFadden1994} implies that $\hat{\boldsymbol{\kappa}}_{\mathrm{S-GMM}}\xrightarrow{p}\boldsymbol{\kappa}^0$.
\end{proof}

\subsection{Bias Analysis  \label{section:sgmm_bias_details}}

This subsection provides technical details and proofs for the bias analysis in Section \ref{sec:synthetic_GMM}.
Assumptions \ref{ass:factor} (Factor Model) and \ref{ass:iv_treated}
(IV Validity for Treated Markets) are stated in full in the main text, as is an abbreviated version of Assumption \ref{ass:balance}
(Pre-Merger Price Balance). We provide the precise statement of
Assumption \ref{ass:balance} in the text below.
All intermediate derivations and results, and the proofs of
Theorems \ref{thm:syn_bias} and \ref{thm:std_bias} are contained here.
We maintain the remaining regularity conditions needed for convergence and use the affine structure of the moment function in Appendix \ref{app:affine} to characterize the resulting biases of synthetic and standard GMM. Our starting point is that the weighted population moment need not be zero at $\bkz$, so Assumption \ref{ass:identification} is not imposed in this subsection.

\subsection*{Step 1.}
\paragraph{Synthetic GMM.}
The Synthetic GMM converges to a pseudo-true value,
\[
  \bksw = \arg\min_{\bk} \muw(\bk)^{\top}\Wmat \muw(\bk).
\]
We have shown previously that when $\muw(\bkz)=\mathbf{0}$ the estimator is consistent, whereas when $\muw(\bkz)\neq\mathbf{0}$ the difference $\bksw-\bkz$ characterizes its asymptotic bias.

Using the affine structure in \eqref{eq:affine_diff}, the weighted population moment can be written as an affine function of the parameters, i.e.,
$\muw(\bk)=\muw(\bkz)+\Gw(\bkz-\bk),$
where $\Gw=\sum_{m,t}\omega_m^* r_t^* \E[\mathbf{B}_{mt}]$ is the
population Jacobian for Synthetic GMM.

The population objective $Q_w(\bk)=\muw(\bk)^{\top}\Wmat\muw(\bk)$ is quadratic in $\bk$. As, $\muw(\bk)=\muw(\bkz)+\Gw(\bkz-\bk)$, its gradient can be written as
$$\nabla_{\bk}Q_w(\bk)=-2\Gw^\top\Wmat\left[\muw(\bkz)+\Gw(\bkz-\bk)\right].$$ At the pseudo-true value $\bksw$, the first-order condition is $\Gw^\top\Wmat\left[\muw(\bkz)+\Gw(\bkz-\bksw)\right]=\mathbf{0}$. Solving gives
\begin{equation}
  \bksw - \bkz
  = (\Gw^{\top}\Wmat\Gw)^{-1}\Gw^{\top}\Wmat \muw(\bkz).
  \label{eq:foc_sgmm}
\end{equation}
To evaluate \eqref{eq:foc_sgmm}, write the weighted population moment in demand-and-supply  blocks:
$\muw(\bkz)=\begin{pmatrix}
\muw^D(\bkz)\\
\muw^S(\bkz)
\end{pmatrix}$.
The demand block is zero by the maintained validity of demand instruments,
$\muw^D(\bkz)=\mathbf{0}$. Potential misspecification arises only in the
supply block, through $\E[\mathbf A_{mt}^S\vartheta_{mt}]$.
Under any weighting scheme, the population moment at the truth decomposes
into treated and control market contributions. By
Assumption \ref{ass:iv_treated}, treated markets contribute zero
regardless of the weights assigned to them, and the demand block is
zero by the maintained validity of demand instruments. Therefore
\begin{align}
  \muw(\bkz) &= \begin{pmatrix}
    \frac{1}{MT}\sum_{m,t}\underbrace{\E[g^D(\mathcal{D}_{mt};\bk^D_0)]}_{=\mathbf 0}\\[4pt]
    \frac{1}{M_T}\sum_{m\in\cT}\sum_{t}r_t^*\underbrace{\E[g^S(\mathcal{D}_{mt};\bk^S_0)]}_{=\mathbf 0}
     +\sum_{m\in\cC}\sum_{t}\omega_m^* r_t^*\E\bigl[\mathbf{A}_{mt}^S\vartheta_{mt}\bigr]
  \end{pmatrix}
  = \begin{pmatrix}\mathbf{0}\\\dtil\end{pmatrix},
  \label{eq:pop_syn_at_truth}
\end{align}
where, using the factor structure in Assumption \ref{ass:factor},
\begin{eqnarray}
  \dtil =\sum_{m\in\cC}\sum_{t}\omega_m^* r_t^*
     \E\bigl[\mathbf{A}_{mt}^S\vartheta_{mt}\bigr]
  = \sum_{m\in\cC}\sum_{t=1}^{T}\omega_m^* r_t^* \delt
   = \sum_{m\in\cC}\sum_{t=1}^{T}\omega_m^* r_t^* \Gamt\mumt \label{eq:delta_tilde}
\end{eqnarray}
is the weighted average invalidity of the IVs for the control pool. Substituting \eqref{eq:pop_syn_at_truth} into \eqref{eq:foc_sgmm}, gives
\begin{equation}
  \bksw - \bkz
  =\begin{pmatrix}\mathbf{0}\\ \mathbf{L}_{S,w}\dtil\end{pmatrix},\qquad
  \mathbf{L}_{S,w}=\bigl(\mathbf{G}_w^{S\top}\mathcal{W}_S\mathbf{G}^S_w\bigr)^{-1}\mathbf{G}_w^{S\top}\mathcal{W}_S,
  \label{eq:bias_syn_block}
\end{equation}
because $\Gw=\operatorname{diag}(\mathbf{B}^D,\mathbf{G}^S_w)$ and $\mathcal{W}=\operatorname{diag}(\mathcal{W}_D,\mathcal{W}_S)$ are block diagonal, so $(\Gw^{\top}\mathcal{W}\Gw)^{-1}\Gw^{\top}\mathcal{W}$ is block diagonal and its demand block multiplies the zero demand moment.

\paragraph{Standard GMM.}

Standard GMM is the special case of Synthetic GMM with uniform weights $\frac{1}{MT}$. Following the same steps as above, but with uniform weights, the weighted invalidity $\dtil$ collapses
to a simple unweighted average over the entire control pool:
\begin{equation}
  \dtil\big|_{\text{uniform}}
  = \sum_{m\in\cC}\sum_{t=1}^{T}
    \frac{1}{M}\cdot\frac{1}{T} \Gamt\mumt
  = \frac{M_C}{M}\cdot\frac{1}{M_C T}
    \sum_{m\in\cC}\sum_{t=1}^{T}\Gamt\mumt
  =: \frac{M_C}{M} \dbar,
  \label{eq:dtil_uniform}
\end{equation}
where $\dbar$
is the average instrument invalidity in the control pool.
Substituting \eqref{eq:dtil_uniform} into the bias
expression \eqref{eq:foc_sgmm}, with uniform-weight replacing
synthetic-weight, gives
\begin{equation}
  \bks - \bkz
  = (\G^{\top}\Wmat\G)^{-1}\G^{\top}\Wmat
    \begin{pmatrix}\mathbf{0}\\\frac{M_C}{M}\dbar\end{pmatrix}
  = \mathbf{L} \frac{M_C}{M} \begin{pmatrix}\mathbf{0}\\\dbar\end{pmatrix}=\frac{M_C}{M}\mathbf L_S\dbar,
  \label{eq:bias_std_block}
\end{equation}
where $\mathbf L:=(\G^{\top}\Wmat\G)^{-1}\G^{\top}\Wmat=(\mathbf L_D,\mathbf L_S)$ is partitioned conformably with the demand-and-supply  moment blocks.

The only difference between the bias term for Synthetic GMM, $\dtil$ in \eqref{eq:delta_tilde}, and for standard GMM, $\dbar$ in \eqref{eq:dtil_uniform}, is the weights.
Under the linear factor model, Synthetic GMM places less weight on observations where $\Gamt\mumt\neq0$, whereas standard GMM weights all observations equally. We formalize this intuition next.
\color{black}{\color{violet!75!black}\textbf{[end of verified port] }}

\subsection*{Step 2.}
\paragraph{Step 2.1. Full Pre-Merger Price Balance.}
We begin with the precise statement of Assumption \ref{ass:balance}, which is stated in abbreviated form in
the main text. The synthetic weight optimizations \eqref{eq:w_opt}--\eqref{eq:r_opt} achieve pre-treatment balance in
residualized prices. In addition, these residualized prices reveal the supply-side factor loadings in the sense that, for pre-merger periods,
$\widetilde P_{mt}=\mumt^\top \mathbf F_t+\eta_{mt}$,
where $\eta_{mt}$ is the remaining idiosyncratic component of residualized
prices, satisfying $\E[\eta_{mt}\mid \mathbf F_t]=0$, uniformly bounded
second moments, and weak dependence across markets and time.\footnote{The
distinction between $\eta_{mt}$ and the supply-side idiosyncratic shock
$\epsilon_{mt}$ in Assumption \ref{ass:factor} is that $\eta_{mt}$ may also
contain the residualized demand-side shock. If residualized prices are
constructed after controlling for $Q_{mt}$, so that the remaining price
residual isolates the supply-side pricing disturbance, then
$\eta_{mt}=\epsilon_{mt}$ under the linear latent factor specification in
Assumption \ref{ass:factor}.}
For all $t\in\fTpre$ and $m\in\cC$, the weights satisfy
\begin{align}
  \left\|\sum_{m\in\cC}\omega_m^*\widetilde{P}_{mt}
    - \frac{1}{M_T}\sum_{m\in\cT}\widetilde{P}_{mt}\right\|
  &= O_p\!\left(\frac{\sigma}{\sqrt{T_0}}\right),
  \label{eq:price_balance}\\[6pt]
  \left\|\sum_{t\in\fTpre}r_t^*\widetilde{P}_{mt}
    - \frac{1}{T_1}\sum_{t\in\fTpost}\widetilde{P}_{mt}
  \right\|
  &= O_p\!\left(\frac{\sigma}{\sqrt{T_0}}\right)
  \label{eq:time_balance}
\end{align}
uniformly in $m\in\cC$, where $\sigma^2=\mathbb{V}(\eta_{mt})$.

Furthermore, for the time-balance condition to translate into balance in
supply-side instrument invalidity, we assume that the time weights balance
the common factors relevant for the instrument--factor covariance. In
particular, letting $ \bar{\mathbf F}:=\frac{1}{T}\sum_{t=1}^{T}\mathbf F_t$ where $\|\mathbf F_t\|<\infty$, the time weights satisfy $\left\|\sum_{t=1}^{T}r_t^*\mathbf F_t-\bar{\mathbf F}\right\|
  =O_p(T_0^{-1/2})$. 
We also assume that the instrument--factor covariance varies with these same
common factors through a bounded linear map. That is, there exists a bounded
linear operator $\mathcal H$ such that $\forall t, \Gamt=\mathcal H(\mathbf F_t)$.\footnote{This condition is not implied by the factor representation of
the cost shock alone. The factor model restricts the cost shock
$\vartheta_{mt}=\mumt^\top \mathbf F_t+\epsilon_{mt}$, whereas
$\Gamt=\E[\mathbf A_{mt}^S\mathbf F_t^\top]$ depends on the joint behavior
of the supply instruments and the common factors. The condition requires
that the part of this covariance relevant for IV invalidity moves with the
same common factors that are balanced by the SDiD time weights.}
Together, these conditions imply $\left\|\sum_{t=1}^{T}r_t^*\Gamt-\frac{1}{T}\sum_{t=1}^{T}\Gamt\right\|=O_p(T_0^{-1/2})$.

We impose the $O_p(\sigma T_0^{-1/2})$ rate to capture the improvement in balance as the pre-merger window grows.
The following lemma formalizes the connection between price balance and
instrument validity balance that drives the main result.

\begin{lemma}
\label{lem:dtil_rate}
Suppose Assumptions \ref{ass:factor}, \ref{ass:iv_treated}, \ref{ass:balance},
\ref{ass:weight_consistency}, and \ref{ass:weight_bounds} hold. Then $\|\dtil\|=O_p(T_0^{-1/2})$. 
\end{lemma}

\begin{proof}
Let $\LamT:=\frac{1}{M_T}\sum_{m\in\cT}\mumt, \widetilde{\LamT}:=\sum_{m\in\cC}\omega_m^*\mumt-\LamT$ and $\bar{\boldsymbol{\Gamma}}:=\frac{1}{T}\sum_{t}\Gamt<\infty$. Then we prove this in three steps. First, we show $\|\widetilde{\LamT}\|=O_p\!\left(\frac{1}{\sqrt{T_0}}\right)$, then we show $\left\|\sum_{t}r_t^*\Gamt-\bar{\boldsymbol{\Gamma}}\right\|=O_p\!\left(\frac{1}{\sqrt{T_0}}\right)$. Then, we show that $\|\dtil\|=O_p(T_0^{-1/2})$.

\medskip\noindent\emph{Step 1: Bounding the residual loading gap
$\|\widetilde{\LamT}\|$.}\quad
Define the pre-merger residualized-price imbalance as $\Delta_t^P:=\sum_{m\in\cC}\omega_m^*\widetilde P_{mt}
-\frac{1}{M_T}\sum_{m\in\cT}\widetilde P_{mt}, t\in\fTpre.$
By Assumption \ref{ass:balance}, residualized prices reveal the same
market-specific factor loadings that determine supply-side instrument
invalidity. Thus, for pre-merger periods, $\widetilde P_{mt}=\mumt^\top\mathbf F_t+\eta_{mt}$, so we get 
$\Delta_t^P=\widetilde{\LamT}^{\top}\mathbf F_t+\bar\eta_t$, and $\bar\eta_t:=\sum_{m\in\cC}\omega_m^*\eta_{mt}
-\frac{1}{M_T}\sum_{m\in\cT}\eta_{mt}$. Stacking over pre-merger periods gives
$\Delta^P=\mathbf F_{\mathrm{pre}}\widetilde{\LamT}+\bar\eta$, where
$\mathbf F_{\mathrm{pre}}=(\mathbf F_t^\top)_{t\in\fTpre}$. Premultiplying by
$T_0^{-1}\mathbf F_{\mathrm{pre}}^\top$ gives $\left(\frac{1}{T_0}\mathbf F_{\mathrm{pre}}^\top\mathbf F_{\mathrm{pre}}\right)\widetilde{\LamT}=\frac{1}{T_0}\mathbf F_{\mathrm{pre}}^\top\Delta^P-\frac{1}{T_0}\mathbf F_{\mathrm{pre}}^\top\bar\eta$.
By Assumption \ref{ass:balance}, the pre-merger balance term and the
weighted residual term on the right-hand side are both
$O_p(T_0^{-1/2})$. Since the pre-merger factor matrix is well conditioned
under Assumption \ref{ass:factor}, we get
$\|\widetilde{\LamT}\|=O_p(T_0^{-1/2})$.

\medskip\noindent\emph{Step 2: Bounding the time-weight approximation error $\left\|\sum_{t}r_t^*\Gamt-\bar{\boldsymbol{\Gamma}}\right\|$.}\quad
By Assumption \ref{ass:balance}, the time weights balance the common factors, i.e., $\left\|\sum_{t}r_t^*\mathbf F_t-\bar{\mathbf F}\right\|=O_p(T_0^{-1/2})$, where $\bar{\mathbf F}:=\frac{1}{T}\sum_{t=1}^{T}\mathbf F_t.$
Also by Assumption \ref{ass:balance}, $\Gamt=\mathcal H(\mathbf F_t)$ for all $t$, where $\mathcal H$ is a bounded linear operator. Therefore, $\sum_{t}r_t^*\Gamt=\sum_{t}r_t^*\mathcal H(\mathbf F_t)=\mathcal H\left(\sum_{t}r_t^*\mathbf F_t\right)$, and $\bar{\boldsymbol{\Gamma}}=\frac{1}{T}\sum_{t}\Gamt=\frac{1}{T}\sum_{t}\mathcal H(\mathbf F_t)=\mathcal H(\bar{\mathbf F})$. Hence, $\sum_{t}r_t^*\Gamt-\bar{\boldsymbol{\Gamma}}=
\mathcal H\left(\sum_{t}r_t^*\mathbf F_t-\bar{\mathbf F}\right)$. Taking norms on both sides, we get $\left\|\sum_{t}r_t^*\Gamt-\bar{\boldsymbol{\Gamma}}\right\|\leq\|\mathcal H\|\left\|\sum_{t}r_t^*\mathbf F_t-\bar{\mathbf F}\right\|$, and because $\|\mathcal H\|<\infty$, we get the desired result $\left\|\sum_{t}r_t^*\Gamt-\bar{\boldsymbol{\Gamma}}\right\|=O_p(T_0^{-1/2})$. 

\medskip\noindent\emph{Step 3: Bounding $\|\dtil\|$.}\quad
Substituting $\sum_{m\in\cC}\omega_m^*\mumt=\widetilde{\LamT}+\LamT$, into $\dtil$ gives $$\dtil=\sum_{m\in\cC}\sum_{t}\omega_m^*r_t^*\Gamt\mumt=\sum_{t}r_t^*\Gamt\left(\widetilde{\LamT}+\LamT\right)=\sum_{t}r_t^*\Gamt\widetilde{\LamT}+\sum_{t}r_t^*\Gamt\LamT={\text{Term 1}}+{\text{Term 2}}.$$

For Term 1, Assumption \ref{ass:weight_bounds} implies $\sum_{t}r_t^*=O(1)$. Also because $\|\Gamt\|\leq\bar\Gamma<\infty$ uniformly in $t$, we get $\|\text{Term 1}\|\leq\sum_{t}r_t^*\|\Gamt\| \|\widetilde{\LamT}\|\leq
\bar\Gamma\left(\sum_{t}r_t^*\right)\|\widetilde{\LamT}\|=O_p(T_0^{-1/2})$, where the last equality follows from Step 1.

For Term 2, add and subtract $\bar{\boldsymbol{\Gamma}}\LamT$ to get $\text{Term 2} =\sum_{t}r_t^*\Gamt\LamT=\left(\sum_{t}r_t^*\Gamt-\bar{\boldsymbol{\Gamma}}\right)\LamT+\bar{\boldsymbol{\Gamma}}\LamT.$
By Assumption \ref{ass:iv_treated}, $\bar{\boldsymbol{\Gamma}}\LamT=\left(\frac{1}{T}\sum_{t}\Gamt\right)
\left(\frac{1}{M_T}\sum_{m\in\cT}\mumt\right)=\frac{1}{M_TT}\sum_{m\in\cT}\sum_{t}\Gamt\mumt=\mathbf 0.$
Therefore, using Step 2 and boundedness of $\|\LamT\|$,
\[
\|\text{Term 2}\|=\left\|\left(\sum_{t=1}^{T}r_t^*\Gamt-\bar{\boldsymbol{\Gamma}}\right)\LamT\right\|\leq
\left\|\sum_{t=1}^{T}r_t^*\Gamt-\bar{\boldsymbol{\Gamma}}\right\|\|\LamT\|=O_p(T_0^{-1/2}).
\]

Combining these bounds we get $\|\dtil\|\leq\|\text{Term 1}\|+\|\text{Term 2}\|=O_p(T_0^{-1/2}).$
\end{proof}

\paragraph{Step 2.2. Bias of Synthetic GMM.}
The rate depends on the quality of the optimization \eqref{eq:w_opt}.
If the fit is poor, the practical bias reduction will be limited even
if $T_0$ is large. Combined with \eqref{eq:bias_syn_block},
$\|\bksw-\bkz\|=O_p(T_0^{-1/2})$, we get that as the pre-merger window grows, the Synthetic GMM optimization finds a weighted control pool whose factor loadings increasingly resemble those of the treated markets, driving $\dtil$ toward zero.
We restate the Theorems  \ref{thm:syn_bias} and \ref{thm:std_bias} from the text with their full assumption lists.

\noindent\textbf{Theorem} \ref{thm:syn_bias}.
\emph{Suppose Assumptions \ref{ass:factor}, \ref{ass:iv_treated}, \ref{ass:balance}, \ref{ass:standard_gmm_start}, \ref{ass:sampling}, \ref{ass:moment_bounds}, \ref{ass:standard_gmm_end}, \ref{ass:weight_consistency}, \ref{ass:asymptotic}, and \ref{ass:weight_bounds} hold. Suppose further that $\mathbf{B}^D=\frac{1}{MT}\sum_{m,t}\E[\mathbf B^D_{mt}]$ has full column rank $d_{\kappa^D}$, that $\mathbf G^S_w=\sum_{m,t}\omega_m^*r_t^*\E[\mathbf B^S_{mt}]$ has full column rank $d_{\kappa^S}$, and that the smallest eigenvalue of $\mathbf G_w^{S\top}\mathcal W_S\mathbf G^S_w$ is bounded away from zero.} Then $\hat{\bk}_{S-GMM}\convp\bksw$, where $\bksw-\bkz$ is given by \eqref{eq:bias_syn_block}, equivalently by \eqref{eq:bias_syn}, and $\|\bksw-\bkz\|=O_p(T_0^{-1/2})$.

\begin{proof}
By Lemma \ref{lem:uniform_lln}, $
\sup_{\bk\in\Sigma}\|\gtil(\bk)-\muw(\bk)\|\convp0,
$ combining it with $\What\convp\Wmat$, gives that the S-GMM objective converges uniformly to the population objective $\muw(\bk)^{\top}\Wmat\muw(\bk)$. Since $\Gw$ has rank $d_\kappa$ and $\Wmat$ is positive definite, $\Gw^\top\Wmat\Gw$ is positive definite, and the population objective has the unique minimizer $\bksw$. The extremum estimator theorem \citep[Theorem 2.1]{NeweyMcFadden1994} therefore gives $\hat{\bk}_{S-GMM}\convp\bksw$. The rate follows from $\|\bksw-\bkz\|=\|\mathbf{L}_{S,w}\dtil\|\leq\|\mathbf{L}_{S,w}\|\cdot\|\dtil\|$. By the assumed lower bound on the smallest eigenvalue of $\mathbf G_w^{S\top}\mathcal W_S\mathbf G^S_w$, Assumptions \ref{ass:moment_bounds} and \ref{ass:weight_bounds}, and positive definiteness of $\mathcal W_S$, the operator norm $\|\mathbf{L}_{S,w}\|$ is bounded. Since $\|\dtil\|=O_p(T_0^{-1/2})$ by Lemma \ref{lem:dtil_rate}, it follows that $\|\bksw-\bkz\|=O_p(T_0^{-1/2})$.
\end{proof}

\paragraph{Step 2.3. Bias of GMM.}
Here, we state and prove the bias for standard GMM.

\noindent\textbf{Theorem} \ref{thm:std_bias}. \emph{(a) Suppose Assumptions \ref{ass:factor}, \ref{ass:iv_treated}, \ref{ass:standard_gmm_start}, \ref{ass:sampling}, \ref{ass:continuity}, \ref{ass:moment_bounds}, \ref{ass:standard_gmm_end}, and \ref{ass:asymptotic} hold. Suppose further that $\G$ has full column rank $d_\kappa$ and that $\mathbf L_S\dbar\neq\mathbf0$. Then $\hat{\bk}_{\mathrm{GMM}}\convp\bks$, where $\bks-\bkz$ is given by \eqref{eq:bias_std_block}, and $\bks\neq\bkz$. (b) In addition, suppose that as $M,T\rightarrow\infty$ according to Assumption \ref{ass:asymptotic}, $\G\rightarrow\G^\infty$, where $\G^\infty$ has full column rank $d_\kappa$, and $\dbar\rightarrow\dbar^\infty$. Let $\mathbf L^\infty:=\left((\G^\infty)^\top\Wmat\G^\infty\right)^{-1}(\G^\infty)^\top\Wmat=(\mathbf L_D^\infty,\mathbf L_S^\infty)$, and suppose that $\mathbf L_S^\infty\dbar^\infty\neq\mathbf0$. Then the asymptotic bias is nonvanishing: there is a constant $b_0>0$ such that $\liminf_{M, T\rightarrow\infty}\|\bks-\bkz\|\geq b_0>0$.}

\begin{proof}
This theorem has two parts. First, we prove part (a). By the same affine uniform-convergence argument used in Lemma \ref{lem:uniform_lln}, now with deterministic uniform weights, it follows that $\sup_{\bk\in\Sigma}\|\gbar(\bk)-\mup(\bk)\|\convp 0$. Combined with $\What\convp\Wmat$, the sample GMM objective converges uniformly to $Q(\bk)=\mup(\bk)^{\top}\Wmat\mup(\bk)$. This is a strictly convex quadratic in $\bk$ with Hessian $2\G^{\top}\Wmat\G$, which is positive definite since $\G$ has rank $d_{\kappa}$ and $\Wmat$ is positive definite. Setting the gradient to zero and solving yields the unique minimizer $\bks$ given by \eqref{eq:bias_std_block}. The extremum estimator theorem \citep[Theorem 2.1]{NeweyMcFadden1994} then gives $\hat{\bk}_{\mathrm{GMM}}\convp\bks$. Finally, from \eqref{eq:bias_std_block}, we have $\bks-\bkz=\frac{M_C}{M}\mathbf L_S\dbar$. As $M_C/M>0$ and $\mathbf L_S\dbar\neq\mathbf0$ by assumption, it follows that $\bks\neq\bkz$.

Next, we prove part (b). From \eqref{eq:bias_std_block}, $\bks-\bkz=\frac{M_C}{M}\mathbf L_S\dbar$. By Assumption \ref{ass:asymptotic}, $M_C/M\rightarrow\iota\in(0,1)$. Since $\G\rightarrow\G^\infty$, $\G^\infty$ has full column rank, and $\Wmat$ is positive definite, continuity of matrix inversion and multiplication gives $\mathbf L_S\rightarrow\mathbf L_S^\infty$. Together with $\dbar\rightarrow\dbar^\infty$, it follows that $\bks-\bkz\rightarrow\iota\mathbf L_S^\infty\dbar^\infty$. Because $\iota>0$ and $\mathbf L_S^\infty\dbar^\infty\neq\mathbf0$, $\left\|\iota\mathbf L_S^\infty\dbar^\infty\right\|=\iota\left\|\mathbf L_S^\infty\dbar^\infty\right\|>0$. Let $b_0:=\frac{\iota}{2}\left\|\mathbf L_S^\infty\dbar^\infty\right\|>0$. Then, for all sufficiently large $M$ and $T$, $\|\bks-\bkz\|\geq b_0$. Hence $\liminf_{M,T\rightarrow\infty}\|\bks-\bkz\|\geq b_0>0$, proving the result.
\end{proof}

\subsection{Placebo Inference with Few Treated Markets\label{sec:few_treated_inference}}
Some of the merger parameters are identified from only a small number of treated markets; see the discussion in Section \ref{sec:data}. For inference on these parameters, we follow \citet{ConleyTaber2011}, while maintaining the assumption that the merger is exogenous. The central idea is to use the large pool of untreated markets to construct a reference distribution for the merger-effect estimates under the null of no merger effect.

The intuition is straightforward. Because only a few markets are treated, conventional large-sample approximations based on the number of treated markets may poorly capture sampling uncertainty. We instead ask how large an estimated merger effect would arise if markets that were in fact untreated were assigned the merger treatment. In each placebo replication, we select the same number of markets as were actually treated, assign them the actual merger date, recompute the SDiD weights under this placebo assignment, and re-estimate the structural model. Repeating this exercise produces a distribution of placebo estimates that reflects the variation in estimated merger effects generated by untreated markets. Because the placebo distribution need not be centered at zero in finite samples, we center both the observed estimate and the placebo estimates at the mean of the placebo distribution when conducting inference. Importantly, we re-estimate the SDiD weights in every placebo replication, so the reference distribution incorporates variation from constructing the synthetic comparison group as well as from estimating the structural parameters.

For each merger and sample, let $\boldsymbol{\psi}$ collect the two primitive merger effects: the efficiency effect and the change in strategic conduct. The two null hypotheses of interest are that (i) the merger does not affect efficiency and (ii) the merger does not affect conduct. We implement the placebo procedure as follows:
\begin{enumerate}
\item Estimate the model using the observed treatment assignment and obtain $\widehat{\boldsymbol{\psi}}$.
\item From the untreated markets, draw $M_T$ markets without replacement and designate them as placebo treated markets, where $M_T$ is the number of treated markets.
\item Assign the merger date to the placebo-treated markets so the placebo assignment has the same treatment date and number of treated markets as the observed merger.
\item Using the placebo assignment, solve the market weights of Equation (\ref{eq:w_opt}).
\item Re-estimate the Synthetic GMM model using the new weights to obtain $\widehat{\boldsymbol{\psi}}^{(b)}$.
\item Repeat Steps 2--5 for $b=1,\ldots, B=100$. The empirical distribution $\{\widehat{\boldsymbol{\psi}}^{(b)}\}_{b=1}^{B}$ provides the reference distribution for inference under the null of no merger effect.
\end{enumerate}

Each placebo replication therefore solves a new synthetic-control problem before re-estimating the structural parameters, so the procedure incorporates variation in estimating the market weights.\footnote{For GMM, we use the same placebo assignments and re-estimate the model in every replication.}
The two merger effects are estimated jointly, so the placebo draws also provide their joint reference distribution. Let $\overline{\boldsymbol{\psi}}=\frac{1}{B}\sum_{b=1}^{B}\widehat{\boldsymbol{\psi}}^{(b)}$ denote the mean of the placebo estimates, and define their covariance matrix $\widehat{\Sigma}_{\psi}=\frac{1}{B-1}\sum_{b=1}^{B}\left(\widehat{\boldsymbol{\psi}}^{(b)}-\overline{\boldsymbol{\psi}}\right)\left(\widehat{\boldsymbol{\psi}}^{(b)}-\overline{\boldsymbol{\psi}}\right)^{\top}$.
Inference on each economic channel can first be conducted separately using the corresponding marginal placebo distribution. For example, for the efficiency effect in the linear model, the two-sided placebo $p$-value for the null of no merger-induced efficiency effect is 
$$\widehat p_{\theta_3}=\frac{1+\sum_{b=1}^{B}\mathbf{1}\!\left\{\left|\widehat\theta_3^{(b)}-B^{-1}\sum_{b=1}^{B}\widehat\theta_3^{(b)}\right|\geq\left|\widehat\theta_3-B^{-1}\sum_{b'=1}^{B}\widehat\theta_3^{(b')}\right|\right\}}{B+1},$$ with an analogous expression for the conduct effect.\footnote{The 95 percent placebo interval of the tables is the set of values $\psi_0$ that this test does not reject at the 5 percent level when $\widehat\theta_3-\bar\theta_3$ is replaced by $\widehat\theta_3-\bar\theta_3-\psi_0$, where $\bar\theta_3=B^{-1}\sum_{b=1}^{B}\widehat\theta_3^{(b)}$: it is $(\widehat\theta_3-\bar\theta_3)\pm q$, with $q$ the $\lfloor 0.05(B+1)\rfloor$-th largest of the placebo deviations $|\widehat\theta_3^{(b)}-\bar\theta_3|$ (the fifth largest when $B=100$) and $B$ the number of draws that return an estimate. The interval is centered at the estimate net of the placebo mean, so it excludes zero exactly when the placebo $p$-value is at most 0.05. For a quantity that is not zero under a placebo assignment, such as a pass-through, the interval is centered at the estimate itself: it is the estimate plus or minus $q$, with $q$ computed in the same way from the placebo draws of that quantity.}

We also test the joint null that the merger affects neither efficiency nor strategic conduct. Because the two effects may be correlated, we use the covariance matrix of the placebo estimates. Define $S=\left(\widehat{\boldsymbol{\psi}}-\overline{\boldsymbol{\psi}}\right)^{\top}\widehat{\Sigma}_{\psi}^{-1}\left(\widehat{\boldsymbol{\psi}}-\overline{\boldsymbol{\psi}}\right),$ and, for each placebo replication, $S^{(b)}:=(\widehat{\boldsymbol{\psi}}^{(b)}-\overline{\boldsymbol{\psi}})^{\top}\widehat{\Sigma}_{\psi}^{-1}(\widehat{\boldsymbol{\psi}}^{(b)}-\overline{\boldsymbol{\psi}}).$
The corresponding placebo $p$-value is $\widehat p_{\mathrm{joint}}=\frac{1+\sum_{b=1}^{B}\mathbf{1}\{S^{(b)}\geq S\}}{B+1}.$

Finally, we conduct inference directly on the merger-induced price effect. Because the price effect is a function of the estimated structural parameters, we recompute it in every placebo replication rather than holding the remaining parameters fixed. In the linear model, $\widehat{\beta}^{RF,(b)}=\frac{\widehat{\theta}_3^{(b)}}{1-\widehat{\widetilde{\gamma}}_1^{(b)}\widehat{\alpha}_1^{(b)}},$ and in the log-linear model, $\widehat{\beta}_{\log}^{RF,(b)}=
\frac{\widehat{\theta}_3^{(b)}-\left(\widehat{\Lambda}_{\mathrm{post}}^{(b)}-\widehat{\Lambda}_{\mathrm{pre}}^{(b)}\right)}{1-\widehat{\alpha}_1^{(b)}\widehat{\theta}_1^{(b)}}.$
For either specification, letting $\widehat\beta^{(b)}$ denote the corresponding placebo price effect, inference on the null of no merger-induced price effect is based on $$\widehat p_{\beta}=\frac{1+\sum_{b=1}^{B}\mathbf{1}\!\left\{\left|\widehat\beta^{(b)}-B^{-1}\sum_{b'=1}^{B}\widehat\beta^{(b)}\right|\geq\left|\widehat\beta-B^{-1}\sum_{b^{''}=1}^{B}\widehat\beta^{(b)}\right|\right\}}{B+1}.$$

\section{Monte Carlo Design and Additional Results\label{app:mc_design_appendix}}

This appendix provides the complete design and additional results for the Monte Carlo experiments in Section \ref{sec:montecarlo}. Every simulation uses artificial panels generated from the linear model of Section \ref{sec:structural_retrospective}. The merger efficiency effect, the conduct change, and all other parameters are specified by the design and are therefore known. No airline data enter the simulations. The purpose of the exercise is to evaluate the estimators and the placebo inference procedure under controlled violations of the supply-side moment conditions. 

\subsection{Design}

\paragraph{Demand, costs, and equilibrium.}
We generate panels of $M_T$ treated and $M_C$ control markets,
observed for $T_0$ pre-merger and $T_1$ post-merger periods.
The $M_T$ markets are treated, and $I_{mt}$ equals one
for these markets after the merger.

Demand follows
$Q_{mt}
=\alpha_0+\alpha_1P_{mt}
+\alpha_2Z_{mt}+\alpha_3Z^{(2)}_{mt}
+\varphi_{mt}$,
where
$(\alpha_0,\alpha_1,\alpha_2,\alpha_3)=(10,-2,1,1)$,
$Z_{mt}$ and $Z^{(2)}_{mt}$ are independent standard normal
demand shifters, and
$\varphi_{mt}\sim {\mathcal N}(0,0.5^2)$.
The second demand shifter provides relevance for the supply
instrument introduced below.
Marginal cost is $C_{mt}=\theta_0+\theta_1Q_{mt}+\theta_2W_{mt}+\theta_3I_{mt}+\vartheta_{mt}$,
where $(\theta_0,\theta_1,\theta_2,\theta_3)
=(2,0.3,0.5,-0.5)$
and $W_{mt}\sim {\mathcal N}(0,1)$.
The corresponding pricing equation following Equation (\ref{eq:systems}) is
$P_{mt}
=\theta_0+\tilde{\gamma}_{mt}Q_{mt}
+\theta_2W_{mt}+\theta_3I_{mt}
+\vartheta_{mt}$,
where
$\tilde{\gamma}_{mt}
=\theta_1-\lambda(I_{mt})/\alpha_1$.
We set pre-merger conduct to $\lambda_{\text{pre}}=0.3$ and the
merger-induced change to $\Delta\lambda=0.15$.
Thus, $\tilde{\gamma}_0=0.45$ and
$\tilde{\gamma}_1=0.525$.
The conduct change is common across treated markets.
Prices and quantities are obtained by solving the demand
and pricing equations jointly in each market and period.

The cost shock has a two-factor structure:
$\vartheta_{mt}=\varpi_{m1}F_{1t}+\varpi_{m2}F_{2t}+\varepsilon_{mt}$,
where $F_{1t}$ and $F_{2t}$ are independent AR(1)
processes with autoregressive coefficient $0.6$ and
unit unconditional variance (innovation variance $0.64$),
and $\varepsilon_{mt}\sim {\mathcal N}(0,0.5^2)$.
The first factor loading, $\varpi_{m1}$, determines whether a market
violates the supply-side moment condition.
It is zero in all treated markets and in valid control
markets, but nonzero in invalid controls.
The second, $\varpi_{m2}$, generates heterogeneity in
prices without directly generating
instrument invalidity.
For control markets, we set $\varpi_{m2}\sim {\mathcal N}(1,0.3^2)$, and for treated markets, we draw
$\varpi_{m2}\sim {\mathcal N}(1+0.3s,(0.3v)^2)$.
For the baseline, we set $(s,v)=(0,1)$, so the treated and
control distributions are the same, and we also consider $s\in\{1,2\}$, and 
$v\in\{0.5,2\}$.

\paragraph{Supply instrument and invalidity.}
The supply instrument is
$A_{mt}
=(\nu_{mt}+dZ^{(2)}_{mt}+cF_{1t})/\sigma_A$,
where $\nu_{mt}$ is independent of the other shocks,
$d=1$, and $\sigma_A$ normalizes the instrument's
variance.
The demand shifter $Z^{(2)}_{mt}$ makes the instrument
relevant, while its common-factor component can make
it endogenous.
In particular, the market-specific violation of the
supply moment condition, as defined in Equation \eqref{eq:bias_syn},  is
$\delta_{mt}:=\mathbb{E}[A_{mt}\vartheta_{mt}]
=c \sigma_F^2 \varpi_{m1}/\sigma_A$,
where $\sigma_F^2$ is the variance of $F_{1t}$.
Hence, $\delta_m=0$ when $\varpi_{m1}=0$.
By construction, the supply moments are valid in
every treated market and in a fraction $1-\pi$ of
control markets.

We select the fraction $\pi$ of invalid controls
according to their distance from the treated markets
in the second factor loading, $\varpi_{m2}$.
Specifically, the invalid controls are the
$\pi M_C$ markets farthest from the treated markets
in this dimension. For these markets, we set $\varpi_{m1}=w_m x_m$,
where $x_m\sim {\mathcal N}\bigl(\mu_I,(\kappa\mu_I)^2\bigr)$
and $\mu_I=0.625\times \mathfrak{D}$, and we choose $\mathfrak{D}$ later. The weight $w_m$ increases with the market's distance
from the treated markets in $\varpi_{m2}$ and is
normalized to average one among invalid controls.
All other markets have $\varpi_{m1}=0$.

This construction links instrument invalidity to
pre-merger price dynamics in two ways.
Markets farther from the treated group in the second
factor loading are more likely to have invalid moments,
and their exposure to the first factor is larger.
Because both factors enter the cost shock, the
heterogeneity associated with invalidity is also
reflected in equilibrium prices.
The synthetic weights can therefore distinguish
more credible controls using pre-merger price
similarity, as required by the mechanism underlying
Theorem \ref{thm:syn_bias}.

Three parameters affect the moment conditions for $\pi M_C$ markets. 
The fraction $\pi$ determines how many controls
have invalid supply moments.
The parameter $\Delta$ is the mean moment violation
among invalid controls, with baseline value $0.6$
and additional values $0.1$ and $0.3$.
We vary $c$ in $\delta_{mt}$ to get the target $\Delta$, and 
its baseline value is $0.48$.
The parameter $\mathfrak{D}\in\{3,1,0.3\}$ governs how visible
the heterogeneity associated with invalidity is
in pre-merger prices.
Finally, $\kappa\in\{0.25,1,2\}$ controls the
dispersion of invalidity across markets, allowing
some violations to have the opposite sign when
dispersion is sufficiently large.

\paragraph{Different Cases.}
The main experiments vary the number of markets,
the share and characteristics of invalid controls,
and the length of the pre-merger period.
Specifically, we consider
$(M_T,M_C)\in\{(3,120),(10,400),(40,120),(40,900)\}$,
$\pi\in\{0,0.3,0.5\}$,
$\mathfrak{D}\in\{3,1,0.3\}$,
$\kappa\in\{0.25,1,2\}$, and
$T_0\in\{8,16,32\}$, holding $T_1=8$.
The baseline uses $(M_T,M_C)=(10,400)$, $\pi=0.3$,
$\mathfrak{D}=3$, $\kappa=1$, $T_1=8$ and $T_0 \in\{8, 16, 32\}$.
The resulting grid contains 51 distinct designs
with $\theta_3=-0.5$ and $\Delta\lambda=0.15$,
and 51 corresponding designs imposing the joint
null of no efficiency effect and no conduct change.
Each design uses $R=500$ Monte Carlo replications
and $B=200$ placebo assignments per replication.

\paragraph{Performance and Inference.}
We evaluate the estimators using the absolute median
bias and root mean squared error (RMSE), defined in
Section \ref{sec:montecarlo}.
Statistics are computed within each design before
being averaged across designs.
For the inference exercises below, we additionally
report rejection rates and coverage of the
placebo-based confidence intervals; see Appendix \ref{sec:few_treated_inference}.\footnote{Because the pricing equation contains two endogenous regressors, we assess the joint relevance of the excluded instruments using the Cragg--Donald statistic \citep{CraggDonald1993}. In the main simulation, the mean statistic
within a design ranges from 62 to 806, with a median of 290, indicating strong first stages.
}
In each Monte Carlo replication, we draw $B=200$ placebo assignments from the control markets, keeping the number of treated markets and
the merger date fixed. For each assignment, we reconstruct the
synthetic weights and re-estimate the model, obtaining a placebo
distribution of estimated merger effects. We use this distribution
to calculate placebo $p$-values and 95 percent confidence intervals.

Following Appendix \ref{sec:few_treated_inference}, the placebo test recenters both the observed estimate and the placebo estimates at the mean of the placebo distribution. The corresponding
confidence interval for a merger effect $\psi$ is $(\hat\psi_r-\bar\psi_r)\pm q_r$, where $\bar\psi_r$ is the mean placebo estimate and $q_r$ is the $\lfloor 0.05(B+1)\rfloor$-th largest absolute deviation of
the placebo estimates from their mean. We measure size by the rejection rate in designs with no merger effects and coverage by the fraction of replications in which
the confidence interval contains the true parameter $\psi_0$.

For the demand slope $\alpha_1$, which is not zero under placebo assignments, we instead center the interval at the estimate, $\hat\alpha_{1,r}\pm q_r$. We also use intervals centered at the estimate,
$\hat\psi_r\pm q_r$, for the coverage results in Tables \ref{tab:mc_firststage}. These intervals have the same width as the recentered intervals
but differ in their centers.

\subsection{Performance}

The central experiment examines how the estimators perform as the
fraction of control markets with invalid supply moments increases.
When all moments are valid ($\pi=0$), both GMM and Synthetic GMM
target the true parameters. As $\pi$ increases, standard GMM
becomes increasingly contaminated by invalid control-market
moments. Synthetic GMM is less sensitive to this contamination
because its weights favor controls with pre-merger price dynamics
similar to those of the treated markets.\footnote{We compared our one-step estimator with two-step GMM, and find that one-step method has better performance, motivating our choice of one-step estimator in themain analysis.}

The main-text results illustrate this difference. For the
efficiency parameter $\theta_3$, the absolute median bias of
standard GMM rises from $0.100$ when all controls are valid to
$0.293$ when 30 percent are invalid and $0.525$ when 50 percent
are invalid. The corresponding biases for Synthetic GMM are
$0.074$, $0.101$, and $0.074$. Synthetic GMM also has substantially
lower RMSE, ranging from 29.1 to 42.8 percent of standard GMM's
RMSE across these designs. Over the full simulation grid, the
RMSEs for the efficiency effect are 1.12 for Synthetic GMM and
3.53 for standard GMM; for the conduct change, they are 0.61
and 1.96, respectively. The aggregate RMSEs are influenced by
a small number of extreme standard-GMM estimates, which is why
the main text emphasizes comparisons by the fraction of
invalid controls.

The weights help explain these results. Synthetic GMM places
greater weight on controls whose pre-merger price paths resemble
those of the treated markets. In our design, these markets also
tend to have smaller violations of the supply moment conditions.
Among the 60 designs with 30 percent invalid controls, invalid
markets receive, on average, only 19.3 percent of the total
control weight. Moreover, the ratio of weighted to unweighted
moment contamination falls from $0.859$ when $\mathfrak{D}=0.3$ to
$0.153$ when $\mathfrak{D}=3$. Thus, the gains are larger when pre-merger
prices provide a clearer signal of instrument invalidity.

\paragraph{When price dynamics are uninformative.}
We next examine whether the gains from Synthetic GMM depend
on pre-merger prices being informative about instrument
invalidity. As a negative benchmark, we construct invalid
supply moments through the idiosyncratic cost shock rather
than the common factor loading. Specifically, we set
$\varpi_{m1}=0$ and replace the supply instrument with $A_{mt}=
\frac{\nu_{mt}+dZ^{(2)}_{mt}+g_m\varepsilon_{mt}}
{\sigma_A}.$ 
We also vary whether invalidity is related to the factor loading that drives pre-merger price similarity.

Table \ref{tab:mc_visibility} reports the share of synthetic
weight assigned to invalid untreated markets. When invalidity is
related to pre-merger price similarity, these markets receive
substantially less than their share of the control
pool. When invalidity operates through the idiosyncratic shock and is unrelated to price similarity, however, invalid controls receive 30.1\% of the weight, essentially their population share. Thus, SDiD weighting does not downweight invalid controls when pre-merger
prices contain no information about their validity.

\begin{table}[t!!]
\centering
\caption{Monte Carlo: Weight Assigned to Invalid Controls
\label{tab:mc_visibility}}
\begin{threeparttable}
\setlength{\tabcolsep}{8pt}
\begin{tabular}{llc}
\toprule
Source of invalidity
& Liked to Price similarity
& Weight on invalid controls \\
\midrule
Factor loading       & Yes & 0.145 \\
Factor loading       & No  & 0.193 \\
Idiosyncratic shock  & Yes & 0.179 \\
Idiosyncratic shock  & No  & 0.301 \\
\bottomrule
\end{tabular}
\begin{figurenotes}
The table reports the share of total synthetic weight
assigned to invalid control markets, which account for
30\% of the total untreated market pool. The experiment uses
$\Delta=0.20$ and 1,000 Monte Carlo replications.
Monte Carlo standard errors range from 0.001 to 0.002.
``Linked to Price similarity'' indicates that invalid
controls are selected according to their distance from
treated markets in the second factor loading. The
experiment imposes a difference of one between treated
and control loadings in this dimension.
\end{figurenotes}
\end{threeparttable}
\end{table}

\paragraph{Placebo inference.}
We next examine the size and coverage of the placebo procedure.
The first question is whether the test rejects too often when
the true merger effects are zero. Table \ref{tab:mc_size}
reports rejection rates under the joint null of no efficiency
effect and no conduct change.
In the main design, where the instruments are strong,
the placebo test has rejection rates at or below the nominal
levels. At a nominal significance level of 5 percent, rejection
rates range from 3.1 to 4.1 percent across the estimators and
parameters considered. The procedure is therefore 
conservative in these designs. Recentering both the observed
estimate and the placebo estimates, as in our proposed
procedure, also performs better than recentering only the
placebo estimates. The latter approach produces rejection
rates as high as 8.6 percent at the 5 percent nominal level.

These results depend on instrument strength. To examine this
dependence, we separately vary the relevance of the excluded
instruments while holding the other principal features of the
design fixed. When the mean Cragg--Donald statistic is below
approximately 6, the placebo test can over-reject. Rejection
rates are mixed around a statistic of 10. At mean statistics
of approximately 18 or higher, the test does not exhibit
systematic over-rejection and is often conservative.
Table \ref{tab:mc_size} summarizes these findings.

\begin{table}[t!!]
\centering
\caption{Monte Carlo: Size of the Placebo Test Under the Null
\label{tab:mc_size}}
\begin{threeparttable}
\setlength{\tabcolsep}{7pt}
\begin{tabular}{lccc}
\toprule
& \multicolumn{3}{c}{Nominal significance level} \\
\cmidrule(lr){2-4}
Procedure & 5\% & 10\% & 1\% \\
\midrule
\multicolumn{4}{l}{\emph{Panel A. Main Simulation Design}} \\
Both estimates recentered (proposed)
    & 0.031--0.041 & 0.067--0.088 & 0.006--0.008 \\
Placebo estimates recentered only
    & 0.046--0.086 & 0.087--0.139 & 0.009--0.031 \\
    &&&\\
\midrule
\multicolumn{4}{l}{\emph{Panel B. Proposed procedure by instrument strength}} \\
First-stage statistic & \multicolumn{3}{l}{Behavior at the 5\% level} \\
\midrule
1.4--6.1   & \multicolumn{3}{l}{Over-rejection in many designs} \\
$\approx$10  & \multicolumn{3}{l}{Mixed results} \\
18--118    & \multicolumn{3}{l}{No systematic over-rejection; often conservative} \\
213        & \multicolumn{3}{l}{Conservative} \\
\bottomrule
\end{tabular}
\begin{figurenotes}
Rejection rates under the joint null of no efficiency effect and
no conduct change. Panel A reports the range across six
estimator-by-parameter averages in the 51 null designs of the
main grid. The mean Cragg--Donald statistic ranges from
approximately 62 to 806 in these designs. Panel B summarizes
the separate first-stage experiment at
$\Delta=0.05$, $(M_T,M_C)=(10,400)$, and $T_0=T_1=8$.
At the 5 percent nominal level, a rejection rate between
0.031 and 0.069 is within two Monte Carlo standard errors
of the nominal rate when $R=500$.
\end{figurenotes}
\end{threeparttable}
\end{table}

\paragraph{Coverage and instrument strength.}
We also examine whether the placebo confidence intervals
contain the true parameters and how estimation accuracy
changes with instrument strength. Across the 51 main-grid
designs with nonzero merger effects, coverage is 94.5 percent
for the efficiency effect and 97.1 percent for the price effect.
Across all 102 designs, the corresponding coverage rates are
95.5 and 96.8 percent.

Table \ref{tab:mc_firststage} examines how estimation accuracy
varies with instrument strength. As the mean Cragg--Donald
statistic increases, the estimates move toward their true
values. For example, when the statistic is approximately 10,
the median efficiency estimate is $0.41$, compared with the
true value of $-0.5$. At a statistic of approximately 116,
the median estimate is $-0.40$, and at approximately 209
it is $-0.42$. The conduct estimates show a similar improvement,
moving toward the true change of $0.15$. These results
underscore the importance of instrument strength for
accurate structural estimation.

\begin{table}[t!!]
\centering
\caption{Monte Carlo: Synthetic GMM and Instrument Strength
\label{tab:mc_firststage}}
\begin{threeparttable}
\setlength{\tabcolsep}{8pt}
\begin{tabular}{lcc}
\toprule
Mean first stage
& Efficiency $\theta_3$
& Conduct change $\Delta\lambda$ \\
& (true $-0.5$)
& (true $0.15$) \\
\midrule
1.4--6.8 & $0.26$ to $0.84$ & $-0.23$ to $-0.59$ \\
10.2  & $0.41$ & $-0.38$ \\
17.8  & $0.20$ & $-0.27$ \\
24.5  & $0.04$ & $-0.16$ \\
32.2  & $-0.00$ & $-0.16$ \\
40.1  & $-0.13$ & $-0.05$ \\
50.0  & $-0.25$ & $0.03$ \\
62.9  & $-0.30$ & $0.03$ \\
116.1 & $-0.40$ & $0.09$ \\
209   & $-0.42$ & $0.11$ \\
\bottomrule
\end{tabular}
\begin{figurenotes}
Synthetic GMM estimates from the first-stage experiment at
$\Delta=0.05$, $(M_T,M_C)=(10,400)$, and $T_0=T_1=8$.
Each design uses $R=500$ Monte Carlo replications.
The first-stage statistic is the mean Cragg--Donald statistic
within a design. Entries report median estimates,
computed as the true parameter plus the median estimation error.
\end{figurenotes}
\end{threeparttable}
\end{table}

\paragraph{Sample size.}
We also examine whether estimation becomes more precise
as the number of markets increases. We first vary the number
of treated markets while keeping all supply moments valid
($\pi=0$). With 10, 20, 40, and 80 treated markets, the
standard deviation of the Synthetic GMM estimate of the efficiency parameter $\theta_3$ 
falls from $0.95$ to $0.65$, $0.51$, and $0.40$, respectively.
The corresponding standard deviations for standard GMM also fall from 
$1.29$, $0.76$, $0.59$, to $0.52$, but is higher than Synthetic GMM. 

Increasing the number of control markets has a smaller effect
on bias. With 40 treated markets,
the median bias of Synthetic GMM is $0.011$ with 120 controls
and $0.019$ with 900 controls. Increasing the pre-merger
window likewise produces little change in median bias. These findings suggest that increasing
the number of treated markets is particularly useful for
precision, while adding controls or pre-merger periods does
not necessarily improve estimation, likely once the available
controls provide a sufficiently close match.

\section{Additional Empirical Results}

This appendix reports the demand and pricing estimates underlying
Section \ref{sec:structural_estimates}, together with additional
results for the merger effects in Table \ref{tab:merger_effects}.
We report standard errors and placebo confidence intervals, examine
how conduct changes with the number of competitors, and decompose
the estimated price effects into their cost and market-power
components.

\subsection{Demand, Elasticities, and Pricing Slopes}
\label{sec:appendix_gmm_full}

Table \ref{tab:conduct_all_markets} reports the GMM estimates of
the demand and pricing parameters for the linear and log-linear
models. The demand equation is estimated without synthetic weights,
so its estimates are common to GMM and Synthetic GMM. The table
also reports implied demand elasticities, pricing slopes,
pass-through rates, and merger price effects.

\paragraph{Demand estimates.}
The estimated demand slope is negative and statistically
significant in all four mergers under both functional forms.
In the linear model, the estimates range from $-0.50$ for
DL--NW to $-1.22$ for AA--US. Evaluated at treated markets'
pre-merger means, the implied elasticities are $-1.80$,
$-1.26$, $-0.39$, and $-2.65$ for DL--NW, UA--CO, WN--FL,
and AA--US, respectively. The log-linear estimates imply
elasticities of $-1.59$, $-1.75$, $-2.93$, and $-2.95$.
The two specifications therefore yield broadly similar
elasticities for three mergers, but differ substantially
for WN--FL.

\paragraph{Pricing estimates.}
The pricing slope measures the response of price to quantity,
holding the observed cost shifters fixed. In the linear model,
this coefficient combines the marginal-cost slope with
pre-merger conduct; it need not equal the marginal-cost slope.
In the log-linear model, conduct enters the pricing intercept,
so the reported slope is the marginal-cost coefficient itself.
The estimated pass-through rates are generally high,
ranging from $0.78$ to $0.93$ in the reported linear fits
and from $0.91$ to $0.94$ in the log-linear model.
The corresponding price effects are reported in the final
column. The merger effects on conduct that vary with the number of competitors,
are examined below.

\begin{table}[ht!]
\centering
\caption{GMM Estimates of Demand and Pricing Parameters}
\label{tab:conduct_all_markets}
\begin{threeparttable}
\small
\setlength{\tabcolsep}{4pt}
\begin{tabular}{lcccccc}
\toprule
& & \multicolumn{2}{c}{Demand elasticity}
& Pricing & Pass- & Price \\
Merger & Demand slope $\alpha_1$ & Control & Treated
& slope & through & effect \\
\cmidrule(lr){3-4}
\midrule
\multicolumn{7}{l}{\emph{Panel A. Linear model}} \\
DL--NW & $-0.499$ ($0.203$) & $-0.62$ & $-1.80$
& $0.086$ ($0.020$) & $0.784$ ($0.107$)
& $-0.038$ ($0.057$) \\
UA--CO & $-0.973$ ($0.403$) & $-1.14$ & $-1.26$
& n/a & n/a & n/a \\
WN--FL & $-0.541$ ($0.203$) & $-0.72$ & $-0.39$
& $0.075$ ($0.028$) & $0.933$ ($0.032$)
& $0.283$ ($0.086$) \\
AA--US & $-1.222$ ($0.300$) & $-1.74$ & $-2.65$
& $0.008$ ($0.021$) & $0.823$ ($0.088$)
& $0.020$ ($0.043$) \\
\addlinespace
\multicolumn{7}{l}{\emph{Panel B. Log-linear model}} \\
DL--NW & $-1.594$ ($0.346$) & \multicolumn{2}{c}{$-1.59$}
& $0.055$ ($0.010$) & $0.919$ ($0.021$)
& $0.007$ ($0.020$) \\
UA--CO & $-1.746$ ($0.455$) & \multicolumn{2}{c}{$-1.75$}
& $0.044$ ($0.016$) & $0.928$ ($0.029$)
& $0.015$ ($0.013$) \\
WN--FL & $-2.926$ ($0.387$) & \multicolumn{2}{c}{$-2.93$}
& $0.036$ ($0.011$) & $0.906$ ($0.026$)
& $0.078$ ($0.018$) \\
AA--US & $-2.946$ ($0.346$) & \multicolumn{2}{c}{$-2.95$}
& $0.020$ ($0.011$) & $0.944$ ($0.029$)
& $0.012$ ($0.018$) \\
\bottomrule
\end{tabular}

\begin{tablenotes}[flushleft]
\item \footnotesize
\emph{Notes.} GMM estimates from the combined sample, with
equal weight on each market-quarter. Standard errors clustered
at the city-market level are in parentheses; standard errors
for derived quantities use the delta method. Demand estimates
are the same for GMM and Synthetic GMM. In Panel A, prices are
measured in hundreds of dollars, and the demand elasticities
are evaluated at the control and treated markets' pre-merger
means. The pricing slope is
$\tilde{\gamma}_C=\theta_1-\lambda_C/\alpha_1$,
and pass-through is $1/(1-\tilde{\gamma}_1\alpha_1)$.
Panel A imposes a common conduct change across treated markets.
In Panel B, $\alpha_1$ is the constant demand elasticity,
the pricing slope is $\theta_1$, and pass-through is
$1/(1-\alpha_1\theta_1)$, and the price effects are in
log points. The marked UA--CO cells are suppressed because
the pricing equation fails the identification diagnostic.
\end{tablenotes}
\end{threeparttable}
\end{table}

\subsection{Merger Effects and Their Decomposition}
\label{sec:appendix_se}

We next report the structural merger effects with their
standard errors and placebo confidence intervals.
Table \ref{tab:conductN_main} focuses on the linear model,
allowing the merger-induced conduct change to depend on
the number of competitors. Table \ref{tab:weighted_system}
shows the price effects under structural and DiD approaches, including the implied price effects for untreated markets.

\paragraph{Conduct and cost effects.}
Table \ref{tab:conductN_main} evaluates the conduct change
at three, four, and five competitors. For DL--NW and AA--US,
the point estimates generally indicate a positive conduct
change that becomes smaller as the number of competitors
increases. For WN--FL, the estimated conduct change is
negative throughout this range. The estimates are less
precise for DL--NW and AA--US, whereas the WN--FL estimates
are statistically different from zero under standard GMM.
As before, the Synthetic GMM estimates for the UA--CO merger are not reported.
The cost estimates show the largest merger-induced increase
for WN--FL. The estimated cost effects for DL--NW and
AA--US are small, while UA--CO has a positive estimate
under standard GMM. The final column combines the cost
and conduct changes into price effect on
treated markets. The stimates are similar
for both methods.

\begin{table}[t!!]
\caption{Structural Merger Effects: Linear Model}
\label{tab:conductN_main}
\hspace{-0.7in}\begin{tabular}{lcccccc}
\toprule
& \multicolumn{3}{c}{Conduct change $\Delta\lambda(N)$}
& Cost effect & Pass- & Price effect \\
\cmidrule(lr){2-4}
Merger & $N=3$ & $N=4$ & $N=5$
& (\$) & through & (\$) \\
\midrule
DL--NW & & & & & & \\
GMM
& $0.121$ & $0.085$ & $0.058$ & $-3.7$ & $0.875$ & $0.5$ \\
& ($0.172$) & ($0.115$) & ($0.075$)
& ($8.5$) & ($0.119$) & ($5.0$) \\
& $[0.015, 0.181]$
& $[0.001, 0.136]$
& $[0.000, 0.091]$
& $[-9.7, 10.5]$
& $[0.730, 1.020]$
& $[-6.0, 9.6]$ \\
S-GMM
& $0.098$ & $0.064$ & $0.038$ & $-2.4$ & $0.913$ & $0.4$ \\
& ($0.162$) & ($0.109$) & ($0.074$)
& ($8.0$) & ($0.132$) & ($4.8$) \\
& $[-0.013, 0.165]$
& $[-0.014, 0.110]$
& $[-0.020, 0.072]$
& $[-7.9, 11.7]$
& $[0.767, 1.059]$
& $[-6.0, 10.0]$ \\
\addlinespace

UA--CO & & & & & & \\
GMM
& $0.247^{\dagger}$ & $0.166$ & $0.103$
& $6.2$ & $0.818$ & $6.2$ \\
& ($0.285$) & ($0.196$) & ($0.126$)
& ($2.9$) & ($0.130$) & ($3.7$) \\
& $[-0.598, 1.144]$
& $[-0.413, 0.782]$
& $[-0.249, 0.481]$
& $[-21.5, 33.4]$
& $[-1.361, 2.996]$
& $[-16.2, 37.1]$ \\
S-GMM$^\spadesuit$
& n/a & n/a & n/a
& n/a & n/a & n/a \\
\addlinespace

WN--FL & & & & & & \\
GMM
& $-0.121$ & $-0.089$ & $-0.065$
& $38.0$ & $0.976$ & $10.3$ \\
& ($0.077$) & ($0.055$) & ($0.039$)
& ($12.1$) & ($0.032$) & ($4.7$) \\
& $[-0.220, -0.004]$
& $[-0.152, -0.015]$
& $[-0.111, -0.011]$
& $[24.9, 48.1]$
& $[0.589, 1.364]$
& $[2.9, 16.5]$ \\
S-GMM$^{\ddagger}$
& $-0.072$ & $-0.053$ & $-0.038$
& $27.2$ & $0.969$ & $10.9$ \\
& ($0.055$) & ($0.039$) & ($0.027$)
& ($10.2$) & ($0.025$) & ($4.4$) \\
& $[-0.175, 0.049]$
& $[-0.125, 0.031]$
& $[-0.085, 0.016]$
& $[14.5, 37.2]$
& $[0.611, 1.327]$
& $[3.6, 17.2]$ \\
\addlinespace

AA--US & & & & & & \\
GMM
& $0.132$ & $0.078$ & $0.036$
& $-0.2$ & $0.693$ & $2.9$ \\
& ($0.176$) & ($0.095$) & ($0.036$)
& ($6.5$) & ($0.091$) & ($3.4$) \\
& $[-0.130, 0.420]$
& $[-0.090, 0.260]$
& $[-0.084, 0.163]$
& $[-14.6, 11.7]$
& $[-0.466, 1.851]$
& $[-7.8, 12.6]$ \\
S-GMM
& $0.145$ & $0.086$ & $0.040$
& $-0.4$ & $0.742$ & $3.2$ \\
& ($0.167$) & ($0.091$) & ($0.036$)
& ($6.2$) & ($0.099$) & ($3.6$) \\
& $[-0.087, 0.409]$
& $[-0.071, 0.262]$
& $[-0.080, 0.169]$
& $[-13.9, 10.6]$
& $[-0.096, 1.580]$
& $[-8.0, 13.8]$ \\
\bottomrule
\end{tabular}
\begin{figurenotes}The table reports estimates from the linear model, estimated separately for each merger. Conduct changes are evaluated at $N\in\{3, 4,5\}$. The cost effect is the average merger-induced change in marginal cost. Pass-through and price effects are averaged across treated markets. Parentheses report cluster-robust delta-method standard errors, and brackets report 95\% placebo confidence intervals based on $B=100$ assignments. Placebo intervals are centered at the estimate net of the mean placebo estimate, except for pass-through intervals, which are centered at the estimate.
$^{\dagger}$: The UA--CO conduct estimate at $N=3$ extrapolates below the range observed in treated markets. $^{\spadesuit}$: UA--CO Synthetic GMM estimates are omitted because the estimated pass-through is unstable. $^{\ddagger}$: The WN--FL Synthetic GMM pricing equation has a Kleibergen--Paap statistic of $4.01$, indicating weak instrument relevance; these estimates should be interpreted with caution.
\end{figurenotes}
\end{table}

\paragraph{Price effects across estimators.}
Table \ref{tab:weighted_system} compares the price effects
from first-difference DiD, synthetic DiD, GMM, and Synthetic
GMM.  The comparisons are organized so that the estimators
within each pair use the same control markets and weights:
standard GMM is compared with conventional DiD, and Synthetic
GMM with synthetic DiD. 
The four approaches give broadly similar conclusions for
DL--NW and AA--US, with relatively small and imprecisely
estimated price effects. WN--FL is the clearest exception:
all four estimators indicate a positive price effect,
although its magnitude varies across specifications.
For UA--CO, the estimates are less stable, consistent
with the weaker identification  discussed above.

\paragraph{Cost and market-power components.}
The linear structural model also separates the total price
effect into changes in marginal cost and market power.
For WN--FL, both GMM and Synthetic GMM attribute a large
positive price effect to higher marginal costs, partly
offset by better conduct. Under Synthetic
GMM, these components are \$26.3 and $-\$15.4$,
respectively, yielding a net price increase of \$10.9.
For the other mergers, the components are
smaller and less precise.
In the log-linear model, we report the conduct gradient. The final panel extends the estimated 
effects to untreated markets. These 
effects differ from the treated-market effects because of 
market characteristics and market structure.

\begin{landscape}
\thispagestyle{empty}
\begin{table}[p]
\caption{Merger Price Effects and Their Decomposition
\label{tab:weighted_system}}
\scalebox{0.92}{
\begin{tabular}{lcccc|cccc}
\toprule
& \multicolumn{4}{c|}{Linear model (\$ per one-way trip)}
& \multicolumn{4}{c}{Log-linear model (log points)} \\
\cmidrule(lr){2-5}\cmidrule(lr){6-9}
& DiD & Structural & Efficiency & Market power
& DiD & Structural & Conduct & ATE \\
Merger
& price effect & price effect & component & component
& price effect & price effect & gradient & price effect \\
\midrule

\multicolumn{9}{l}{\emph{DL--NW}} \\

GMM
& $-1.4$ & $0.5$ & $-3.2$ & $3.7$
& $-0.004$ & $0.007$ & $-0.118$ & \\
& ($6.0$) & ($5.0$) & n/a & n/a
& ($0.026$) & ($0.020$) & ($0.064$) & \\
& $[-12.3, 8.1]$ & $[-6.0, 9.6]$
& $[-8.9, 10.3]$ & $[-7.8, 9.9]$
& $[-0.055, 0.040]$ & $[-0.026, 0.042]$
& $[-0.108, 0.012]$ & \\

S-GMM
& $3.5$ & $0.4$ & $-2.1$ & $2.6$
& $0.012$ & $0.000$ & $-0.107$ & $-0.002$ \\
& & ($4.8$) & ($7.2$) & ($5.6$)
& & ($0.020$) & ($0.066$) & ($0.020$) \\
& $[-2.6, 10.8]$ & $[-6.0, 10.0]$
& $[-8.1, 12.0]$ & $[-8.7, 8.8]$
& $[-0.020, 0.049]$ & $[-0.028, 0.033]$
& $[-0.097, 0.024]$ & $[-0.028, 0.030]$ \\

\addlinespace[0.8em]
\multicolumn{9}{l}{\emph{UA--CO}} \\

GMM

& $0.4$ & $6.2$ & $5.1$ & $1.1$

& $-0.009$ & $0.015$ & $-0.085$ & \\

& ($2.4$) & ($3.7$) & n/a&n/a

& ($0.010$) & ($0.013$) & ($0.078$) & \\

& $[-15.6, 14.8]$ & $[-16.2, 37.1]$

& $[-23.0, 45.9]$ & $[-36.1, 34.1]$

& $[-0.087, 0.062]$ & $[-0.045, 0.077]$

& $[-0.271, 0.063]$ & \\

S-GMM$^\spadesuit$

& $5.5$ & n/a & n/a & n/a

& $0.015$ & $0.010$ & $-0.088$ & $0.026$ \\

& & & & &

& ($0.011$) & ($0.079$) & ($0.020$) \\

& $[-5.9, 17.7]$ & & &

& $[-0.037, 0.073]$ & $[-0.048, 0.071]$

& $[-0.275, 0.058]$ & $[-0.038, 0.090]$ \\
\addlinespace[0.8em]
\multicolumn{9}{l}{\emph{WN--FL}} \\

GMM
& $6.4$ & $10.3$ & $37.1$ & $-26.8$
& $0.041$ & $0.078$ & $-0.181$ & \\
& ($2.3$) & ($4.7$) & n/a & n/a
& ($0.013$) & ($0.018$) & ($0.082$) & \\
& $[-2.7, 14.3]$ & $[2.9, 16.5]$
& $[23.2, 47.9]$ & $[-34.6, -17.2]$
& $[-0.003, 0.079]$ & $[0.047, 0.104]$
& $[-0.272, -0.116]$ & \\

S-GMM$^{\ddagger}$
& $8.4$ & $10.9$ & $26.3$ & $-15.4$
& $0.076$ & $0.065$ & $-0.129$ & $0.078$ \\
& & ($4.4$) & ($10.2$) & ($9.8$)
& & ($0.020$) & ($0.063$) & ($0.023$) \\
& $[1.4, 13.4]$ & $[3.6, 17.2]$
& $[13.4, 36.4]$ & $[-23.6, -5.3]$
& $[0.046, 0.098]$ & $[0.035, 0.090]$
& $[-0.220, -0.064]$ & $[0.044, 0.105]$ \\

\addlinespace[0.8em]
\multicolumn{9}{l}{\emph{AA--US}} \\

GMM
& $-0.2$ & $2.9$ & $-0.2$ & $3.1$
& $0.000$ & $0.012$ & $0.006$ & \\
& ($6.8$) & ($3.4$) & n/a & n/a
& ($0.033$) & ($0.018$) & n/a & \\
& $[-11.4, 9.6]$ & $[-7.8, 12.6]$
& $[-18.8, 13.1]$ & $[-16.8, 27.4]$
& $[-0.054, 0.047]$ & $[-0.022, 0.043]$
& $[-0.078, 0.108]$ & \\

S-GMM
& $4.8$ & $3.2$ & $-0.3$ & $3.5$
& $0.010$ & $0.012$ & $0.000$ & $0.012$ \\
& & ($3.6$) & ($4.6$) & ($3.0$)
& & ($0.017$) & ($0.048$) & ($0.017$) \\
& $[-3.6, 12.6]$ & $[-8.0, 13.8]$
& $[-17.0, 12.7]$ & $[-13.7, 23.9]$
& $[-0.026, 0.043]$ & $[-0.021, 0.043]$
& $[-0.077, 0.097]$ & $[-0.024, 0.045]$ \\

\bottomrule
\end{tabular}
}
\begin{figurenotes}
Merger effects estimated on the combined sample.
The linear model reports dollars per one-way trip; the
log-linear model reports log points. DiD estimates use
the same market weights as the corresponding GMM estimates.
Structural price effects are averages over treated markets.
In the linear model, the efficiency and market-power
components sum to the structural price effect.
The log-linear conduct gradient is
$\psi_N/(1-\alpha_1\theta_1)$.
ATE (Average Treatment Effects) apply the structural estimates to
untreated markets. Parentheses report cluster-robust
delta-method standard errors, where available; brackets
report 95 percent placebo intervals, recentered by the
mean placebo estimate. Synthetic DiD uses placebo
inference only. $^\spadesuit$: We do not report Synthetic GMM estimates for the UA-CO merger, and n.a. indicates unavailable standard errors.
$^{\ddagger}$ The WN--FL linear S-GMM pricing first-stage
statistic is $4.01$.
\end{figurenotes}

\end{table}
\end{landscape}

\end{document}